%% file: Asynch_Final_version.tex
\documentclass[12pt,draftcls, onecolumn]{IEEEtran}

\usepackage{times}
\usepackage{amsmath,dsfont, bm}
\usepackage{amssymb,amsthm}
\usepackage{epsfig,verbatim}
\usepackage{subfigure}
\usepackage{setspace}
\usepackage{color}
\usepackage{cite}
\usepackage{epstopdf}
\usepackage{graphics}
\usepackage{accents}
\usepackage{acronym}
\usepackage[bookmarks,breaklinks, colorlinks,pdfencoding=auto, psdextra]{hyperref}
\usepackage{soul}
\usepackage{enumitem}
\usepackage{mathrsfs}
\usepackage{bookmark}

\newtheorem{definition}{Definition}
\newtheorem{theorem}{Theorem}
\newtheorem*{theorem*}{Theorem}
\newtheorem{corollary}{Corollary}
\newtheorem{proposition}{Proposition}
\newtheorem{lemma}{Lemma}
\newtheorem{remark}{Comment}

\newif\ifFullVersion
\FullVersionfalse 
\FullVersiontrue 

\newif\ifIncludeConditionUniformk
\IncludeConditionUniformkfalse

\newif\ifcomments
\definecolor{CmtColor}{rgb}{0,0.6,1}

\input{Preamble.tex}
 \IEEEoverridecommandlockouts
\ifFullVersion
\title{On the Capacity of Communication Channels with Memory and Sampled Additive Cyclostationary Gaussian Noise:\\
Full Version with Detailed Proofs
}
\else
\title{On the Capacity of Communication Channels with Memory and Sampled Additive Cyclostationary Gaussian Noise
}
\fi

\vspace{-0.5cm}

\author{
\IEEEauthorblockN{\vspace{-0.2cm} Ron Dabora, \IEEEmembership{Senior Member, IEEE},  and Emeka Abakasanga\\
}

\thanks{The authors are with the Department of ECE, Ben-Gurion University, Be'er-Sheva 8410501, Israel (e-mail: daborona@bgu.ac.il; abakasan@post.bgu.ac.il). Ron Dabora is currently a Visiting Fellow at the Department of Electrical and Computer Engineering, Princeton University, Princeton, NJ, USA.  This work was supported by the Israel Science Foundation under Grant 584/20.}

\vspace{-1.0cm}

}

\begin{document}
\maketitle
\pagestyle{plain}
\thispagestyle{plain}

\begin{abstract}
  In this work we study the capacity of interference-limited channels with memory. These channels model non-orthogonal communications scenarios, such as the non-orthogonal multiple access (NOMA) scenario and underlay cognitive communications, in which the interference from other communications signals is much stronger than the thermal noise. 
  Interference-limited communications is expected to become a very common scenario in future wireless communications systems, such as 5G, WiFi6, and beyond.  As communications signals are inherently cyclostationary in continuous time (CT), then after sampling at the receiver, the discrete-time (DT) received signal model contains the sampled desired information signal with additive sampled CT cyclostationary noise. The sampled noise can be modeled as either a DT cyclostationary process or a DT {\em almost}-cyclostationary process, where in the latter case the resulting  channel is not information-stable. In a previous work we characterized the capacity of~this model for the case in which the DT noise is {\em memoryless}. In the current work we come closer to~practical scenarios by modelling the resulting DT noise as a {\em finite-memory} random process. The presence of memory requires the development of a new set of tools for analyzing the capacity of channels with additive non-stationary noise which has memory. Our results show, for the first time, the relationship between memory, sampling frequency synchronization and capacity, for interference-limited communications.  The insights from our work provide a link between the analog and the digital time  domains, which has been missing in most previous works on capacity analysis. Thus, our results can help improving spectral efficiency and suggest optimal transceiver designs for  future communications paradigms.

\end{abstract}

\vspace{-0.4cm}
\section{Introduction}
\label{sec:Intro}

We consider maximizing the information rates in interference-limited communications, which is a communications scenario in which message decoding is impeded by another communications signal, instead of by the commonly-studied thermal noise. Interference-limited communications has been attracting much interest in recent years; One major reason is the emergence of \ac{noma} as a major paradigm for 5G communications \cite{dai2015non}. Another important motivation is that interference-limited communications corresponds to multiple existing communications scenarios, including, for example, \ac{dsl}, in which crosstalk is limiting the rate of information \cite{campbell1983cyclostationary}, and underlay cognitive communications, in which the secondary user is the major source of interference to the primary user \cite{Skoglund:2012}.

Since communications signals are man-made, then they inherently possess cyclostationary statistics \cite[Ch. 1.3]{gardner1994cyclostationarity}, which follows as the signal generation process repeats at every symbol interval. Consequently, when communications is limited by interference, the corresponding \ac{ct} channel is modeled as a linear channel with additive  \ac{wscs} noise.  In modern communications, the receiver first samples the received \ac{ct} signal in order to facilitate digital processing. When the sampling interval at the receiver is commensurate with the period of the \ac{ct} \ac{wscs} interference process, a situation referred to in this work as {\em synchronous sampling},  the resulting sampled \ac{dt} interference is also \ac{wscs}. This channel model was extensively analyzed in previous works: The capacity of \ac{ptp} \ac{dt} channels with a finite memory and with \ac{acgn}  was derived in \cite{shlezinger2015capacity} for the case in which the channel input is subject to a time-averaged per-symbol power constraint.  Capacity characterization in \cite{shlezinger2015capacity}  was obtained via both a time-domain approach and a frequency-domain approach. Subsequently, the capacity of \ac{dt} \ac{mimo} channels with finite-memory \ac{acgn} was derived in \cite{shlezinger2016capacity}, the secrecy capacity of \ac{dt} finite-memory channels with \ac{acgn} was derived in \cite{shlezinger2017secrecy}, and bounds on the capacity of \ac{dt} channels with non-Gaussian \ac{wscs} noise were presented in \cite{shlezinger2018BBPLC}. Algorithmic aspects of reception in the presence of \ac{acgn} have also been studied:  In \cite{shlezinger2014frequency} a receiver structure which uses the periodicity of the noise correlation function for noise cancellation was presented, and in \cite{shlezinger2017LMS} optimal adaptive filtering  based on least-mean-squares adaptation for \ac{dt} jointly 
\ac{wscs} processes was studied.


\label{pg:intro_approx}
We note that, practically, the frequency of an oscillator cannot be deterministically set due to its inherent physical properties, see, e.g., \cite{OscillatorReport}. Thus, even though in the system design,   symbol clocks are typically set to nominal values given by finite-precision decimal numbers, see, e.g., \cite{5Gfreq:ISWCS2014}, then, as the interference and the \ac{soi} are clocked by physically separate oscillators, there is no reason to assume that their actual symbol intervals are related by a rational factor, due to the clocks' inherent frequency variability.  This is the main motivation for the model\footnote{\label{footnote:clock} It is  noted that while our model is closer to practicality than the models in previous works, still it is assumed that the variations of the  clocks' frequencies around their respective actual  values in practice, can be ignored. A complete model would  account also for the impact of such variations on the statistics of the interference, but this is outside the framework of the current analysis.} considered in the current work.
When the sampling interval at the receiver is incommensurate with the period of the \ac{ct} \ac{wscs} interference process, a situation referred to in this work as {\em asynchronous sampling}, the resulting sampled \ac{dt} interference is no longer \ac{wscs}, but rather it is a \ac{wsacs} random process \cite[Sec. 3.9]{gardner2006cyclostationarity}. As \ac{wsacs} processes are {\em non-stationary}, the resulting \ac{dt}  channel is generally not information-stable, namely, the  conditional distribution of the channel output given the input does not behave ergodically \cite{dobrushin1959general}. As a consequence, standard information-theoretic tools (e.g., based on joint typicality) cannot be applied in the capacity characterization of such channels. It is noted that, as in practice, a receiver synchronizes its  sampling rate with the symbol rate of the {\em desired} information signal, rather than with the symbol rate of the interference, asynchronous sampling is necessarily a frequent situation in practical systems, and thus, analysis of scenarios with asynchronous sampling carries practical, as well as theoretical, importance.

While communications with synchronous sampling was extensively analyzed, communications scenarios with asynchronous sampling have not been treated until recently. In \cite{shlezinger2019capacity}, we took a first step towards the capacity analysis of asynchronously-sampled interference-limited Communication Channels by considering the {\em memoryless} case. In this context, a \ac{dt} memoryless interference process is obtained by sampling a \ac{ct} finite-memory \ac{wscs} process with a sampling interval that is greater than its correlation length. Thus, the correlation function of the resulting \ac{dt} process is either a periodically time-varying or an almost periodically time-varying,  scaled Kornecker's delta functions, which, for Gaussian processes implies that different samples are independent. It follows that \cite{shlezinger2019capacity} {\em restricts the shape of the \ac{ct} correlation function as well as restricts the sampling rate to be low}, which restricts the information rate carried by the \ac{soi}. 
\label{pg:explain_memoryless}
For this scenario,   \cite{shlezinger2019capacity} derived 
a limiting expression for the capacity. As the channel is not information-stable, analysis was carried out within the framework of  information spectrum, leading to a capacity expressed as the limit-inferior of a sequence of capacities corresponding to synchronously-sampled \ac{ct} channels with \ac{acgn}. The work \cite{shlezinger2019capacity} presented several interesting insights: First, it was shown that when sampling is {\em synchronous},  capacity depends on the sampling interval and on the sampling phase,
even when the sampling interval is smaller than half the period of the noise correlation function.
Another important insight obtained  from \cite{shlezinger2019capacity} is that  when sampling is  {\em asynchronous},  capacity does not vary with the sampling rate or the sampling phase. Finally, it was observed that for some synchronous sampling rates, capacity in higher than the capacity obtained with asynchronous sampling rates arbitrarily close to the corresponding synchronous sampling rates. This means that practically, capacity of  sampled \ac{ct} interference-limited channels should be computed assuming {\em asynchronous} sampling, to avoid a false notion of a high capacity which hinges on an impractically accurate sampling frequency synchronization between the receiver and the interference. The impact of sampling frequency synchronization was subsequently studied in \cite{abakasanga2020rate} for the dual problem of compressing a \ac{dt} memoryless Gaussian random source process, obtained by asynchronously sampling a \ac{ct} \ac{wscs} Gaussian source process. The \ac{rdf} for this scenario was derived for the low distortion regime, as a limit of \ac{rdf}s obtained by synchronously sampling the \ac{ct} source process. It was observed that asynchronous sampling can result in higher compression rates than those obtained for synchronous sampling, mirroring the conclusions of \cite{shlezinger2019capacity} on the channel capacity. 

The relationship between the analog domain and the digital domain has also attracted attention from additional aspects, as part of the research effort to accurately characterize the information rates for communications over  {\em physical} channels: The work of \cite{EldarGoldsmith2013} considered sampling of a \ac{ct} \ac{lti} additive stationary noise channel, and showed that sampling rates higher than the Nyquist rate do not facilitate increase in capacity.
The work of \cite{HanShamai:CTCapacity2021}, provided a quantitative analysis of the rate of convergence of the  mutual information between the message and the sampled (i.e., digital) \ac{awgn} channel outputs, to the \ac{ct} (i.e., analog) channel's mutual information, with and without feedback,  under certain conditions.

In the current work we extend the scenario considered in the previous work of \cite{shlezinger2019capacity} by analyzing the capacity of  sampled \ac{ct} interference-limited channels, in which the interfering \ac{ct} process is a {\em correlated} \ac{wscs} process, and the sampling interval is   {\em allowed to be shorter than the correlation length} of the interference. 
\label{pg: explain_diff_2}
Therefore, the scenario considered in the current work {places restrictions on the shape of the \ac{ct} noise correlation function, while allowing sampling intervals shorter than the correlation length}. In contrast, \cite{shlezinger2019capacity} restricts  both  the shape of the \ac{ct} noise correlation function and  requires the sampling interval to be longer than the correlation length, in order to obtain an impulse-shaped \ac{dt} lag profile. One important consequence of this difference is that samples of the \ac{dt} interference process in the current scenario may be statistically dependent, while in \cite{shlezinger2019capacity} they must be independent.
When sampling is synchronous, we arrive at the model studied in \cite{shlezinger2016capacity}, thus, the focus of the current work is on asynchronous sampling. The studied setup provides a  connection between the analog model and the digital model obtained after sampling at the receiver,  when sampling results in a non-stationary \ac{dt} channel model with memory -- a situation which has a practical relevance for current and future communications~setups, but has not been analyzed previously.

{\bf {\slshape Main Contributions}:} In this work we analyze the fundamental rate limits for DT channels with  {\em correlated} \ac{wsacs} Gaussian noise having a {\em finite correlation length}, arising from sampling the output of \ac{ct} channels with \ac{acgn}. 
Since additive \ac{wsacs} Gaussian noise channels are {\em not information-stable}, it is not possible to employ standard information-theoretic arguments in the study of their capacity,  and we resort to information-spectrum characterization of the capacity \cite{han2003information}, within which we derive a new set of tools for the capacity analysis of \ac{dt} channels with sampled  finite-memory cyclostationary noise. 
We first observe that due to non-stationarity, the distribution function of the sampled noise process depends on the sampling time offset w.r.t. the period of the \ac{ct} noise correlation function, referred to in this work as the {\em sampling phase}. 
\label{pg:tranmission delay} 
Then, we obtain a general expression for the capacity when {\em transmission delay is not allowed}, namely when the transmitter must start transmitting the next message immediately upon completion of the transmission of the current message. Finally, for the case in which the correlation function decreases sufficiently fast with the lag, and the transmitter is allowed to delay the transmission of the next message by a finite and bounded delay, s.t. the optimal sampling phase is attained for subsequent message transmissions, a situation we refer to as {\em transmission delay is allowed}, then  capacity can be expressed as the limit-inferior of a sequence of capacities corresponding to \ac{dt} \ac{acgn} channels with finite memory,
such that the correlation function of the sequence of \ac{dt} \ac{wscs} noise processes approaches the correlation function of the \ac{dt} \ac{wsacs} noise process as the sequence index increases. 
Our characterization leads to important observations on the relationship between channel memory, sampling frequency synchronization and the achievable rate:
We show that for synchronous sampling, capacity varies with the sampling rate and the sampling phase. It is then  numerically demonstrated that when the power of the white thermal noise at the receiver is much smaller than the power of the interference, then an increase in the sampling rate results in higher capacity values. This follows as at higher sampling rates, the \ac{psd} of the \ac{dt} noise varies in the frequency domain, thereby  facilitating a better allocation of the transmit power across the noise spectrum.

{\bf {\slshape Organization}:} The rest of this paper is organized as follows: Section~\ref{sec:Preliminaries}  sets the mathematical notations  and quantities applied in this study. Section \ref{sec:Cap_Async}  presents the problem formulation and discusses the initial channel state;
Section \ref{subsec:Capacity_with_knowledge_Initial_state} states the capacity characterization for asynchronously-sampled \ac{acgn} channels with finite memory when transmission delay is not allowed. Section \ref{sec:cap_char_stationary} characterizes the capacity when a finite  and bounded transmission delay is allowed.  Section~\ref{sec:Simulations} presents numerical results to demonstrate the impact of the different scenario parameters on capacity. Lastly, Section~\ref{sec:Conclusions}  concludes the paper. The proofs of the theorems are detailed in the appendices.


\section{Preliminaries}
\label{sec:Preliminaries}

\subsection{Notations}
\label{subsec:Pre_Notations}
We use upper-case letters, e.g., $X$, to denote \acp{rv},  lower-case letters, e.g., $x$, to denote deterministic values,  and calligraphic letters, e.g., $\mathcal{X}$, to denote sets.
The sets of real numbers, rational numbers, non-negative integers and integers are denoted by $\mR$, $\mQ$, $\mN$, and $\mZ$ respectively, where $\mN^+$ denotes the set of positive integers. Sans-Serif font is used for denoting matrices, e.g., $\mathsf{B}$,  where the element at the $i$-th row and the $j$-th column of $\mathsf{B}$ is denoted with  $(\mathsf{B})_{i,j}$, $i, j\in\mN$. For $k, l\in\mNp$
we denote the all-zero $k \times l$ matrix with $\mathsf{0}_{k\times l}$, all-zero $k\times k$ square matrix with $\mathsf{0}_{k}$ and  the $k\times k$ identity matrix with $\myMat{I}_k$. 
For a $k\times k$ matrix $\Cmat$ we use $\maxEig\{\Cmat\}$ to denote its maximal eigenvalue,  $\Lambda_i^{(k)}\{\Cmat\}$ to denote its $i$-th ordered eigenvalue in descending order, $0\le i \le k-1$, $\Tr\left\{\Cmat\right\}$ to denote its trace, and  $\Det\big(\Cmat\big)$ to denote its determinant. We use $\mathrm{rank}(\Bmat)$ to denote the rank of a matrix $\Bmat$ and $\mathrm{range}(\Bmat)$ to denote its column range.
Column vectors are denoted with boldface letters, where lower-case letters denote deterministic vectors, e.g., ${\bf{x}}$, and upper-case letters denote random vectors, e.g., ${\bf X}$;  the $i$-th element of ${\bf{x}}$, $i \in \mN$, is denoted with $({\bf{x}})_i$, and for $a,b \in \mN$, $b>a$, we write $x_a^b\triangleq \big[({\bf{x}})_a, ({\bf{x}})_{a+1}, ..., ({\bf{x}})_b \big]^T$, where we also denote  $x^{(b)} \equiv x_0^{b-1}$. 
We use $X\sim\cdf{X}$ to denote that the \ac{cdf} of the  \ac{rv} $X$ is $\cdf{X}$. Specifically, $\Xvec\sim \dsN(\xvec_0,\Cmat)$ denotes a real Gaussian random vector $\Xvec$ with mean $\xvec_0$ and covariance matrix $\Cmat$. For the \acp{rv} $X$ and $Y$ we use $X\independent Y$ to denote that they are statistically independent. 
Transpose, Euclidean norm, 1-norm, $\infty$-norm, stochastic expectation,  differential entropy, and mutual information are denoted by 
$(\cdot)^T$, $\left\|\cdot\right\|$, $\left\|\cdot\right\|_1$, $\left\|\cdot\right\|_\infty$, $\E\{ \cdot \}$, $h(\cdot)$,  and $I(\cdot; \cdot)$,  respectively, and we define $a^+ \triangleq \max\left\{0,a\right\}$.
A square $k\times k$ real matrix $\mathsf{A}$ is called positive definite, denoted $\mathsf{A} \succ 0$ (positive semidefinite, resp., denoted $\mathsf{A} \succcurlyeq 0$) if for any $x^{(k)}\in\mR^k$, $x^{(k)}\ne\alzer{k\times 1}$, it holds that $(x^{(k)})^T\cdot \Amat \cdot x^{(x)} >0$ ($(x^{(k)})^T\cdot \Amat \cdot x^{(x)} \ge 0$, resp.).
We use $\EqDist{}$ to denote equality in distribution,  $\log(\cdot)$ to denote the base-$2$ logarithm, and $\ln(\cdot)$ to denote the natural logarithm.
Lastly, for any sequence ${\bf y}[i]$, $i \in \mN$, and  $b\in\mNp$, we use ${\bf y}^{(b)}$  to denote the column vector obtained by stacking the first $b$ sequence elements $\left[\big( {\bf y}[0]\big) ^T,\ldots,\big( {\bf y}[b-1]\big) ^T\right]^T$.

\subsection{Wide-Sense Cyclostationary Processes}
\label{subsec:wscs}
We next review some preliminaries from the theory of cyclostationarity, beginning with the definition of a wide-sense cyclostationary process:

\begin{definition}[A wide-sense cyclostationary process {\cite[Def. 17.1]{giannakis1998cyclostationary}, \cite[Pg. 296]{gardner1994cyclostationarity}}]
\label{Def:wscs}
A scalar stochastic process $\{X(t)\}_{t\in \mathcal{T}}$, where $\mathcal{T} = \mZ$ or $\mathcal{T} = \mR$, is referred to as {\em\ac{wscs}} if both its mean and its correlation function are periodic with respect to $t \in \mathcal{T}$ with some period $T_p\in \mT$:
$\mu_X(t)\triangleq\E\{X(t)\} = \mu_X(t + T_p)$ and $c_X(t,\tau)\triangleq \E\{X(t + \tau)X(t)\} = c_X(t+T_p,\tau)$,  $\forall t,\tau \in \mySet{T}$.
\end{definition}

Next, we consider \ac{wsacs} random processes, recalling first the definition of an almost-periodic function:
\begin{definition}[An almost-periodic function {\cite[Defs. 10, 11]{GUAN20131165}, \cite[Ch. 1.2]{napolitano2012generalizations}}] 
A function $f(t), t\in \mT$, where $\mathcal{T} = \mZ$ or $\mathcal{T} = \mR$, is referred to as {\em almost-periodic} if for every $\epsilon > 0$, there exists a number $l(\epsilon) \in \mT$, $l(\epsilon) > 0$, such that for any $t_0\in \mT$, there exists $\tau_{\eps} \in \big(t_0,t_0+l(\epsilon)\big)$, such that
\begin{equation*}
    \sup_{t\in\mT}|f(t+\tau_{\eps})-f(t)|<\epsilon.
\end{equation*}

\end{definition}

\begin{definition}[A wide-sense almost-cyclostationary process {\cite[Def. 17.2]{giannakis1998cyclostationary}, \cite[Pg. 296]{gardner1994cyclostationarity},\cite[Ch. 1.3]{napolitano2012generalizations}}] 
 A scalar stochastic process $\left\{X(t)\right\}_{t\in \mathcal{T}}$, where $\mathcal{T} = \mZ$ or $\mathcal{T} = \mR$, is called {\em\ac{wsacs}} if its mean and its autocorrelation function are almost-periodic functions with respect to $t\in \mathcal{T}$. 
\end{definition}

\subsection{Sampling of CT WSCS Random Processes}
\label{Sampled_wscs}
\begin{figure}
    \centering
    	{\includegraphics[width=1\columnwidth]{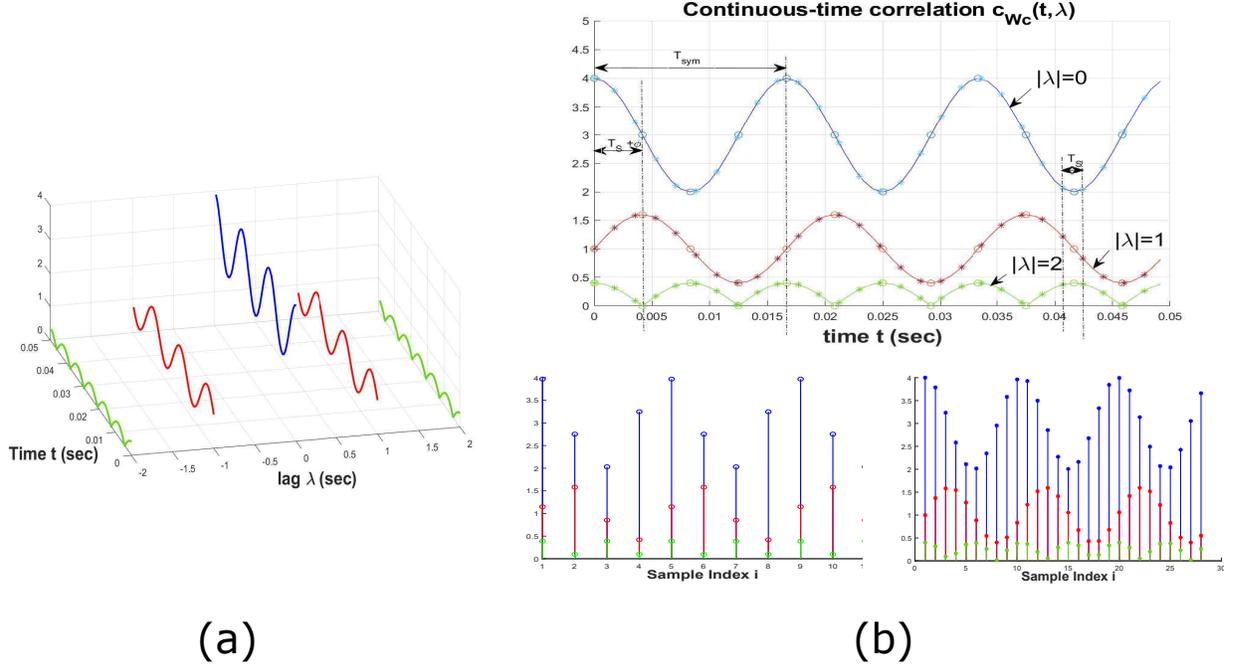}}
		\caption{(a). 3D visualization of the correlation function periodic for a \ac{ct} \ac{wscs} random process at  three lag values: $|\lambda|=[0,1,2]$; (b). Illustration of the DT correlation function obtained by sampling at different sampling rates and offsets. $'\mbox{o}'$ marks  synchronous sampling, with $\frac{\Tsamp}{T_{\rm sym}}=\frac{1}{4}\in\mQ$, and $'*'$ marks asynchronous sampling, with $\frac{\Tsamp}{T_{\rm sym}}=\frac{1}{3\pi}\notin\mQ$.
		}
		\label{fig:sampled_var}		
	\end{figure}

To facilitate the application of digital processing, the received signal is sampled at the receiver. Consider a \ac{dt} random process $X_{\Tsamp, \tau_0}[i]$, $i \in \mZ$, obtained by sampling the \ac{ct} \ac{wscs} random process $X(t)$, which has a period of $T_p$,  with a sampling interval of $\Tsamp$ and at a sampling phase of $\tau_0\in [0, T_p)$, i.e., $X_{\Tsamp, \tau_0}[i] \triangleq X (i \cdot \Tsamp + \tau_0)$. In the following, we demonstrate that contrary to sampled stationary processes, for \ac{ct} cyclostationary processes, the values of $\Tsamp$ and $\tau_0$ have a significant impact on the statistics of the resulting sampled process $X_{\Tsamp, \tau_0}[i]$. 
Consequently, the common practice of applying stationary theory to such scenarios can lead to erroneous results, e.g., \cite{Gardner:Ptifalls87}.
As an example, we illustrate in Fig \ref{fig:sampled_var}(a) the correlation function of a \ac{ct} \ac{wscs} random process at
three \ac{ct} lag values: $|\lambda|= 0, 1, 2$ $[\mbox{sec}]$, where the period in $t$ is 
$T_{\rm sym}=1/60$ $[\mbox{sec}]$. The bottom plots in Fig \ref{fig:sampled_var}(b) depict two sampling scenarios: In the first case, shown in the bottom-left plot, the \ac{ct} signal is sampled at a sampling interval of $\Tsamp=\frac{T_{\rm sym}}{4}$ and the sampling phase is $\tau_0 = \frac{\Tsamp}{2\pi}$. This is depicted by the $'\mbox{o}'$ markers on the \ac{ct} correlation function at the top figure in (b). Observe that in this case, the correlation function at the three lag values, $|\lambda|=0,1,2$, depicted by the blue, red and green plots, respectively, is periodic in \ac{dt}. 
As stated in Section \ref{sec:Intro}, such sampling, which maintains the periodicity of the statistics  in \ac{dt}, is referred to as {\em synchronous sampling}, and the resulting \ac{dt} process is \ac{wscs}. The bottom-right plot of Fig \ref{fig:sampled_var}(b) depicts the \ac{dt} correlation function obtained for $\tau_0=0$ and $\Tsamp=\frac{T_{\rm sym}}{3\pi}$, represented by the $'*'$ markers on the \ac{ct} correlation function at the top figure in (b). Hence, $\frac{\Tsamp}{T_{\rm sym}}$ is an irrational number, and  the \ac{dt} correlation function is not periodic but is almost-periodic at all lags, contrary to the first case. Therefore, the resulting \ac{dt} random process is not \ac{wscs} but \ac{wsacs}\cite[Sec. 3.9]{gardner2006cyclostationarity}. As stated in Section \ref{sec:Intro} such a sampling scenario is referred to as {\em asynchronous sampling}. 

This example  clearly demonstrates that when sampling \ac{ct} \ac{wscs} processes, slight variations in the sampling interval and the sampling phase may result in significantly different statistics for the sampled processes. As explained in Section \ref{sec:Intro}, such sampling scenarios exist in many Communication Channels, e.g., in interference-limited communications, in which the noise component corresponds to a sampled \ac{ct} \ac{wscs} process. The consequence of the variability of the statistics of sampled \ac{ct} \ac{wscs} processes is that channel capacity of a \ac{dt} channel obtained by sampling the output of a \ac{ct} additive  \ac{wscs} noise channel strongly depends on the actual sampling rate. In a recent study, \cite{shlezinger2019capacity}, we characterized the capacity of such \ac{dt} channels assuming the sampled additive noise is Gaussian and {\em memoryless}. This study aims to generalize the capacity characterization to \ac{dt} channels with additive Gaussian noise  with a finite memory, where the noise processes arise from  the sampling of \ac{ct} finite-memory \ac{wscs} processes representing communications signals, as developed in the subsequent sections.


\section{Model and Problem Formulation}
\label{sec:Cap_Async}
In this section we derive the considered channel model and detail the different assumptions on the communications scenario. We begin with the mathematical definition of the setup.

\begin{definition}
    \label{def:Channel}
    {\em A finite-memory \ac{dt} real-valued random process} $W[i]$, $i\in \mZ$, with memory $\Memd\in \mNp$, satisfies that $\forall i\in \mZ$,
	\begin{align}\label{EqnDef:memory}
        &\dsE\big\{W[i]\cdot W[i-\lambda]\big\}=\dsE\big\{W[i]\big\}\cdot \dsE\big\{W[i-\lambda]\big\}, \quad \forall |\lambda|>\Memd.  
	\end{align}
\end{definition}
Such a model is appropriate for processes representing  single-carrier digitally modulated signals with \ac{isi}, where the memory of the process is determined by the finite length of the overall \ac{cir} (i.e., the \ac{cir} which accounts for the pulse shape, propagation through the physical medium and analog filters in the signal paths at the transmitter and at the receiver). This model is also appropriate  for processes representing  \ac{ofdm} modulated signals, as in such processes the lack of correlation follows from the finite duration of the \ac{ofdm} symbol, where the \ac{cp} interval and receiver processing induce statistical independence between the received samples corresponding to different \ac{ofdm} symbols.

\begin{definition}
	\label{def:Code}
	An $\left[R, l \right]$ {\em code} with rate $R\in\mR^{++}$ and blocklength $l \in \mNp$ consists of:
	{\em 1)} A message set $\mathcal{U} \triangleq \{1,2,\ldots,2^{lR}\}$; 
	{\em 2)} An  encoder $e_l$ which maps a message $u \in \mySet{U}$ into a codeword $\xin_{u}^{(l)} = \Big[\xin_{u}\left[0\right],\xin_{u}\left[1\right],\ldots,\xin_{u}\left[l-1\right]\Big]$; and 	{\em 3)}  A decoder $d_l$ which maps the sequence of channel outputs, denoted  $y^{(l)}$, into a message $\hat{u} \in \mySet{U}$.

\end{definition}
The set $\{\xin_{u}^{(l)}\}_{u =1}^{2^{lR}}$ is referred to as the {\em codebook} and the message $u$ is  selected uniformly and independently from $\mySet{U}$. Note that as the noise process is generally non-stationary, then the probability distribution of the channel output sequence, denoted $Y^{(l)}$, depends on an initial channel state $s_0\in\mS_0$, where $\mS_0$ is the set of all possible initial channel states. 
The set $\mS_0$ will be explicitly stated in the context of this work in Sec. \ref{subsec:initial_state}.
The average probability of error when the initial channel state is $s_0$ is defined as:
\begin{equation*}
    \label{eqn:def_avgError}
    P_{e}^{l}(s_0) =  \frac{1}{2^{lR}}\sum\limits_{u = 1}^{2^{lR}} \Pr \left( \left. {{d_l }\left( \Yi^{(l)}\right) \ne u} \right|U \!= \! u, s_0\right).
\end{equation*}

\begin{definition}
	\label{def:Rate}
	A rate $\Rs\in\mR^{++}$ is {\em achievable} if for every $\eta_1,\eta_2 > 0$, $\exists l_0 \in\mNp$ such that $\forall l > l_0$ there exists an $\left[R, l \right]$ code which satisfies 
	\begin{subequations}
		\label{eqn:def_Rs}
		\begin{equation}
		\label{eqn:def_Rs1}
		\mathop {\sup }\limits_{s_0 \in \mS_0}  P_e^{l}(s_0) < \eta_1,
		\end{equation}
		and
		\begin{equation}
		\label{eqn:def_Rs2}
		R \geq \Rs - \eta_2.
		\end{equation}
	\end{subequations}
	{\em Capacity} is defined as the supremum over all achievable rates.
\end{definition}
Lastly, we recall the definition of the limit-inferior in probability \cite[Def.  1.3.2]{han2003information}:
\begin{definition}
	\label{def:pliminf}
	The {\em limit-inferior in probability} of a sequence of real \acp{rv} $\{\zk{k}\}_{k \in \mNp}$ is defined as
	\begin{equation}
	\label{eqn:pliminf}
	{\rm p-}\mathop{\lim\!\inf}\limits_{k \rightarrow \infty} \zk{k} \triangleq \sup\left\{\alpha \in \mR \big| \mathop{\lim}\limits_{k \rightarrow \infty}\Pr \left(\zk{k} < \alpha \right) = 0   \right\}
	\triangleq \alpha_0.
	\end{equation}
\end{definition}
\noindent Hence, $\forall \talpha > \alpha_0$, $\exists \eps>0$, such that there exist countably many $k\in\mNp$ for which  $\Pr(Z_k<\talpha)>\eps$.

\smallskip
As was stated in \cite{shlezinger2019capacity}, see also \cite[Pg. VIII]{han2003information}, the quantity ${\rm p-}\mathop{\lim\!\inf}\limits_{k \rightarrow \infty} \zk{k}$ is well-defined even when the sequence of \acp{rv} $\{\zk{k}\}_{k \in \mNp}$ does not converge in distribution. This makes the limit-inferior in probability applicable to the analysis of scenarios in which methods based on the law of large numbers cannot be applied, e.g., when non-stationary and non-ergodic processes are considered \cite{han1993approximation}.
We note, however, that the application of Def. \ref{def:pliminf} in information-theoretic analysis typically results in expressions which are very difficult to compute. 

\subsection{Problem Formulation}
\label{subsec:problem_formulation}
Consider a real-valued zero-mean \ac{ct} \ac{wscs} {Gaussian} random process $\Wc(t)\in\mR$, whose autocorrelation function, $\Cwc(t,\lambda) \triangleq \E\big\{\Wc(t+\lambda)\Wc(t)\big\}$, is continuous in $t$ and in $\lambda$, periodic in $t$ with a period $\Tc\in\mR^{++}$ and has a finite correlation length $\Memc \in\mR^{++} $, i.e., $\Cwc(t,\lambda)  =  \Cwc(t + \Tc,\lambda)$, $\forall t,\lambda \in \mR$, and $\Cwc(t,\lambda)  = 0$ for any $|\lambda| \ge \Memc > 0$. 

Since the autocorrelation function $\Cwc(t,\lambda)$ is continuous and periodic, it is sufficient to  characterize its properties only over a compact interval $\mT_0 \in \mR$ where $\mT_0=[t_0, t_0+\Tc]$, for some arbitrary $t_0\in\mR$; it follows that $\Cwc(t,\lambda)$ is bounded and uniformly continuous with respect to time $t\in \mT_0$ and lag $\lambda \in \mR$ \cite[Ch. III, Thm. 3.13]{amann2005analysis}.

The process $\Wc(t)$ is sampled at a finite, positive sampling interval $\Tsamp(\eps)$ such that $\Tc = (\Td + \myEps)\cdot \Tsamp(\eps)$ where $\Td \in \mNp$ and $\myEps \in [0,1)$, resulting in the \ac{dt}  random process $\Weps[i]\triangleq W_c(\tau_0+i\cdot \Tsamp(\eps))$, $i\in\mZ$. In this study we focus on the case in which the sampling interval is smaller than the correlation span of the \ac{ct} process $\Wc(t)$, hence, letting $\tau_0\in\mR$ denote the sampling phase corresponding to index $i=0$, then $\Weps[i]$ is  a zero-mean Gaussian random process whose autocorrelation function is given by:
\begin{align}
   \!\!\! \Cweps^{\{\tau_0\}}[i,\Delta]
    &\equiv \E\Big\{\Weps[i+\Delta]\cdot \Weps[i]\Big|\tau_0\Big\} \notag \\
    &\triangleq \E\left\{\!\Wc\!\left(\frac{(i\!+\!\Delta)\cdot \Tc}{\Td + \myEps}\!+\!\tau_0\right)\!\cdot\! \Wc\!\left(\frac{i \cdot \Tc}{\Td\! +\! \myEps}\!+\!\tau_0\right)\!\right\}
    \!=\! \Cwc\!\left(i \cdot\frac{\Tc}{\Td\! +\! \myEps}\!+\!\tau_0,\Delta \cdot \frac{\Tc}{\Td\! +\! \myEps} \right)\!.
    \label{eqn:AnsycAutocorr}
\end{align}
It follows that $\Cweps^{\{\tau_0\}}[i,\Delta] = 0$ for all $|\Delta| >\left\lceil \frac{(\Td + 1) \cdot\Memc}{\Tc} \right\rceil \triangleq \Memd < \infty$, hence, the correlation length of $\Weps[i]$ is finite.     In the following we say that $\Weps[i]$ has a {\em finite memory} $\Memd<\infty$, referring to  the finite correlation length of the sampled process $\Weps[i]$. Due to the fact that $\Weps[i]$ is a sampled physical noise process, it can be assumed that the correlation matrix corresponding to any sequence length is positive definite, see elaboration in Comment \ref{rem:CovMatIsPosDef}.

Next, observe that from \eqref{eqn:AnsycAutocorr}, see also \cite{shlezinger2019capacity}, it follows that when $\eps\in\mQ^{++}$, i.e., $\exists u,v \in \mNp$ s.t. $\myEps = \frac{u}{v}$, then the process $\Weps[i]$ is \ac{wscs} with a period which is equal to $p_{u,v}\triangleq \Td \cdot v + u$. Such a sampling scenario corresponds to {\em synchronous sampling}, for which capacity was characterized in \cite{shlezinger2015capacity}. On the other hand, when $\myEps$ is irrational ($\myEps \notin \mQ$), the resulting \ac{dt}  process $\Weps[i]$ is \ac{wsacs} \cite[Sec. 3.9]{gardner2006cyclostationarity}, which corresponds to {\em asynchronous sampling}. 
In this work, we consider \ac{dt} channels  with real-valued additive, finite-memory \ac{wsacs} Gaussian noise. 
Letting $\Xscal[i]$ and $\Yeps[i]$ denote the real-valued channel input and output, respectively, at time $i\in\mZ$,
the input-output relationship for the transmission of a sequence of $l\in \mNp$ channel inputs is given by:
\begin{equation}
    \label{eqn:AsnycModel1}
    \Yeps[i] = \Xscal[i] + \Weps[i], \qquad i \in \{0,1,\ldots,l - 1\},
\end{equation}
where the subscript $\eps$ is retained to indicate the synchronization mismatch between the period of the \ac{ct} noise and the sampling interval at the transmitter. 
The noise process $W_{\eps}[i]$ has a temporal correlation length of $\Memd$ samples. 
 The channel input, $X[i]$, is subject to a per-codeword  power constraint $P$,
\begin{equation}
    \label{eqn:AsyncConst1}
    \frac{1}{l}\sum_{i=0}^{l-1}\big(\xin_{u}\left[i\right]\big)^2 \le P, \qquad u\in\mU.
\end{equation}
Lastly, it is assumed that the input and the noise in  \eqref{eqn:AsnycModel1} satisfy the independence assumption:

\smallskip
\noindent {\em Assumption 1}: $\big\{\Xscal[i]\big\}_{i\in\mZ}$ is independent of $\big\{\Weps[i]\big\}_{i\in\mZ}$.
\smallskip

The channel model in \eqref{eqn:AsnycModel1} is particularly relevant for modelling channels in which $\Wc(t)$ is a digitally modulated interfering communications signal, and is thus a \ac{ct} \ac{wscs} process with a period which is equal to its symbol duration \cite[Sec. 5]{amann2005analysis}. For example, when $\Wc(t)$ is an \ac{ofdm} modulated signal with a sufficiently large number of subcarriers, then it can be modeled as a Gaussian process \cite{wei2010convergence}, and the correlation function is periodic with a period which is equal to the duration of an \ac{ofdm} symbol.
As another example, when $\Wc(t)$ is a linearly modulated \ac{qam} signal with a partial response pulse shaping, then, when the \ac{isi} spans a sufficiently long interval, it follows that $\Wc(t)$ is modeled as a Gaussian process, see \cite[Sec. III-A]{Metzger87:ISI}. Here, again the correlation function is periodic, where the period is equal to the duration of an information symbol. In both examples, the process $\Wc(t)$  has a finite correlation length. 
In such interference-limited setups, as the sampling rate at the receiver is generally not synchronous  with the symbol rate of the interferer, then the resulting \ac{dt} interfering signal can be modeled as a finite-memory \ac{wsacs} process, giving rise to the channel input-output relationship in \eqref{eqn:AsnycModel1}. {\em Our goal is to characterize the capacity of the channel \eqref{eqn:AsnycModel1} subject to the power constraint \eqref{eqn:AsyncConst1}.} 
We also note that the model of Eqn. \eqref{eqn:AsnycModel1} has previously been used in the analysis of communications systems, e.g., in \cite{LiElia2015ARmodelling}, which studied feedback capacity for stationary,
finite-dimensional Gaussian channels, hence, the current model adds the non-stationary noise characteristics to the model considered in \cite{LiElia2015ARmodelling} (without feedback).

The analysis in this work relies also on the following two assumptions:

\smallskip
\noindent {\em Assumption 2}: The transmitter (Tx) and the receiver (Rx), are both assumed to know the \ac{ct} noise correlation function $\Cwc(t,\lambda)$. 
\smallskip

Note that, as explained in detail in the next subsection, non-stationarity of the sampled noise statistics implies that knowledge of $\Cwc(t,\lambda)$ is not sufficient for maximizing the rate, which is in contrast to the situation  for \ac{dt} channels with additive stationary noise.
Therefore, it is also assumed that 

\smallskip
    \noindent {\em Assumption 3}: The propagation delay between the transmitter and receiver is negligible compared to the period and to the maximal slope of the (uniformly continuous) noise correlation function.

This assumption implies that the receiver can attain perfect sampling time synchronization with the transmitter, as well as that both the transmitter and the receiver can identify the temporal phase within the period of the noise correlation function at any time instant. This is referred to in the following as {\em perfect Tx-Rx timing synchronization}. Note that the sampling interval used by the transmitter and receiver is not synchronized with the period of the noise correlation function in the sense that their ratio is an irrational number.

\subsection{An Example Scenario: Lowpass Channel with ACGN}
\label{subsec:AnotherExample}
As another motivating example for the \ac{dt} model of Eqn. \eqref{eqn:AsnycModel1}, consider a  \ac{ct} baseband channel model with memory, in which the received signal is given by $Y(t)=h(t)*X(t)+W(t)$, where $'*'$ denotes the linear convolution, and $W(t)$ is a \ac{wscs} Gaussian random process with a finite memory (i.e., an interfering communications signal).  For simplicity of the discussion assume that the channel can be approximated as a first-order stable lowpass filter with a \ac{tf}
\[
    H(s)=\frac{1}{s+a}, \qquad  a\in \mR^{++}, \;\Re\{s\} > -a.
\]
Letting $X[m]$ denote the \ac{dt} information sequence at a rate of $\frac{1}{T_s}$, the overall received \ac{ct} signal component can be modeled as
\[
    \mathop{\sum}\limits_{m=-\infty}^{\infty} X[m]\delta(t-mT_s)* h(t)= \mathop{\sum}\limits_{m=-\infty}^{\infty} X[m]\cdot h(t-mT_s).
\]
Thus, sampling at intervals of $T_s\in\mR^{++}$, we obtain the following \ac{dt} relationship between $X[m]$, $Y[n]\triangleq Y(n\cdot \Tsamp)$, $W[n]\triangleq W(n\cdot \Tsamp)$ and $h[n]\triangleq h(n\cdot T_s)$:
\[
    Y[n]=\mathop{\sum}\limits_{m=-\infty}^{\infty} X[m]\cdot h[n-m]+W[n],
\]
which is a real-valued linear, time-invariant \ac{dt} channel with memory, where $W[n]$ is a \ac{dt} Gaussian random process with a finite memory. Observe that the equivalent \ac{dt} \ac{cir} is obtained from the \ac{cir} of the \ac{ct} channel by sampling  (i.e., via the so-called impulse-invariance method, see, e.g., \cite[Sec. 11.3.2.2]{giannakis1998cyclostationary}), which results in a \ac{dt} \ac{tf} of the form:
\begin{equation*}
    H(z)=\frac{1}{1-e^{-a\cdot T_s}\cdot z^{-1}}, \quad |z|>e^{-a\cdot T_s}.
\end{equation*}
Since $\alpha \triangleq e^{-a\cdot T_s}<1$ for all considered $T_s$, then $H(z)$ corresponds to a stable and causal channel with a stable and causal inverse. Such a channel model is very popular in  communications, see, e.g., \cite{giannagik1998ARMA}.
The zero-forcing equalizer for this model is the highpass filter whose \ac{tf} is obtained by taking the inverse of $H(z)$:  $H_{ZF}(z)=1-\alpha\cdot z^{-1}$, and its impulse response is $h_{ZF}[n]=\delta[n]-\alpha\cdot \delta[n-1]$. Observe that after filtering $Y[n]$ with $h_{ZF}[n]$, the resulting interference process, namely, $W[n]*h_{ZF}[n]$, is a random process with a finite memory, hence, after zero-forcing equalization, the resulting overall \ac{dt} channel  is appropriately modeled via Eqn. \eqref{eqn:AsnycModel1}. 
In particular, when the interference is a single-carrier \ac{qam} signal, filtering effectively increases the duration of the \ac{cir}, hence, filtering increases the interference's memory. Note that when the interference is an \ac{ofdm} signal, then, if the length of the overall \ac{cir} is shorter than the length of the \ac{cp} (which should hold by design), then subsequent symbols remain independent.




\subsection{The Initial Channel State}
\label{subsec:initial_state}

Note that the channel model in Eqn. \eqref{eqn:AsnycModel1} with the input power constraint of Eqn.  \eqref{eqn:AsyncConst1} corresponds to an additive Gaussian noise channel, where the Gaussian noise is non-stationary. Due to Gaussianity, the distribution of the noise is completely characterized by the first two moments. As the noise has a zero-mean, then non-stationarity of the noise manifests itself via the fact that its  correlation matrix is not Toeplitz. It is noted that such a general model was discussed in \cite{cover1989gaussian}, subject to  an {\em average sum-power} constraint. In this context we make the following observation: In order to transmit a message, the transmitter has to select the respective codeword. As the sampled noise correlation function is generally non-periodic, then the noise correlation matrices corresponding to  {\em different message intervals} may be {\em completely different}. Accordingly, to facilitate analysis of such channels, it is necessary to introduce an initial state variable which identifies the noise correlation function  present during the transmission of the message. For the case of  sampled cyclostationary noise, the initial state corresponds to the relative location of the first sample of the \ac{dt} noise correlation function within the period of the \ac{ct} noise correlation function, which, for the subsequent discussion is denoted with $\tau_0$. 
Due to the periodicity of  $\Cwc(t,\lambda)$ in $t\in\mR$, it is enough to consider $\tau_0\in[0,\Tc)$, hence, the initial state space $\mS_0$ is the interval $[0,\Tc)$.
In this context, it is noted that the models  \cite{cover1989gaussian} and of \cite{verdu1994general} do not consider an initial channel state in the model while pertaining to be relevant to non-stationary channels. This implies that the models in \cite{cover1989gaussian} and \cite{verdu1994general} make a {\em hidden assumption} that while the channel is non-stationary, {\em there exists a synchronization mechanism that sets the channel statistics to be identical for subsequent message intervals}. 

Moreover, as the correlation function of the optimal input is a function of the sampled noise correlation function, it follows that knowledge of the noise correlation function $\Cwc(t,\lambda)$ alone at the transmitter is {\em not sufficient} for obtaining the optimal performance, and knowledge of the initial state $\tau_0$ for each message is required. Due to {\em Assumption 3}, this knowledge is available in our setup.
This knowledge requirement is not necessary when the additive noise is stationary.

%
\section{Capacity Characterization when Transmission Delay is not Allowed}
\label{subsec:Capacity_with_knowledge_Initial_state}

When the initial state $\tau_0$ is known at the transmitter, it is able to adapt the statistics of the codebook such that the achievable rate is maximized. When {\em  transmission delay is not allowed}, then once a message transmission has been completed,  the subsequent message is transmitted immediately, without further delay,  at the sampling phase $\tau_0$ present at the time the transmission of the current message has finished.
Due to  {\em Assumption 3}, of perfect Tx-Rx timing synchronization, it follows that both the receiver and the transmitter know the  $\ac{dt}$ noise correlation function at every message transmission. Letting $\Xinkopt$ denote the input process which maximizes the mutual information between $\Xink$ and $\Yepsk$ when the sampling phase is $\tau_0$, i.e., maximizes $I(\Xink;\Yepsk|\tau_0)$ , we obtain the following capacity characterization:
\begin{theorem}
\label{Thm:Capacity_tau0_at_Tx_No_Adap}
    Consider the channel \eqref{eqn:AsnycModel1} with power constraint \eqref{eqn:AsyncConst1}, when the transmitter can identify the sampling phase within a period of the \ac{ct} noise correlation function, $\tau_0\in [0,\Tc)$, and is allowed to adapt its information rate and codebook accordingly. If no transmission delay is  allowed, then capacity is given by 
    \[
        C_\eps = \frac{1}{\Tc}\int_{\tau_0=0}^{\Tc}
        C_\eps(\tau_0)\mbox{\em d}\tau_0,
    \]
    where $C_\eps(\tau_0)\triangleq\liminfk \frac{1}{k} I(\Xinkopt;\Yepsk|\tau_0)$,
    as long as the maximizing input  $\Xinkopt$ is Gaussian, with a distribution which depends on $\tau_0$ and  satisfies  $\frac{1}{k}\Tr\Big\{\corrmatxeopt(\tau_0)\Big\}\le P$ and  $\frac{1}{k^2}\Tr\Big\{\big(\corrmatxeopt(\tau_0)\big)^2\Big\}\mathop{\longrightarrow}\limits_{k\rightarrow\infty}0$.
\end{theorem}
\begin{IEEEproof}
    The proof is provided in Appendix \ref{Appndx:Capacity_General}.
\end{IEEEproof}

\bigskip
\begin{remark}[The Requirement on the Trace of the Squared Input Correlation Matrix]
    \label{rem:power_const_NoDelay}
    {\em We note that the condition  $\frac{1}{k^2}\Tr\Big\{\big(\corrmatxeopt(\tau_0)\big)^2\Big\}\mathop{\longrightarrow}\limits_{k\rightarrow\infty}0$ is introduced 
    in order to satisfy the per-codeword power constraint \eqref{eqn:AsyncConst1}. Such a condition was also considered in \cite[Sec. VIII]{cover1989gaussian}. 
    }
\end{remark}

\smallskip
\begin{remark}[Capacity with an Average Power Constraint]
    \label{rem:Cap_Avg_Power_Constr}
    {\em Instead of the per-codeword power constraint  \eqref{eqn:AsyncConst1} one may consider a more-relaxed average sum-power constraint, as in, e.g., \cite[Eqn. (7)]{hirt1988capacity}, \cite[Eqn. (7)]{cover1989gaussian}:
    \begin{equation}
        \label{eqn:Avg_sum_power_constr}
        \frac{1}{l}\sum_{i=0}^{l-1}\dsE_{U}{|x_U[i]|^2}\le P.
    \end{equation}
    With such a constraint then Thm. \ref{Thm:Capacity_tau0_at_Tx_No_Adap} holds {\em without} requiring the consideration of $\Tr\Big\{\big(\corrmatxeopt(\tau_0)\big)^2\Big\}$. This follows as any codebook  of length $k$, generated randomly according to a Gaussian distribution $\Xinkopt \sim \dsN\Big(\alzer{k\times 1}, \corrmatxeopt(\tau_0)\Big)$ such that $\frac{1}{k}\Tr\Big\{\corrmatxeopt(\tau_0)\Big\} = P-\delta$, where $\delta>0$ is arbitrarily small, will satisfy
    \begin{align}
        \frac{1}{k}\sum_{i=0}^{k-1}\dsE_{U}\big\{|x_{U,{\rm opt}}[i]|^2|\big\}& \stackrel{(a)}{=} \frac{1}{2^{kR}}\sum_{u=0}^{2^{kR}-1}\frac{1}{k}\sum_{i=0}^{k-1} |x_{u,{\rm opt}}[i]|^2\notag\\
        & \stackrel{(b)}{=} \frac{1}{2^{kR}}\sum_{u=0}^{2^{kR}-1}\omega_u\notag\\
        & \stackrel{(c)}{\mathop{\longrightarrow}\limits_{k\rightarrow\infty}} \dsE\{\Omega_1\} \qquad (\mbox{in probability})\notag\\
        & \le P,\notag
    \end{align}
    where (a) follow by the uniform selection of codewords for transmission; in (b) we consider the realizations $\omega_u\triangleq \frac{1}{k}\sum_{i=0}^{k-1} |x_{u,{\rm opt}}[i]|^2$: Note that for different indexes $u\in\mU$, the realizations $\omega_u$ are generated  independently using the same multivariate distribution for all messages $u\in\mU$. The expectation of the generating \ac{rv}, $\dsE\{\Omega_1\}$, is equal to 
    \begin{align*}
        \dsE\{\Omega_1\} & = \dsE\Big\{\frac{1}{k}\sum_{i=0}^{k-1} |X_{1,{\rm opt}}[i]|^2|\Big\}\\
        & =\frac{1}{k}\Tr\Big\{\corrmatxeopt(\tau_0)\Big\}\le P, \qquad u\in\mU.
    \end{align*}
    Step (c) follows by the weak law of large numbers \cite[Sec. 7.4]{grimmett2001probability},  as the mean of  $2^{kR}$ independent realizations of the \ac{iid} \acp{rv} $\{\Omega_u\}_{u\in\mU}$  converges in probability to its expectation.
    
    Then, we can conclude that the corresponding ${\rm p-}\mathop{\lim\!\inf}\limits_{k \rightarrow \infty} \zkeps{k}{\eps}\Big( \cdf{\Xinkopt|\tau_0}|\tau_0\Big)$, defined in \eqref{eqn:Relationship_PliminfZkeps_and_MIeps}, is achievable by considering the proof of the direct part of \cite[Thm. 3.2.1]{han2003information}, as there is no need to restrict the selected codewords when generating them according to the distribution $\dsN\Big(\alzer{k\times 1}, \corrmatxeopt(\tau_0)\Big)$. This follows as for sufficiently large $k$ the codebooks generated according to this Gaussian distribution  satisfy the average constraint \eqref{eqn:Avg_sum_power_constr} with a probability arbitrarily close to $1$, as $k$ increases. Thus, subject to \eqref{eqn:Avg_sum_power_constr}, the optimal input for Thm. \ref{Thm:Capacity_tau0_at_Tx_No_Adap} is 
    $\Xinkopt\sim \dsN\Big(\alzer{k\times 1}, \corrmatxeopt(\tau_0)\Big)$,  with $\frac{1}{k}\Tr\Big\{\corrmatxeopt(\tau_0)\Big\} \le P$.
    }
\end{remark}
\color{black}
\bigskip

When codebook adaptation is allowed but the rate has to be fixed, the following corollary is immediate:
\begin{corollary}
    \label{corr:No rate adapt}
    Consider the channel \eqref{eqn:AsnycModel1} with power constraint \eqref{eqn:AsyncConst1}, when the transmitter can identify the sampling phase within a period of the \ac{ct} noise correlation function, $\tau_0\in[0,\Tc)$, and is allowed to adapt its codebook accordingly. If the message rate has to be fixed, and no transmission delay is allowed, then capacity is given by 
    \[
        C_\eps = \min_{\tau_0\in[0,\Tc)} \liminfk \frac{1}{k} I(\Xinkopt;\Yepsk|\tau_0),
    \]
    as long as the maximizing input $\Xinkopt$ is Gaussian, with a  distribution which depends on $\tau_0$ and satisfies  $\frac{1}{k^2}\Tr\Big\{\big(\corrmatxeopt\big)^2\Big\}\mathop{\longrightarrow}\limits_{k\rightarrow\infty}0$. 
\end{corollary}

%

\section{Capacity Characterization When Transmission Delay is Allowed}
\label{sec:cap_char_stationary}

In this section we consider a transmission scenario in which the transmitter is allowed to delay the transmission of the next message such that it would  begin at the optimal sampling phase within the period of the noise correlation function. In such a scenario, capacity can be expressed via a sequence of capacities of \ac{dt} additive \ac{wscs} Gaussian noise channels.

\subsection{Approaching the Relationship \texorpdfstring{\eqref{eqn:AsnycModel1}}{(5)} via a Sequence of DT ACGN Channels}
\label{subsec:Cap_Main}
To characterize the capacity of the channel \eqref{eqn:AsnycModel1}, we define for each $n \in \mNp$ a rational number $\myEps_n \triangleq \frac{\lfloor n \cdot \myEps \rfloor}{n}$ and a corresponding \ac{dt} process $\Wn[i] \triangleq \Wc \left(\frac{i \cdot \Tc}{\Td + \myEps_n} + \tau_0 \right)$, $i\in\mZ$, $\tau_0\in [0,\Tc)$. As follows from the  discussion in Sec. \ref{subsec:problem_formulation}, the \ac{dt} process $\Wn[i]$  is a {\em zero-mean \ac{wscs} Gaussian random process} with period $\Td_n = \Td \cdot n + \lfloor n \cdot \myEps \rfloor$. 
Note that as
\[
    \Memd \triangleq \left\lceil\frac{(\Td + 1) \cdot\Memc}{\Tc}\right\rceil \ge \left\lceil \frac{(\Td +\myEps_n ) \cdot\Memc}{\Tc}\right\rceil,
\]
then the correlation length of the noise $\Wn[i]$ can be set to $\Memd$ for all $n \in \mNp$, hence, the noise process $\Wn[i]$ has a finite memory of $\Memd$.

Next, we define a channel with input $X[i]$ and output $\Yn[i]$ via the input-output relationship:
\begin{equation}
    \label{eqn:AsnycModel2}
    \Yn[i] = X[i] + \Wn[i],
\end{equation}
where the channel input is subject to the per-codeword power constraint \eqref{eqn:AsyncConst1}.
The channel \eqref{eqn:AsnycModel2} is an additive noise channel with correlated, finite-memory \ac{wscs} Gaussian noise $\Wn[i]$, whose period is $\Td_n$. The capacity of the channel \eqref{eqn:AsnycModel2} was explicitly derived in \cite[Thm. 1]{shlezinger2016capacity},
by  transforming the \ac{dt} channel \eqref{eqn:AsnycModel2}  into a \ac{mimo} channel via the \ac{dcd} \cite[Sec. 17.2]{giannakis1998cyclostationary}. For blocklengths which are integer multiples of $\Td_n$, the \ac{dcd} transforms the process $\Wn[i]$  into an equivalent $\Td_n$-dimensional stationary process $\tilde{\myVec{W}}_{n}^{\{\Td_n\}}[i]$, such that \big($\tilde{\myVec{W}}_{n}^{\{\Td_n\}}[i]\big)_b=W_n[i\cdot p_n + b]$, $0\le b \le p_n-1$. We define the correlation matrix for sampling phase $\tau_0$ as $\Cmat_{\tilde{\myVec{W}}_{n}^{\{\Td_n\}}}[\tau;\tau_0]\triangleq \E\bigg\{\tilde{\myVec{W}}_{n}^{\{\Td_n\}}[i+\tau]\cdot\left(\tilde{\myVec{W}}_{n}^{\{\Td_n\}}[i]\right)^T\bigg|\tau_0\bigg\}$. From the finite correlation length of the process $\Wn[i]$, it follows that for all $n$ such that $p_n >\Memd$,  $\left(\Cmat_{\tilde{\myVec{W}}_{n}^{\{\Td_n\}}}\left[\tau;\tau_0\right]\right)_{k_1, k_2}=0$, $\forall |\tau|>1$, $\forall k_1, k_2 \in \{0,1,\ldots,\Td_n-1\}$, see \cite[Sec. IV]{shlezinger2015capacity}.
Next, for all $\theta\! \in\! \left[-\pi,\pi\right)$, define the $\Td_n \!\times\! \Td_n$ matrix $\Cmat'_{\tilde{\myVec{W}}_{n}^{\{\Td_n\}}}(\theta;\tau_0)\! \triangleq\!  \sum\limits_{\tau=-1}^1\Cmat_{\tilde{\myVec{W}}_{n}^{\{\Td_n\}}}\left[\tau;\tau_0\right]e^{-j\theta \tau}$, let $\{\TLambda_{k,n} (\theta;\tau_0)\}_{k=0}^{\Td_n -1}$ be the eigenvalues of $\left(\Cmat'_{\tilde{\myVec{W}}_{n}^{\{\Td_n\}}}(\theta;\tau_0)\right)^{-1}$, and let  $\TDelta^{\{\Td_n;\tau_0\}}$ be the unique solution to
\begin{equation}
    \label{eqn:MainThmCst}
    \frac{1}{2\pi \cdot \Td_n}\sum\limits_{k=0}^{\Td_n -1} \int\limits_{\theta = -\pi}^{\pi} \left(\TDelta^{\{\Td_n;\tau_0\}} - \left(\TLambda_{k,n} (\theta;\tau_0)\right)^{-1}\right)^+ {\rm d}\theta =  \PCst.
\end{equation}
Then, the capacity of the channel  \eqref{eqn:AsnycModel2}, denoted $\Capacity_n(\tau_0)$,  is given  as \cite[Thm. 1]{shlezinger2016capacity}:
\begin{equation}
    \label{eqn:Cn_tau0_def}
    \Capacity_n(\tau_0) = \frac{1}{4\pi \cdot \Td_n}\sum\limits_{k=0}^{\Td_n -1} \int\limits_{\theta = -\pi}^{\pi} \Bigg(\log \left(\TDelta^{\{\Td_n;\tau_0\}} \cdot {\TLambda_{k,n} (\theta;\tau_0)}\right)\Bigg)^+ {\rm d}\theta \quad [\mathrm{bits\; per \; channel\; use}].
\end{equation}
Note that the capacity  $\Capacity_n(\tau_0)$ generally depends on the initial sampling phase $\tau_0$.
Then, maximizing over the initial sampling phase we define
\begin{equation}
    \label{eqn:DefCn}
    \Capacity_n = \max_{\tau_0\in [0,\Tc]}\Capacity_n(\tau_0).
\end{equation}
With the aid of \eqref{eqn:MainThmCst}-\eqref{eqn:DefCn}, we subsequently obtain a characterization for the capacity of the asynchronously-sampled channel \eqref{eqn:AsnycModel1}, denoted $\Capacity_\myEps$, when transmission delay of up to $\Memd + \Tc$  between subsequent messages is allowed. This is stated in the following theorem: 
\begin{theorem}
	\label{thm:AsycCap}
	Consider the channel \eqref{eqn:AsnycModel1} with power constraint \eqref{eqn:AsyncConst1}, when the transmitter can identify the sampling phase within a period of the noise correlation function, $\tau_0\in[0,\Tc)$,  and may delay its message transmission time by up to $\Memd + \Tc$ time units. If the noise correlation function $\Cwc(t,\tau)$, characterized in Section \ref{subsec:problem_formulation}, satisfies
    \begin{equation}
        \label{eqn:ConditionThmUnifConv_2}
        \mathop{\min}\limits_{0 \leq t \leq \Tc}\bigg\{\Cwc(t,0)-2\Memd\cdot\mathop{\max}\limits_{|\lambda|>\frac{\Tc}{p+1}}
        \big\{|\Cwc(t, \lambda)|\big\}\bigg\}\geq \gamma_1>0,
    \end{equation}
 and the power constraint $P$ satisfies
 \begin{equation}
 \label{eqn:cond_P}
        P >  \mathop{\max}\limits_{t\in [0,\Tc]}\Big( \Cwc(t,0)+\Memd\cdot\mathop{\max}\limits_{|\lambda|>\frac{\Tc}{p+1}}\big\{|\Cwc(t, \lambda)|\big\}\Big),
 \end{equation}
    then, for any fixed value of $\eps\in (0,1)$, $\eps \notin \mQ$, capacity is given by
	\begin{equation}
	    \label{eqn:capacity-lim}
	    \Capacity_\myEps = \mathop{\lim\!\inf}\limits_{n \rightarrow \infty} \Capacity_n,
	\end{equation}
	where $C_n$ is obtained via \eqref{eqn:MainThmCst}-\eqref{eqn:DefCn}. Furthermore, Gaussian inputs are optimal.
\end{theorem}
				
\begin{IEEEproof}
 The proof is detailed in Appendix \ref{app:Proof2}.
\end{IEEEproof}

\begin{remark}[On the Capacity for Synchronous Sampling]
{\em When $\eps\in\mQ^{++}$, i.e.,  $\eps = \frac{u}{v}$ for some $u, v \in \mNp$, then the \ac{dt} noise process  $\Weps[i]$ is \ac{wscs} with a period which is equal to $p_{u,v} = \Td \cdot v + u$. As noted in Sections \ref{sec:Intro} and \ref{Sampled_wscs}, such a sampling scenario corresponds to {\em synchronous sampling}, whose capacity, when $\tau_0$ is given, was characterized in \cite{shlezinger2015capacity}, \cite{shlezinger2016capacity} and is given by \eqref{eqn:MainThmCst}$-$\eqref{eqn:Cn_tau0_def}, where $p_n$ is replaced by $p_{u,v}$,  $\tilde{\myVec{W}}_{n}^{\{\Td_n\}}[i]$ is replaced by  $\tilde{\myVec{W}}_{u,v}^{\{p_{u,v}\}}[i]$, and the quantities appearing in the statement of \eqref{eqn:MainThmCst}$-$\eqref{eqn:Cn_tau0_def} are replaced by appropriate corresponding quantities. Note that for such an $\eps$, then when $n=b\cdot v$, $b\in\mNp$, we have that $\eps_n=\frac{u}{v}$, consequently, for $\eps\in\mQ^{++}$ it follows that $\liminfcn \le C_{\frac{u}{v}}$. Yet, from the upper bound in \eqref{appndxEqn:UpperBound} it holds that $C_{\eps}\le \liminfcn$, hence for  $\eps\in\mQ^{++}$ we immediately obtain $C_{\eps}=\liminfcn$.
}
    
\end{remark}

\begin{remark}[Elaboration on Condition \eqref{eqn:ConditionThmUnifConv_2}]
    \label{rem:cond_strict_diag}
{\em    Condition \eqref{eqn:ConditionThmUnifConv_2} guarantees that for any sequence of $k$ samples of the noise process, denoted  $\Wnk\equiv\big\{\Wn[i]\big\}_{i=0}^{k-1}$, and for any sampling phase $\tau_0\in [0,\Tc)$, the correlation matrix, denoted $\corrmatwn(\tau_0)$, is \ac{sdd} \cite[Eqn. (4)]{ahlberg1963convergence}. 
To see this, recall that by definition, $\big(\corrmatwn(\tau_0)\big)_{u,v}  \triangleq  \dsE\big\{W_{n}[u]\cdot W_{n}[v]\big|\tau_0\big\} \equiv \disccorrn^{\{\tau_0\}}[v,u-v]$, $0\le u, v\le k-1$.
The diagonal dominance can be verified by noting that for any $0 \le u \le k-1$ it holds that
\begin{align}
    \label{eqn:diagonal_dominance}
    &\Big|\big(\corrmatwn(\tau_0)\big)_{u,u}\Big|-\mathop{\sum}\limits_{v=0, v\neq u}^{k-1}\Big|\big(\corrmatwn(\tau_0)\big)_{u,v}\Big|\notag\\
    &\quad = \Big| \dsE\big\{\big(W_{n}[u]\big)^2\big|\tau_0\big\}\Big|-\mathop{\sum}\limits_{v=0, v\neq u}^{k-1}\Big| \dsE\big\{W_{n}[u]\cdot W_{n}[v]\big|\tau_0\big\}\Big|\notag\\
    &\quad \stackrel{(a)}{=}\Cwc\left(u\cdot \frac{\Tc}{p+\epsilon_n}+\tau_0,0\right)-\mathop{\sum}\limits_{v=0, v\neq u}^{k-1}\left|\Cwc\left(u\cdot \frac{\Tc}{p+\epsilon_n}+\tau_0,(v-u)\cdot\frac{\Tc}{p+\epsilon_n}\right)\right|\notag\\
    &\quad \geq \mathop{\min}\limits_{0 \leq t \leq \Tc}\bigg\{\Cwc(t,0)-2\Memd\cdot\mathop{\max}\limits_{|\lambda|>\frac{\Tc}{p+1}}\big\{|\Cwc(t, \lambda)|\big\}\bigg\}\geq \gamma_1>0,
\end{align}
where (a) follows from the definition of the autocorrelation function \eqref{eqn:AnsycAutocorr}, since for the real-valued random process $\Wn[i]$, $i\in\mN$ we can write $\dsE\big\{W_{n}[u]\cdot W_{n}[v]\;\big|\tau_0\big\} = \dsE\big\{W_{n}[v]\cdot W_{n}[u]\;\big|\tau_0\big\}$,
$0\le u, v \le k-1$.
        
Thus, condition \eqref{eqn:ConditionThmUnifConv_2} guarantees that the correlation decreases sufficiently fast as the lag increases, such that a strictly diagonally dominant noise correlation matrix is obtained for any $n, k \in \mNp$. This facilitates upper bounding the eigenvalues of the inverse noise correlation matrix, see Appendix \ref{Appndx:NoiseCorrelationConvergence}.
It is noted, however, that condition \eqref{eqn:ConditionThmUnifConv_2} is stricter than the actual requirement, which is more involved to state analytically: In fact, from step (c) in the derivation of Eqn. \eqref{eqn:UpperBndMaxEig}, it is only required that for every $n\in\mN^+$ sufficiently large, as well as for $\eps$, the correlation matrices $\corrmatwn(\tau_0)$ and $\corrmatwe(\tau_0)$ are \ac{sdd} for all $\tau_0\in[0,\Tc]$. In the simulations in Sec. \ref{sec:Simulations} we directly verify the \ac{sdd} condition.
}
\end{remark}

\begin{remark}[Elaboration on Condition \eqref{eqn:cond_P}]
\label{rem:cond_P}
{\em 
    The lower bound on the power $P$ guarantees that for any sequence of $k$ noise samples, the eigenvalues of the corresponding noise correlation matrix are smaller than $P$. Then, when in Step (b) in the derivation of \eqref{eqn:upper bound on Cn},  waterfilling is applied over the eigenvalues of the noise correlation matrix, see, e.g., \cite[Eqn. (15)-(16)]{cover1989gaussian}, it follows that power is allocated to all eigenvalues.  This facilitates the bounding of the \ac{wscs} channel capacity by the mutual information of any segment of length $k$, where $k$ is sufficiently large, up to an arbitrarily small error.
    }
\end{remark}

\begin{sloppypar}
\begin{remark}[Intuition from Stationary Analysis Does Not Apply Here]
{\em	We emphasize that while it seems intuitive that the limit in \eqref{eqn:capacity-lim} holds, our result shows that for this limit to hold, additional conditions on the noise statistics are required. This highlights the fact that when considering non-stationary channels, then  intuition based on stationary processes may lead to incorrect perceptions. In the current work, we obtain capacity characterization when the noise correlation decays sufficiently fast. 	If this is not the case, 
it is not possible to uniformly bound the difference between the mutual information expressions corresponding to the channels \eqref{eqn:AsnycModel1} and \eqref{eqn:AsnycModel2}, subject to \eqref{eqn:AsyncConst1}, and consequently, showing the interchangeability of the limits in $n$ (the approximation index) and in $k$ (the sequence length) becomes an involved task.
We also require $P$ to be sufficiently large to allow relating the mutual information of a finite segment of length $k$ and capacity. 
Let $\cdf{X}$ denote the \ac{cdf} of the \ac{rv} $X$. and consider, for example, the limit $\liminfk \frac{1}{k}I(\Xink;\Yepsk|\tau_0)$ in Corollary \ref{corr:No rate adapt}. In Lemma \ref{lem:tightness} we show that $ \limn \frac{1}{k}I(\Xnkopt;\Ynk|\taunk\opt)=\frac{1}{k} I(\Xinkopt;\Yepsk|\tauepsk\opt)$, where $(\cdf{\Xnkopt},\taunk\opt)$ and $(\cdf{\Xinkopt},\tauepsk\opt)$ maximize the \ac{lhs} and \ac{rhs} respectively. Following the proof of Lemma  \ref{lem:tightness} it is straightforward to conclude that 
\[
     \liminfk \frac{1}{k} I(\Xinkopt;\Yepsk|\tauepsk\opt) =  \liminfk \limn \frac{1}{k}I(\Xnkopt;\Ynk|\taunk\opt).
\]
However, since the convergence of the sequence $\Big\{ \frac{1}{k}I(\Xnkopt;\Ynk|\taunk\opt)\Big\}_{n\in\mNp}$ is generally not uniform in $k\in\mNp$, it is not possible to switch the order of the limits on the \ac{rhs}, and we cannot relate $\liminfk \frac{1}{k} I(\Xinkopt;\Yepsk|\tauepsk\opt)$ and $C_n$.  This lack of uniform convergence follows as we show in the proof of Lemma \ref{lem:tightness} that  the distance $\Big|\frac{1}{k} I(\Xinkopt;\Yepsk|\tauepsk\opt) -   \frac{1}{k}I(\Xnkopt;\Ynk|\taunk\opt)\Big|$ is proportional to the distance $\zeta[i]\triangleq \Big| \Cwc\left(\frac{i \cdot\Tc}{\Td + \myEps} + \tauepsk\opt,\frac{\Delta \cdot\Tc}{\Td + \myEps} \right) - \Cwc\left(\frac{i \cdot\Tc}{\Td + \eps_n} + \taunk\opt,\frac{\Delta \cdot\Tc}{\Td + \eps_n} \right)\Big|$, $0\le i \le k-1$. For a fixed $n\in\mNp$, we obtain that $\zeta[i]$ periodically increases and decreases over the range $0\le i \le k-1$. Then, as $k$ increases, this distance may increase up to the maximal magnitude of the correlation function, and as consequence the mutual information expressions do not converge
as $k$ increases. Thus, to keep this distance bounded as $k$ increases, $n$ has to increase as well, which implies that convergence in $n$ is not uniform in $k$.
	} 
\end{remark}
\end{sloppypar}

\begin{remark}[Relationship with the Work of Cover and Pombra]
{\em	In \cite{cover1989gaussian}, the  capacity of additive Gaussian noise channels with and without feedback was considered.
By analyzing the distribution of a quadratic form in Gaussian \acp{rv}, it is shown in \cite{cover1989gaussian}  that the asymptotic equipartition property applies to nonergodic Gaussian processes. While in  Appendix \ref{Appndx:Capacity_General} we also analyze a quadratic form in Gaussian \acp{rv}, it is emphasized that the analysis for our situation is considerably more involved than for the situation in  \cite[Sec. V]{cover1989gaussian}, as in our scenario the weighting matrix is not the inverse of the correlation matrix of the Gaussian vector, and moreover, it is an indefinite matrix. The resulting \ac{rv} is thus {\em a weighted sum} of chi-square \acp{rv}, which is not distributed as a chi-square \ac{rv} with a higher degrees-of-freedom, differently from \cite{cover1989gaussian}. In fact, there is no explicit expression for the \ac{pdf} of the above resulting \ac{rv}, which necessities the use of a completely different set of arguments in the analysis in Appendix \ref{Appndx:Capacity_General}. It is also noted that the information-spectrum framework, which was introduced several years after the work of \cite{cover1989gaussian}, has not been applied, as far as we know, to the capacity analysis of channels with additive non-stationary Gaussian noise.
Lastly, note that  in the achievability proof in \cite{cover1989gaussian}, the exponent of the probability of decoding error depends on the blocklength, see \cite[Eqns. (66)-(67)]{cover1989gaussian}. Thus, it is not clear how  it is possible to conclude a vanishing probability of error for a given rate $C_{n, {\rm FB}}$ in the asymptotic as the blocklength increases to infinity, using the arguments in \cite[Sec, VII]{cover1989gaussian} without additional conditions.

} 

\end{remark}

\begin{remark}[Evaluating Capacity in the Presence of Multiple Interferers]
    \label{comment:MultipleInterfers} 
   
{\em      The setup  in Sec. \ref{subsec:problem_formulation} considered the case of a single interferer. We note that when multiple interferers are present at fixed locations and when the channels between the interferers and the receiver are invariant, then the aggregate interference  is a \ac{ct} \ac{wscs} Gaussian process. 
In addition to Gaussianity of the aggregate interference, we note that, in order to apply the scheme derived in the proof of Thm. \ref{thm:AsycCap}, the transmitter should acquire and synchronize with the noise correlation function.  In this context we may consider two possible scenarios: In the first scenario, referred to as {\em partial coordination},  the receiver and transmitter can obtain (e.g., through a control channel) the signal parameters of each interferer (e.g., modulation type, symbol duration, pulse shape for single carrier or subcarrier frequencies for \ac{ofdm}). With these parameters, the receiver and transmitter can obtain the \ac{ct} correlation function of each interferer. Then, to obtain the aggregate \ac{ct} correlation function, the receiver needs to inform the transmitter the delays at which each interferer is received. In multi-interferers scenarios in which this is feasible, then the approach of Thm. \ref{thm:AsycCap} can be applied. In the second scenario, referred to as {\em uncoordinated interferers}, the transmitter and receiver each need to independently obtain the correlation function of the aggregate \ac{ct} interference. In such a case, as the relative delays from each interferer to the receiver and to the transmitter are different, then the estimated aggregate correlation function will likely be different between the transmitter and  the receiver. Therefore, in such a scenario,   multiple uncoordinated interferers cannot be handled via the scheme of Thm. \ref{thm:AsycCap}.}
\end{remark}


\section{Numerical Examples and Discussion}
\label{sec:Simulations}

In this section we use numerical evaluations to derive insights from the analytic capacity characterization of Thm. \ref{thm:AsycCap}.
First, in Subsection \ref{subsec:Sim_capCS}, we consider the evolution of $\Capacity_n(\tau_0)$ w.r.t the index $n$ and the impact of the sampling phase $\phi=\tau_0/\Tc\in[0,1)$ on capacity. Next, in Subsection \ref{subsec:Cap_sampRate}, we study the variations of the capacity of the sampled \ac{dt} channel for different sampling rates and different sampling phases. We also compare the capacity results with the capacity obtained for additive {\em memoryless} \ac{wscs} Gaussian noise channels having the same noise power and signal power.

\begin{figure}
	 \centering
	 {\includegraphics[width=0.85\linewidth]{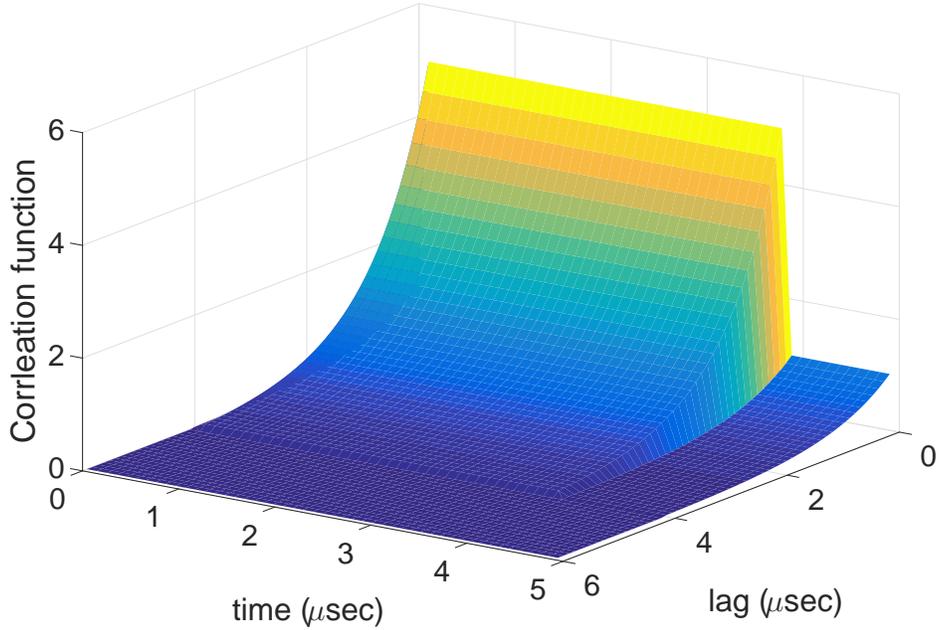}}
	\caption{The correlation function $\Cwc(t,\lambda)$ of Eqn. \eqref{correlation_fxn} at positive lags $\lambda \ge 0$, with normalized sampling time offset $\phi=0$ and $\DC=0.75$.}
	\label{fig:corrprof}		
\end{figure}

To model the correlation function of the \ac{ct} \ac{wscs} noise we define a periodic pulse function, $\Pi_{\DC,\Trise}(t)$, having a rise/fall time of $\Trise=0.01$, a period of $1$, and a \ac{dc} of $\DC$, which is varied in the range $0\leq \DC \leq 0.75$; hence, $\Pi_{\DC,\Trise}(t)=\Pi_{\DC,\Trise}(t+1)$ $\forall t \in \mR$, and for $t\in[0,1)$ the pulse function is expressed mathematically as:
\begin{equation}
	\label{eqn:PerWaveFunc}
	\Pi_{\DC,\Trise}(t) = \begin{cases}
	\frac{t}{\Trise} & t \in [0,\Trise] \\
	1	&				t \in (\Trise, \DC+\Trise ) \\
	1 - \frac{t-\DC-\Trise}{\Trise} & t \in[\DC +\Trise, \DC +2\cdot \Trise] \\
	0 & t\in (\DC +2\cdot \Trise, 1).
	\end{cases} 
	\end{equation}
Let the period of the \ac{ct} correlation function $\Cwc(t,\lambda)$ be 
$\Tc=5$ [$\mu$sec].
Then, given a normalized sampling time offset $\phi \in [0, 1)$, we express the time-varying variance, $\Cwc(t,0)$, as
\[
    \Cwc(t,0)=1+4\cdot \Pi_{\DC,\Trise}\left(\frac{t}{\Tc}-\phi\right). \nonumber
\]
For our setup, the correlation length of the noise process in \ac{ct} is set to $\Memc=4$ [$\mu$sec] and the temporal correlation is modeled as a decaying exponential function for all lags $|\lambda|\leq \Memc$, 
i.e., the correlation at any  lag $\lambda>0$ is given by 
\begin{equation}
\label{correlation_fxn}
    \Cwc(t,\lambda)=\begin{cases}
    e^{-\lambda \cdot10^6}\cdot \Cwc(t,0) & ,0\le\lambda\leq \Memc \\
    0 & ,\lambda > \Memc
    \end{cases},
\end{equation}
and for $\lambda<0$ we use $\Cwc(t,\lambda)=\Cwc(t+\lambda,-\lambda)$.
This correlation function is depicted in Fig. \ref{fig:corrprof} for a single period, $0\le t \le \Tc$, $\DC=0.75$ and $\phi=0$, and $0\le\lambda\le 6$.

	\subsection[Convergence of Cn to Ceps]{Convergence of $\{\Capacity_n(\tau_0)\}_{n \in \mNp}$}
	\label{subsec:Sim_capCS}

	As stated in Theorem \ref{thm:AsycCap}, if the correlation function of the \ac{dt} noise  satisfies the condition \eqref{eqn:ConditionThmUnifConv_2} and the power satisfies condition \eqref{eqn:cond_P}, 
	then the capacity with asynchronous sampling, $\Capacity_\eps$, is equal to the limit-inferior of a sequence of capacities  corresponding to synchronous sampling, $\{\Capacity_n(\tau_0)\}_{n \in \mNp}$. For evaluating this sequence, we set the following parameter values: 
	$\eps=\frac{\pi}{7}$, 
	$p=2$, $\DC \in \{0.45,0.75\}$, $\phi \in \left\{0, \frac{\pi}{20}\right\}$, and the input power constraint $P=10$. First, we evaluate $\Capacity_n(\tau_0)$ using $\eqref{eqn:MainThmCst}$--$\eqref{eqn:Cn_tau0_def}$ for each $n$ and then normalize it by its respective sampling interval $\Tsamp(\eps_n)\triangleq \frac{\Tc}{p+\eps_n}$ to obtain $\Capacity_n(\tau_0)$ in \ac{bps}.
	We note that $\Capacity_n(\tau_0)$ can be evaluated irrespective of condition \eqref{eqn:ConditionThmUnifConv_2}, yet to conclude about $\Capacity_{\eps}$, either \eqref{eqn:ConditionThmUnifConv_2} or the \ac{sdd} condition have to be verified as discussed in Comment \ref{rem:cond_strict_diag}, in addition to condition \eqref{eqn:cond_P}. Recall the definition of $\eps_n$: $\eps_n=\frac{\lfloor n \cdot \myEps \rfloor}{n}\rightarrow\eps$ as $n\rightarrow \infty$; then, it follows that the sampling interval $\Tsamp(\eps_n)$ converges to $\Tsamp(\eps)\triangleq \frac{\Tc}{p+\eps}$ as $n$ increases. We recall that since $\eps_n$ is rational, then the resulting \ac{dt} sampled noise is \ac{wscs} with a fundamental period of $p_n=p\cdot n+\lfloor n \cdot \myEps \rfloor$. 
		\begin{figure}
	    \centering
	    \begin{minipage}{0.55\textwidth}
			\hspace{-0.8cm}{\includegraphics[width=1.05\columnwidth]{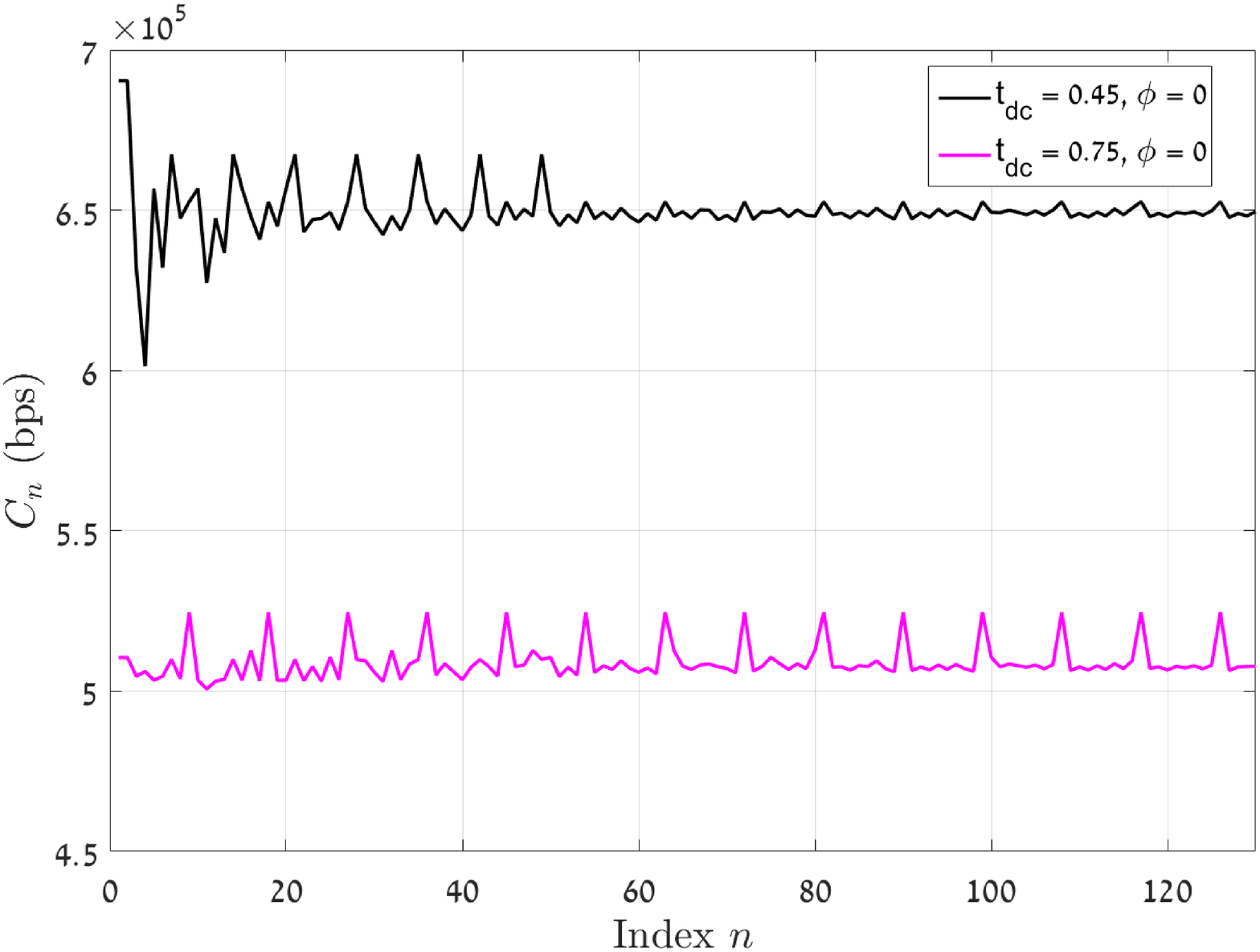}}
			\caption{$C_n(\tau_0)$ versus $n$, for  $\tau_0=0$.
			}
			\label{fig:Cn_vs_n_0}		
		\end{minipage}
		\begin{minipage}{0.55\textwidth}
			\hspace{-1.2cm}{\includegraphics[width=1.05\linewidth]{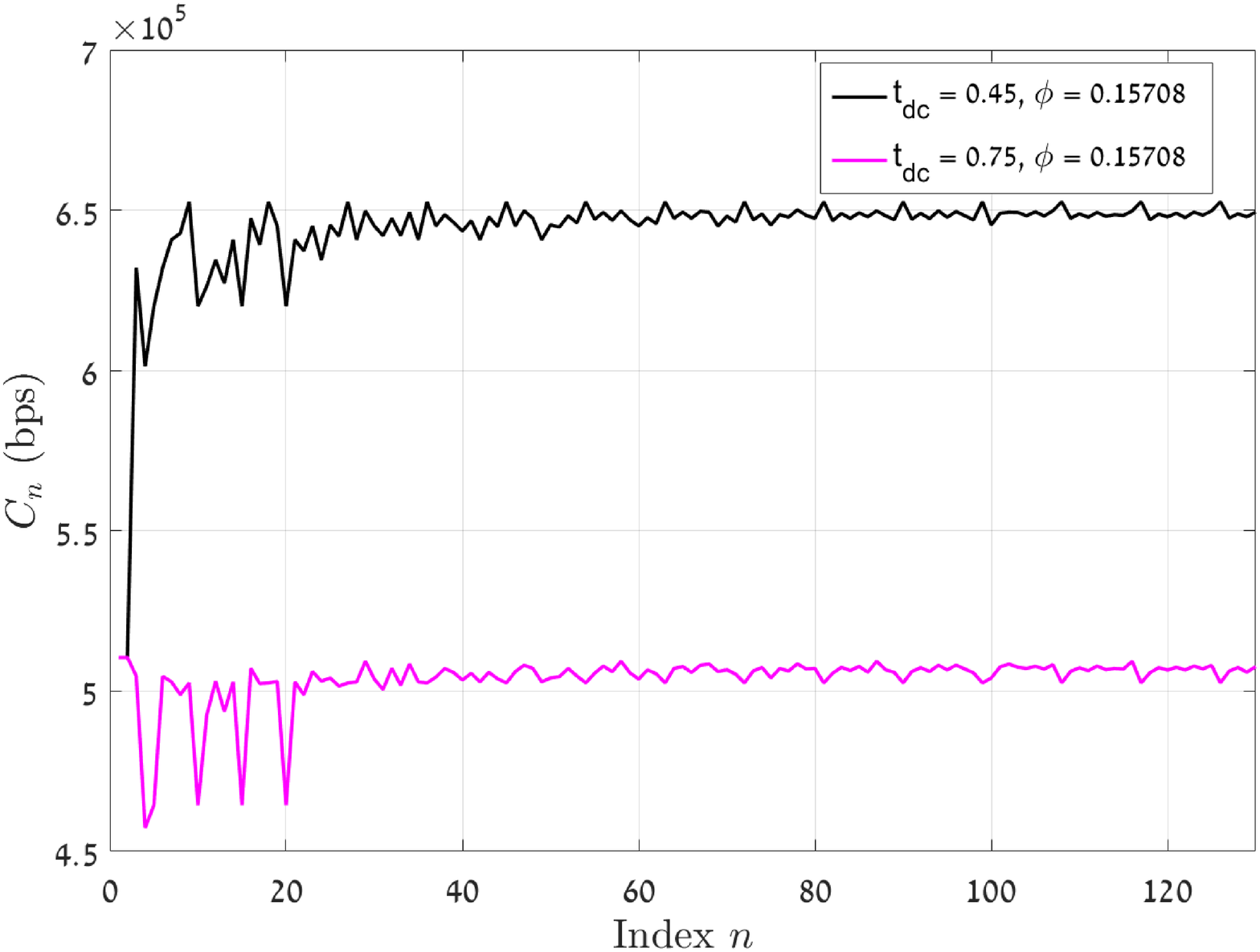}}
			\caption{$\Capacity_n(\tau_0)$ versus $n$,  $\tau_0=\frac{\pi}{20}\Tc$.
			}
			\label{fig:Cn_vs_n_0375}
		\end{minipage}
	\end{figure}
	
	 Figs. \ref{fig:Cn_vs_n_0} and \ref{fig:Cn_vs_n_0375} depict $\Capacity_n(\tau_0)$ for  normalized sampling time offsets of $0$ and $\frac{\pi}{20}$ respectively, for both considered $\DC$ values, where	 $n=\{1,2,...,130\}$.  We observe from the figures that capacity is lower when $\DC$ is higher. This can be explained by the fact that the time-averaged noise power increases as $\DC$ increases. We also observe that the variations in the capacity $\Capacity_n(\tau_0)$ are more pronounced at smaller $n$. This is because at smaller $n$, the resulting fundamental period of the \ac{dt} noise correlation function, $p_n$, consists of  only a few samples, which are sparsely spaced across the period of the \ac{ct} noise correlation function. Then, for the smaller values of $n$, as $n$ varies, the sampling interval varies significantly, and consequently, the values of the sampled noise correlation function may significantly vary as well.  
	 At higher $n$, (i.e., higher $p_n$), it is observed that, as expected, the sequence $\{\Capacity_n(\tau_0)\}_{n\in \mNp}$ does not converge to a limiting value, since the limiting noise process $\Weps[i]$ is non-stationary. However, for sufficiently large $n$, the variations of $\Capacity_n(\tau_0)$ as $n$ increases, seem to follow a regular pattern. This can be explained by noting that for higher $n$, as $n$ increases, the variations of the sampling instances of the CT noise correlation function become smaller, and accordingly, the values of the sampled correlation function do not vary significantly with $n\in\mNp$. 
	 
	 It is also observed from both  Figs. \ref{fig:Cn_vs_n_0} and \ref{fig:Cn_vs_n_0375} that at the smaller values of $n$, the nature of the variations in $\Capacity_n(\tau_0)$ is highly dependent on $\phi=\tau_0/\Tc$. For example, with $\DC=0.45$ and at $n\in[2,25]$ the  value of $\Capacity_n(0)$ is within the range  $\Capacity_n(0) \in [0.601, 0.690]$ \ac{Mbps}  and $\Capacity_n(\frac{\pi}{20}\cdot\Tc) \in [0.510,0.652]$ \ac{Mbps}; with $\DC=0.75$ and $n\in[2,25]$ then $\Capacity_n(0) \in [ 0.501, 0.525]$ \ac{Mbps}  and $\Capacity_n(\frac{\pi}{20}\cdot\Tc) \in [0.457, 0.510]$ \ac{Mbps}. 
	 At higher $n$, these capacity variations become periodic within a constant range.  
	 
	 Figs.  \ref{fig:Cn_vs_n_0} and \ref{fig:Cn_vs_n_0375} also clearly demonstrate that  the capacity with {\em synchronous} sampling may depend on the {\em sampling phase} $\phi$ (i.e.,  the values of $\Capacity_n(\phi\cdot\Tc)$ as $n$ increases  may depend on $\phi$). Note that the {\em capacity} with  {\em asynchronous} sampling, which is the limit-inferior of  $\{\Capacity_n\}_{n\in\mNp}$, is {\em independent} of the {\em sampling phase}.
	 This  is in agreement with engineering intuition: Since with asynchronous sampling the resulting \ac{dt} process is \ac{wsacs}, it is reasonable that capacity should be affected mainly by the \ac{dc} and not by the sampling phase. In the setup of Thm. \ref{thm:AsycCap} this follows as the transmitter may delay the transmission of a message to start at the optimal phase, which is also known at the receiver (via knowledge of the autocorrelation function), thereby facilitating Tx-Rx coordination. 
	 Numerically, the limit-inferior of $\Capacity_n(\phi\cdot\Tc)$ for $\DC=0.45$ was evaluated at $0.648$ \ac{Mbps}
	 for both $\phi=0$ and  $\phi=\frac{\pi}{20}$, and for $\DC=0.75$ it was evaluated at $0.503$ \ac{Mbps} for both values of $\phi$.

\begin{figure}
	    \centering
	    \begin{minipage}{0.49\textwidth}
		\hspace{-0.99cm}{\includegraphics[width=1.15\columnwidth]{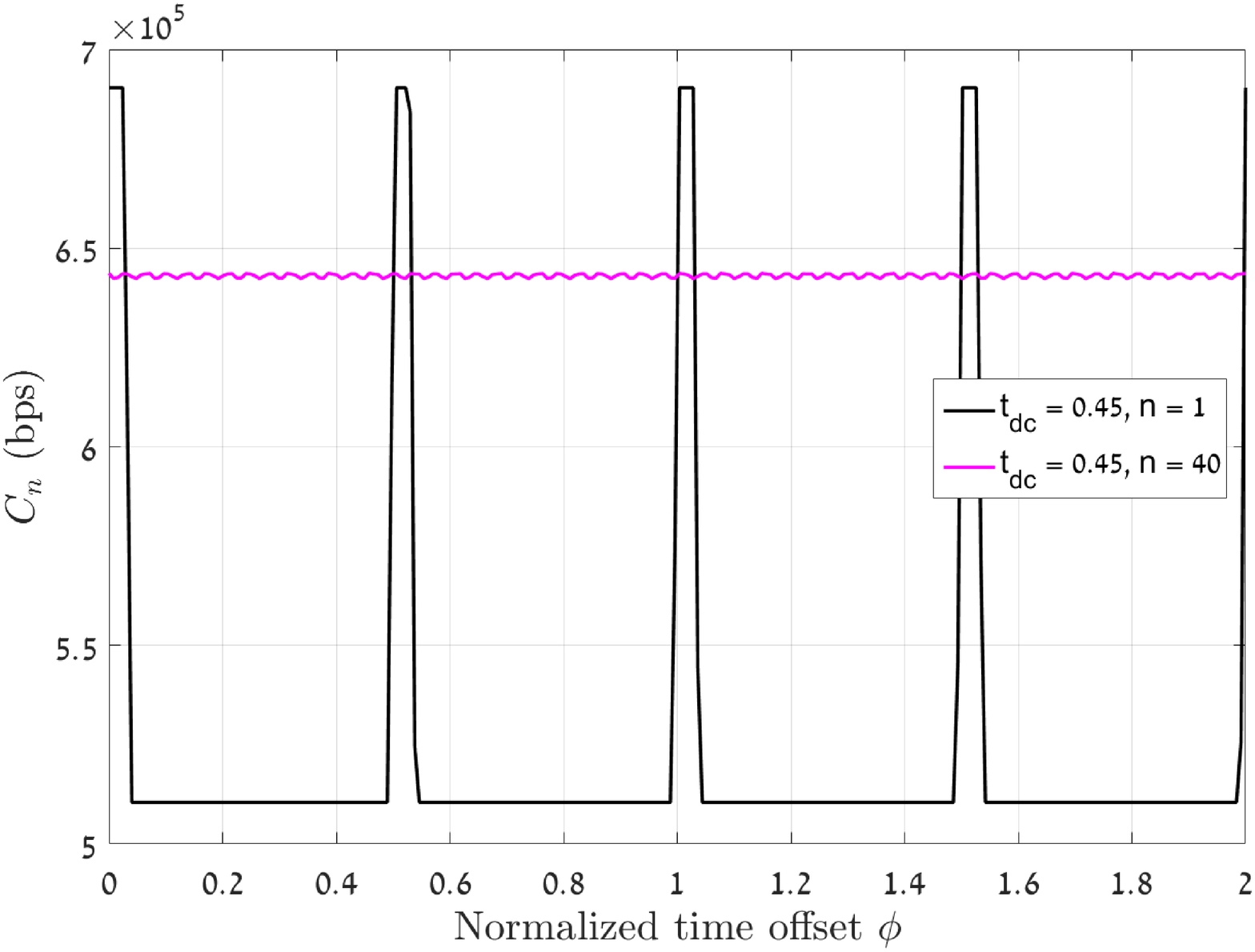}}
			\caption{$C_n(\tau_0)$ versus $\phi=\tau_0/\Tc$; $\DC=0.45$.
			}
			\label{fig:Cn_vs_offset47}		
		\end{minipage}
		\begin{minipage}{0.49\textwidth}
			\hspace{-0.7cm}{\includegraphics[width=1.15\columnwidth]{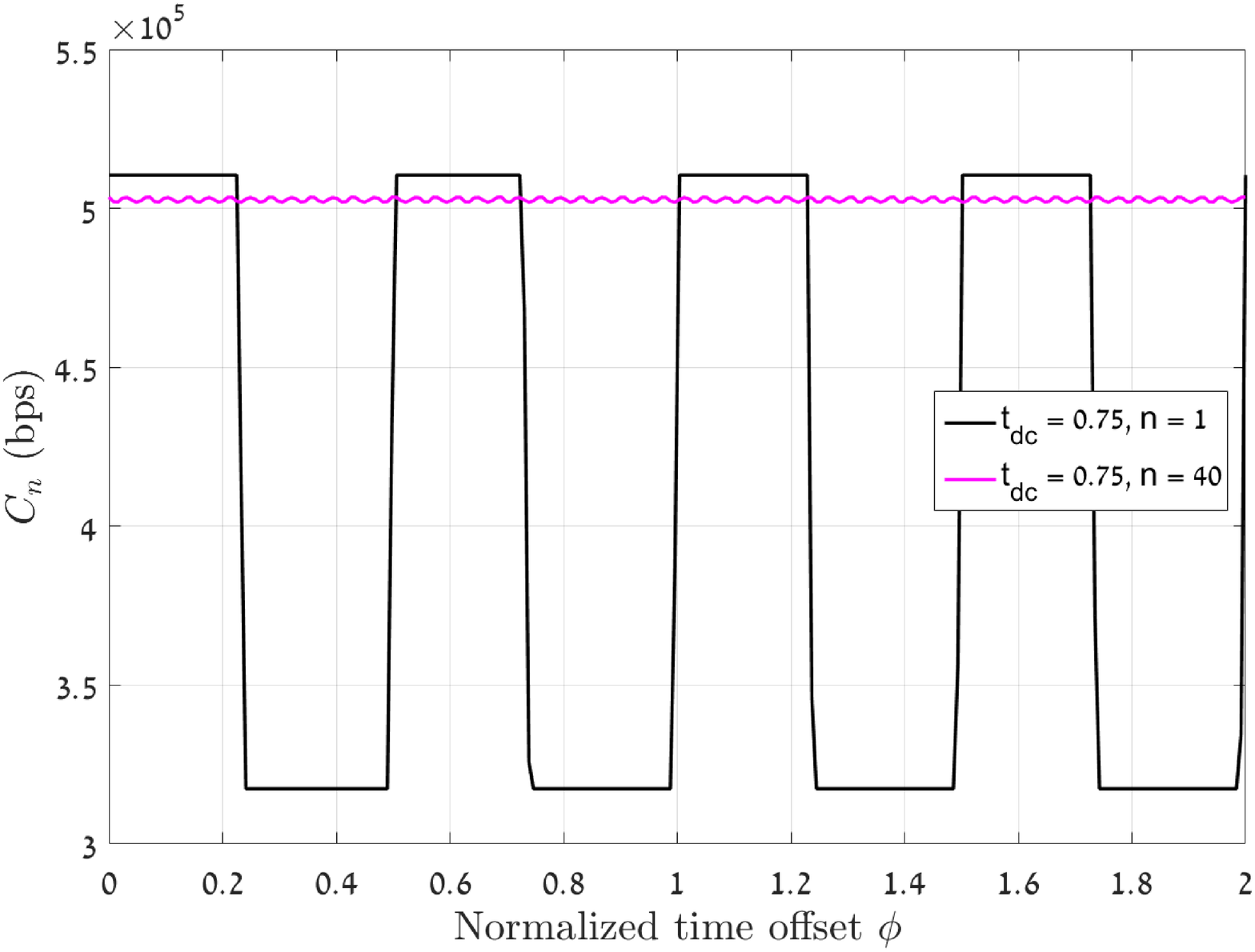}}
			\caption{$C_n(\tau_0)$ versus $\phi=\tau_0/\Tc$; $\DC=0.75$.
			}
			\label{fig:Cn_vs_offset75}
		\end{minipage}
	\end{figure}

	 To further illustrate this behaviour, Figs. \ref{fig:Cn_vs_offset47} and \ref{fig:Cn_vs_offset75} depict the capacity values for sampling time offsets $\phi \in [0,2]$ for two values of the approximation indices $n$: At $n=1$ ($p_n = p = 2$) we observe  significant variations of $\Capacity_1(\phi\cdot\Tc)$ for both \ac{dc} values, $45\%$ and $75\%$. On the other hand, at a sufficiently high $n$, e.g., $n=40$ ($p_n=97$), we observe that capacity $\Capacity_{40}(\phi\cdot\Tc)$  varies very little with $\phi$ for both \ac{dc} values. This follows as at longer periods the correlation function of the process $\Wn[i]$ more closely resembles $\Cwc(t,\lambda)$.
	 We also observe that the variations are periodic in all setups, which is expected, as for a fixed and finite $n\in\mNp$, the \ac{dt} noise process $W_n[i]$ is  \ac{wscs}, thus, its \ac{dt}  correlation function will repeat identically after a single period shift (i.e., an integer value of $\phi$).

	\subsection{Variations of Capacity with the Sampling Rate}
	\label{subsec:Cap_sampRate}

\begin{figure}
	    \centering
	    \begin{minipage}{0.5\textwidth}
			\hspace{-0.9cm}{\includegraphics[width=1.1\linewidth]{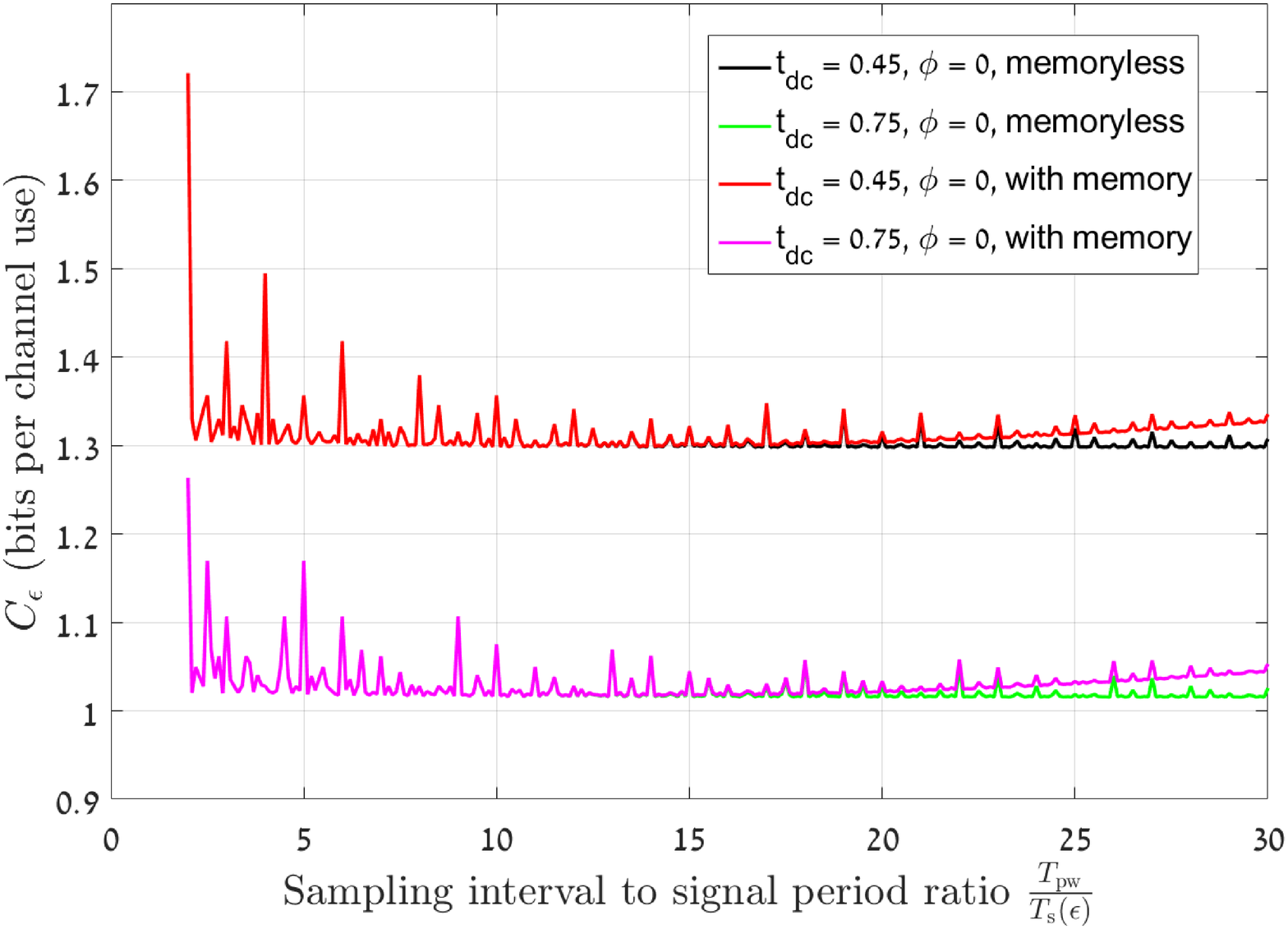}}
			\caption{$\Capacity_{\frac{u}{v}}(\tau_0)$ versus $\frac{\Tc}{\Tsamp(\eps)}$ for offset $\tau_0=0$.
			}
			\label{fig:C_vs_freq_0}		
		\end{minipage}
		\begin{minipage}{0.5\textwidth}
			\hspace{-0.7cm}{\includegraphics[width=1.1\linewidth]{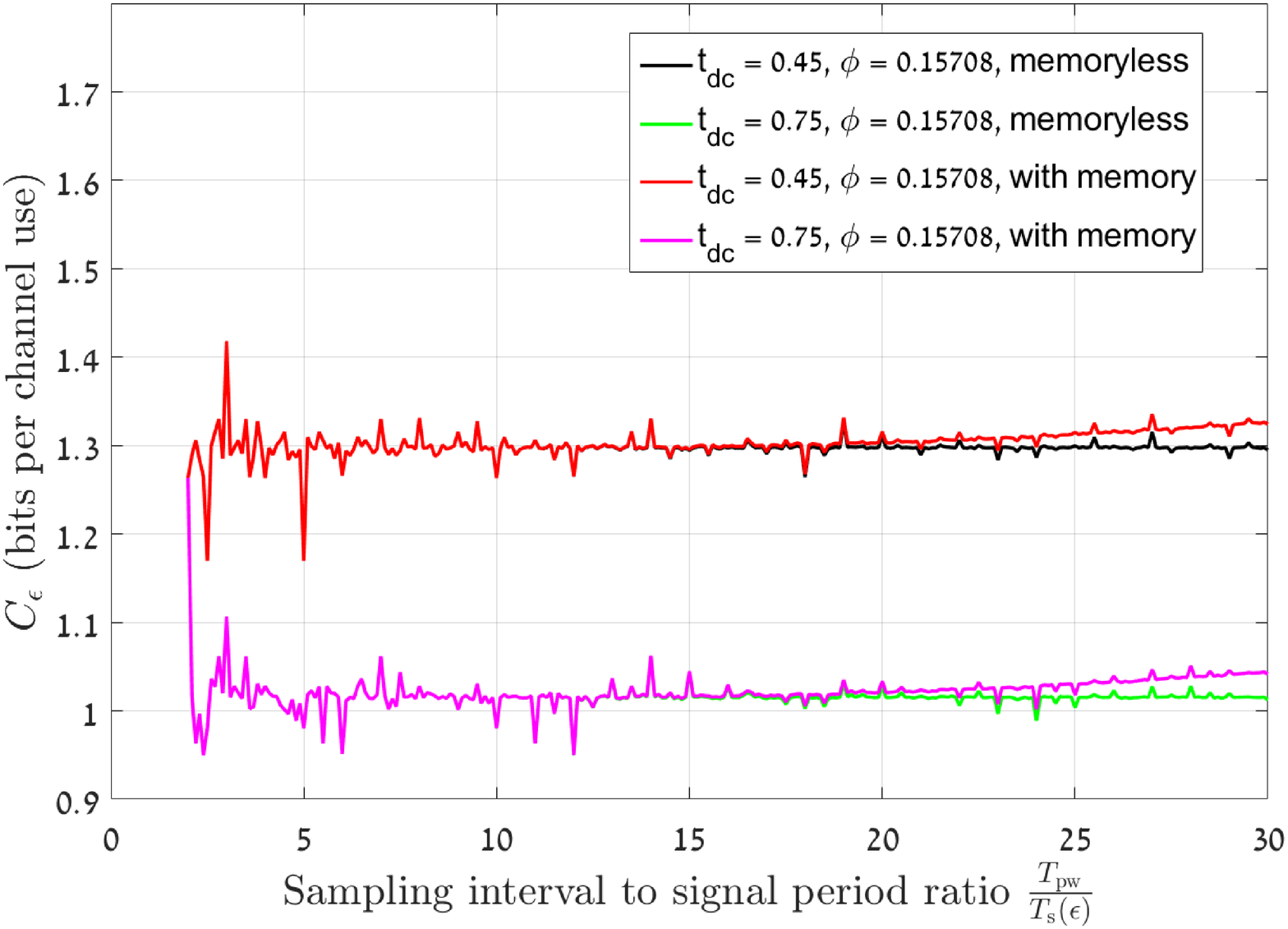}}
			\caption{$\Capacity_{\frac{u}{v}}(\tau_0)$ versus $\frac{\Tc}{\Tsamp(\eps)}$ for offset $\tau_0=\frac{\pi}{20}\Tc$.
			}
			\label{fig:C_vs_freq_0375pi}
		\end{minipage}
	\end{figure}

	Next, we examine how variation of the sampling rate affects the capacity of \ac{dt} channels obtained by sampling \ac{ct} channels with additive \ac{wscs} Gaussian noise having a finite memory, and compare their capacity with that of  \ac{dt} channels with memoryless sampled noise  having the same variance as the noise with finite memory.  The results are depicted in Figs. \ref{fig:C_vs_freq_0} and \ref{fig:C_vs_freq_0375pi}, which present the evaluated capacity (in bits per channel use) for $\phi=0$ and for $\phi=\frac{\pi}{20}$, respectively, for sampling intervals in the range $2\leq \frac{\Tc}{\Tsamp(\eps)}\leq 30$. Recall from the problem formulation in Section \ref{subsec:problem_formulation} that $\frac{\Tc}{\Tsamp(\eps)}=p+\eps$ where $p \in \mNp$ and $\eps \in [0,1)$. In Figs. \ref{fig:C_vs_freq_0} and \ref{fig:C_vs_freq_0375pi} we plot the capacity values $\Capacity_\eps$ for synchronous sampling, i.e., when $\eps$ can be written as $\eps=\frac{u}{v}$, $u,v\in\mNp$, and hence, the fundamental period of the noise statistics is given by $p_{u,v}=p\cdot v +u$. Recall that in this case, capacity depends on $\tau_0$, thus we denote  $C_{\eps}\equiv C_{\frac{u}{v}}(\tau_0)$, yet, this dependence becomes weaker as the period $p_{u,v}$ increases. To highlight the transition from memoryless channels to channels with memory, we use $10^7$ instead of $10^6$ in the power of the exponential function in \eqref{correlation_fxn}.

	For both figures, we observe an increase in the capacity (in bits per channel use) as the sampling rate increases. In addition, we note that when the values of $p$ and $\eps=\frac{u}{v}$ result in a smaller value of the period $p_{u,v}$, the capacity varies significantly, as can be seen by the peaks and dips in both the memoryless Gaussian noise plot and the plot for Gaussian noise with a finite memory; it is also observed that the variations are different for different sampling time offsets. On the other hand, when the period $p_{u,v}$ is large, the capacity approaches the asynchronous-sampling capacity and the peaks/dips  notably reduce. Moreover, at the longer periods, it is observed that capacity values are very similar for both sampling time offsets, which is reasonable when approaching the asynchronous sampling situation. Finally, it is evident from the figures that a slight change in the sampling rate can result in a significant change in the capacity. 
	As an example, consider the  plots for the finite-memory noise in Figs. \ref{fig:C_vs_freq_0} and \ref{fig:C_vs_freq_0375pi}, at $\frac{\Tc}{\Tsamp(\eps)}=5$ (i.e., $p_{u,v}=5$, which is a relatively small period) and  $\phi=0$: The capacity values for the noise with finite memory are $1.356$ and $1.170$ bits per channel use, for $\DC=45\%$ and $\DC=75\%$, respectively. However, when the sampling rate changes to $\frac{\Tc}{\Tsamp(\eps)}=5.2$, these values change to $1.302$ and $1.039$, respectively.
	The impact of the sampling phase is  more pronounced at smaller $\frac{\Tc}{\Tsamp(\eps)}$: For $\frac{\Tc}{\Tsamp(\eps)}=5$, the capacities at $\phi=\frac{\pi}{20}$ for $\DC=45\%$ and $\DC=75\%$ are  $1.170$ and $0.980$ bits per channel use, respectively, which are very different from the respective values at $\phi=0$ noted above. Lastly, 
	consider  $\frac{\Tc}{\Tsamp(\eps)}=23.2$, i.e. $p_{u,v}=116$, which is a relatively long  period for the \ac{dt} correlation function. For this sampling rate, the capacities (with memory) for $\DC=45\%$ and $\DC=75\%$ are  $1.298$ and $1.015$ bits per channel use, respectively, for both $\phi=\frac{\pi}{20}$ and for  $\phi=0$. 
	
	Another property observed from Figs. \ref{fig:C_vs_freq_0} and \ref{fig:C_vs_freq_0375pi}  is that at relatively low sampling rates (i.e., when $\frac{\Tc}{\Tsamp(\eps)}$ is smaller, e.g., $\frac{\Tc}{\Tsamp(\eps)}<12$), the sampled channels for memoryless noise and for noise with a  finite memory have approximately the same capacity. As the sampling rate is increased, it is observed that  
	the sampled channel with a finite-memory noise has a higher capacity than the sampled memoryless channel, whose capacity does not vary much with the sampling rate variation. This is explained by observing that as the sampling rate increases, the sampled noise for the case of finite-memory CT noise begins to exhibit noise correlation, which can be utilized to increase capacity via waterfilling. 
	\label{pg:waterfilling}
	As an Example, at $\DC=75\%$, $\phi=\frac{\pi}{20}$ and $\frac{\Tc}{\Tsamp(\eps)}=8.5$ the capacity is $1.014$ bits per channel use for both the finite-memory and the memoryless cases. However, as $\frac{\Tc}{\Tsamp(\eps)}$ increases, e.g., at $\frac{\Tc}{\Tsamp(\eps)}=29$, and $\phi=\frac{\pi}{20}$, the capacity is $1.310$ bits per channel use for the channel with a finite-memory Gaussian noise, whereas it is $1.285$ bits per channel use for the channel with memoryless Gaussian noise. Observe that this gap, in favor of the channel with sampled finite-memory noise  widens as the sampling rate is further increased. That said, we note that while the model considered \eqref{eqn:AsnycModel1} does not account for additive thermal noise (since $\Cwc(t,\lambda)$ has finite values), then 
	at higher sampling rates, the impact of the thermal noise should also be accounted for in addition to the interference, as higher sampling rates are associated with higher receiver bandwidths. Accounting for the thermal noise will limit the capacity increase observed in the figures.

	Our numerical evaluations reveal a very interesting phenomenon that should be considered  when designing communications systems: It is observed that capacity is greatly dependent upon the precise value of the sampling rate. It is thus recommended to take the asynchronous capacity as the practical capacity value, even if the analytical capacity value due to the nominal sampling rate used in the system design is higher. 
	The results also imply that increasing the sampling rate can increase the capacity even when the sampling rate is higher than the Nyquist rate\footnote{See \cite[Ch. 12.4.3]{gardner1990introduction} for elaboration on the Nyquist rate for bandlimited \ac{wscs} processes.
	}. This observation stands in contrast to the observation in \cite{EldarGoldsmith2013}, which studied linear, time-invariant channels with stationary Gaussian noise. Intuitively, this follows as in the current scenario, sampling is applied to a two-dimensional periodic function, hence it is not enough to be able to identify the temporal correlation profile, but also the periodicity of the correlation function, which may require higher sampling rates.
	

\section{Conclusions}
\label{sec:Conclusions}

In this work we analyzed the capacity of additive Gaussian noise channels obtained by sampling \ac{ct} channels with additive \ac{wscs} Gaussian noise, focusing on the scenario in which the sampled noise is non-stationary. We first explained that in this case, maximizing the information rate requires Tx-Rx time synchronization w.r.t the correlation function of the \ac{ct} noise, and it is not sufficient to have both the transmitter and  the receiver know the noise correlation function without such synchronization.
Subsequently, we derived a general capacity characterization when transmission delay is not allowed. 
Finally, we considered the scenario in which transmission delay of up to sum of the noise memory the noise period is allowed, for which we obtained a limiting capacity expression derived using original bounds on the optimal mutual information density rate of the channel. We then used the limiting expression to examine the impact of the combination of channel memory and sampling on the information rates of the resulting \ac{dt} channel, and presented novel insights arising from this examination. This work is another step in the study of the relationship between sampling and capacity, which provides a much needed missing link between the analog domain models and the respective digital models obtained after sampling.

\setcounter{equation}{0}
\renewcommand{\theequation}{\thesection.\arabic{equation}}
\numberwithin{proposition}{section}
\numberwithin{lemma}{section}
\numberwithin{corollary}{section}
\numberwithin{remark}{section}
\numberwithin{equation}{section}

\begin{appendices}

\section{Proof of Thm. \ref{Thm:Capacity_tau0_at_Tx_No_Adap}}
\label{Appndx:Capacity_General}

We consider the mutual information density rate for the channel \eqref{eqn:AsnycModel1}: Let $\tau_0\in [0,\Tc)$ denote the  sampling phase within a period of the correlation function of the \ac{ct} noise process $W_{\sfc}(t)$. For a given $k\in\mNp$ and a given  $\tau_0\in [0,\Tc)$, let   $\cdf{\Xink|\tau_0}\equiv\cdf{\Xepsk|\tau_0}\left(\xepsk|\tau_0\right)$ denote the \ac{cdf} of the random vector $\big\{\Xscal[i]\big\}_{i =0}^{k-1}$, which is the channel input process when transmission begins at the sampling  phase $\tau_0$. Recall that the transmitter is aware of $\tau_0$, hence, it can choose its codebook accordingly. Furthermore, the transmission scheme appends each codeword with $\Memd$ zeros, and the receiver discards the last $\Memd$ received channel outputs for each message reception. Thus, the received channel output sequences for different messages are statistically independent.
In a similar manner as in  \cite[Appendix A]{shlezinger2015capacity}, it follows that for sufficiently large $k$, this assumption does not affect the capacity.
Lastly, define the random variable corresponding to the mutual information density rate for this transmission as (see \cite[Lemma 7.16]{GraybookEntropy}):
\label{pg:DefOfZkeps}
\begin{equation}
	\zkeps{k}{\eps}\big( \cdf{\Xink|\tau_0}|\tau_0\big) \triangleq \frac{1}{k}\log \frac{\pdf{\Yepsk | \Xepsk,\tau_0}\big( \Yepsk \big| \Xepsk, \tau_0\big) }{\pdf{\Yepsk|\tau_0}\big( \Yepsk|\tau_0\big) }.\notag
\end{equation}
	

\begin{sloppypar}
Note that by \cite{cover1989gaussian}, when $\tau_0$ is given, then the mutual information for each $k\in\mNp$, for the additive Gaussian noise channel \eqref{eqn:AsnycModel1}, $\frac{1}{k}I\big(\Xink; \Yepsk|\tau_0\big)$,
is maximized, subject to an average sum-power constraint, by a Gaussian random input vector. 
We now analyze $\zkeps{k}{\eps}\big( \cdf{\Xink|\tau_0}|\tau_0\big)$ when  $\big\{\Xscal[i]\big\}_{i =0}^{k-1}$ is distributed according to the Gaussian distribution which maximizes $\frac{1}{k}I\big(\Xepsk; \Yepsk|\tau_0\big)$ 
subject to the constraint $\frac{1}{k}\sum_{i=0}^{k-1}\dsE\big\{(X[i])^2\big\}\le P$. In Lemma \ref{lemma:PowerConstrSatisfied}, at the end of the proof, we will show that such a random codebook generation process results in the constraint \eqref{eqn:AsyncConst1} satisfied with a probability which is arbitrarily close to $1$, as long as $\{X[i]\}_{i=0}^{k-1}$ satisfies the trace constraint which appears in the statement of the theorem. 
Let  $\Xinkopt$ denote a random  process generated according to this maximizing input distribution,  let $\corrmatye(\tau_0)$, $\corrmatwe(\tau_0)$, and $\corrmatxeopt(\tau_0)$ denote the correlation matrices of $\Yepsk$, $\Wepsk$ and of $\Xinkopt$, respectively, {\em when $\tau_0$ is given},
and recall  the definition of the correlation matrix $\corrmatwe(\tau_0)$:
\begin{eqnarray}
\label{eqn:corrmatdefe}
        \left(\corrmatwe(\tau_0)\right)_{u,v} & \triangleq & \dsE\big\{W_{\myEps}[u]\cdot W_{\myEps}[v]\big|\tau_0\big\} \equiv \disccorr^{\{\tau_0\}}[v,u-v],
\end{eqnarray}
for $(u,v) \in \mK \times \mK$, see Eqn. \eqref{eqn:AnsycAutocorr}. 

\bigskip
\begin{remark}
    {\em
    \label{rem:CovMatIsPosDef}
    Note that as $\Weps[i]$  is a sampled {\em physical} noise process then {\em the matrix $\corrmatwe(\tau_0)$ has a full rank}. This follows as if $\corrmatwe(\tau_0)$ does not have a full rank, then by the definition of a multivariate Normal \acp{rv}, see \cite[Def. 16.1]{JacodProtter:ProbabilityEssentials}, we obtain that at least one element in the vector $\Wepsk$ is {\em identically equal} to a linear combination of the other elements. Such a linear relationship can be used to design a linear transformation at the receiver which completely eliminates the noise at one or more time indexes of the received sequence, leading to an infinite capacity value, which  naturally does not correspond to physical scenarios.
    }
\end{remark}

Consider the scalar \ac{rv}  $V_{k,\eps}(\tau_0)\triangleq k\cdot\zkeps{k}{\eps}\Big( \cdf{\Xinkopt|\tau_0}|\tau_0\Big)$:
\ifFullVersion   
	\begin{align}
	    V_{k,\eps}&\triangleq \log \frac{\pdf{\Yepsk | \Xepsk}\big( \Yepsk \big| \Xepsk\big) }{\pdf{\Yepsk}\big( \Yepsk \big) } \notag\\
	    &= \log\left(\pdf{\Yepsk | \Xepsk}\Big(\Yepsk \big| \Xepsk\Big)\right)-\log \left(\pdf{\Yepsk}\Big( \Yepsk \Big) \right)\notag\\
		&=\log\left(\pdf{\Wepsk}\Big( \Yepsk - \Xepsk\Big)\right)-\log \left(\pdf{\Yepsk}\Big( \Yepsk \Big) \right). \notag
	\end{align}
	Given that $\pdf{\Yepsk}\left( y^{(k)} \right) $ and $\pdf{\Wepsk}\Big( y^{(k)} - x^{(k)}\Big)$ are Gaussian \acp{pdf}, then $ V_{k,\eps}$ can be explicitly stated as:
	\begin{align}
	    V_{k,\eps}&=\frac{1}{2}\log\left(\frac{\Det\left(\corrmatye\right)}{\Det\left(\corrmatwe\right)}\right)+\frac{\log(e)}{2}\left(\Yepsk\right)^T\left(\corrmatye\right)^{-1}\left(\Yepsk\right)\notag\\
	    &\hspace{2cm} -\frac{\log(e)}{2}\left(\Yepsk-\Xepsk\right)^T\left(\corrmatwe\right)^{-1}\left(\Yepsk-\Xepsk\right).
	\end{align}
\fi  
\ifFullVersion 
	Similarly, we have that 
	\begin{align}
    	\label{eqn:vepsk}
	    V_{k, \eps}&=\frac{1}{2}\log\left(\frac{\Det\left(\corrmatye\right)}{\Det\left(\corrmatwe\right)}\right)+\frac{\log(e)}{2}\left(\Yepsk\right)^T\left(\corrmatye\right)^{-1}\left(\Yepsk\right)\notag\\
	    &\hspace{2cm} -\frac{\log(e)}{2}\left(\Yepsk-\Xinkopt\right)^T\left(\corrmatwe\right)^{-1}\left(\Yepsk-\Xepsk\right).
		\end{align}
\fi    
Next, consider the scalar \ac{rv} $\tilde{V}_{k,\eps}(\tau_0)$:
\begin{eqnarray}
    \!\!\!\!\!\!\!\tilde{V}_{k,\eps}(\tau_0)&\triangleq &
    \big(\Yepsk\big)^T\Big(\corrmatye(\tau_0)\Big)^{-1}\Yepsk -\big(\Yepsk-\Xinkopt\big)^T\Big(\corrmatwe(\tau_0)\Big)^{-1}\big(\Yepsk-\Xinkopt\big)\notag \\
    \ifFullVersion 
	        & \EqDist{}&\left(\Yepsk\right)^T\left(\corrmatye\right)^{-1}\left(\Ynk\right) -\left(\Wepsk\right)^T\left(\corrmatwe\right)^{-1}\left(\Wepsk\right)\notag \\
	\fi  
	 &\EqDist{}&\big(\Xinkopt+\Wepsk\big)^T\Big(\corrmatxeopt(\tau_0)+\corrmatwe(\tau_0)\Big)^{-1}\big(\Xinkopt+\Wepsk\big)\notag\\ 
    &  & \qquad -\big(\Wepsk\big)^T\Big(\corrmatwe(\tau_0)\Big)^{-1}\big(\Wepsk\big)\notag\\
    \ifFullVersion   
	    &=&\left[\!\begin{array}{cc}
	        \Xinkopt \\
	        \Wepsk 
	    \end{array}\!\right]^T\left[\mathsf{I}_k \quad \mathsf{I}_k\right]^T\left(\corrmatxeopt(\tau_0)+\corrmatwe(\tau_0)\right)^{-1} \left[\mathsf{I}_k \quad \mathsf{I}_k\right]\left[\begin{array}{cc}
	        \Xinkopt \\
	        \Wepsk 
	    \end{array}\right] \notag \\
	    & &\hspace{2cm}-\left[\!\begin{array}{cc}
	        \Xinkopt \\
	        \Wepsk 
	    \end{array}\!\right]^T\left[\mathsf{0}_k \quad \mathsf{I}_k\right]^T\left(\corrmatwe(\tau_0)\right)^{-1} \left[\mathsf{0}_k \quad \mathsf{I}_k\right]\left[\begin{array}{cc}
	        \Xinkopt \\
	        \Wepsk 
	    \end{array}\right]\notag\\
    \fi 
    &=&\!\!\!\!-\!\!\left[\begin{array}{cc}
        \Xinkopt \\
        \Wepsk 
    \end{array}\right]^T\!\!\left[\begin{array}{cc}
        \mathsf{I}_k & \mathsf{I}_k \\
        \mathsf{0}_k & \mathsf{I}_k
    \end{array}\right]^T\!\!\left[\begin{array}{cc}
        \!-\left(\corrmatxeopt(\tau_0)\!+\!\corrmatwe(\tau_0)\right)^{-1} & \mathsf{0}_k \\
         \mathsf{0}_k & \!\!\left(\corrmatwe(\tau_0)\right)^{-1}
    \end{array}\right]\!\! \left[\begin{array}{cc}
        \mathsf{I}_k & \mathsf{I}_k \\
         \mathsf{0}_k & \mathsf{I}_k
    \end{array}\right] \!\!\left[\begin{array}{cc}
        \Xinkopt \\
        \Wepsk 
    \end{array}\right]
    \ifFullVersion,\fi
    \notag
\end{eqnarray}
\ifFullVersion
    where we note that 
	\begin{equation*}
	    \left[\begin{array}{cc}
	       \mathsf{I}_k & \mathsf{I}_k \\
	        \mathsf{0}_k & \mathsf{I}_k
	    \end{array}\right] \left[\begin{array}{cc}
		    \Xinkopt \\
		    \Wepsk 
		\end{array}\right]=\left[\begin{array}{cc}
		    \Xinkopt +\Wepsk \\
		    \Wepsk 
		\end{array}\right].
	\end{equation*}
\fi
Using $\tilde{V}_{k,\eps}(\tau_0)$ we can write $V_{k,\eps}(\tau_0)\EqDist{}\frac{1}{2}\log\left(\frac{\Det\big(\corrmatye(\tau_0)\big)}{\Det\big(\corrmatwe(\tau_0)\big)}\right)+\frac{\log(e)}{2}\tilde{V}_{k,\eps}(\tau_0)$. Define next the matrix
\begin{align*}
    \tilde{\Cmat}_{\eps}^{(k)}(\tau_0)&\triangleq \left[\begin{array}{cc}
    \mathsf{I}_k & \mathsf{0}_k \\
         \mathsf{I}_k & \mathsf{I}_k
    \end{array}\right]\left[\begin{array}{cc}
        -\left(\corrmatxeopt(\tau_0)+\corrmatwe(\tau_0)\right)^{-1} & \mathsf{0}_k \\
         \mathsf{0}_k & \left(\corrmatwe(\tau_0)\right)^{-1}
    \end{array}\right] \left[\begin{array}{cc}
        \mathsf{I}_k & \mathsf{I}_k \\
         \mathsf{0}_k & \mathsf{I}_k
    \end{array}\right]\notag \\
    &=\left[\begin{array}{cc}
        -\left(\corrmatxeopt(\tau_0)+\corrmatwe(\tau_0)\right)^{-1} & -\left(\corrmatxeopt(\tau_0)+\corrmatwe(\tau_0)\right)^{-1}\\
         -\left(\corrmatxeopt(\tau_0)+\corrmatwe(\tau_0)\right)^{-1} & \;\;\;\;\;\;-\left(\corrmatxeopt(\tau_0)+\corrmatwe(\tau_0)\right)^{-1}\!\!\!+\left(\corrmatwe(\tau_0)\right)^{-1}
    \end{array}\right].
\end{align*}
With this definition we can express $\tilde{V}_{k,\eps}$ as $\tilde{V}_{k,\eps}=-\left[\begin{array}{cc}
    \Xinkopt \\
    \Wepsk 
\end{array}\right]^T \tilde{\Cmat}_{\eps}^{(k)}(\tau_0)\left[\begin{array}{cc}
    \Xinkopt \\
    \Wepsk 
\end{array}\right] $. 
Note that the matrix $\tilde{\Cmat}_{\eps}^{(k)}(\tau_0)$ is a real, symmetric, full-rank,  {\em indefinite} matrix, which is {\em different from the inverse correlation matrix of the Gaussian vector $\Big[ \big(\Xinkopt\big)^T, \big(\Wepsk\big)^T \Big]^T$}, hence, it is not possible to apply a simple decomposition as was done in, e.g., \cite[Sec. V]{cover1989gaussian}, to express the distribution of $\zkeps{k}{\eps}\big( \cdf{\Xinkopt|\tau_0}|\tau_0\big)$.
		    
\end{sloppypar}	
		    
As generally $\Cmat_{\Xinkopt\Wepsk}(\tau_0)\triangleq\left[\begin{array}{cc}
    \corrmatxeopt(\tau_0) & \alzer{k} \\
    \alzer{k} & \corrmatwe(\tau_0)
    \end{array}\right]$ 
may not be a full-rank matrix, then let $\mathrm{rank}\Big(\Cmat_{\Xinkopt\Wepsk}(\tau_0)\Big)=2k-\ttnepsk$, where $\ttnepsk\in\mN$ denotes the number of degenerate elements of $\Xinkopt$. We can now write the distribution of the Gaussian random vector $\left[\big(\Xinkopt\big)^T, \big(\Wepsk\big)^T\right]^T$, separating the degenerate and the non-degenerate components, as follows \cite[Sec. III-A--III-B]{BaktashCovEst:TVT2017}: First, decompose $\Cmat_{\Xinkopt\Wepsk}(\tau_0)$ as
\begin{equation}
    \label{eqn:Cxw_decomp}
        \Cmat_{\Xinkopt\Wepsk}(\tau_0)=
    \left[\begin{array}{cc}\mathsf{P}_{\eps,k}^{\mbox{\scriptsize CW}}(\tau_0) & \mathsf{P}_{\eps,k}^{0}(\tau_0) \end{array}\right]\left[\begin{array}{cc}
    \Cmat_{\tilde{X}^{(k)}_{\opt}\Wepsk}(\tau_0) & \;\;\alzer{(2k-\ttnepsk)\times \ttnepsk} \\
        \alzer{\ttnepsk\times (2k-\ttnepsk)} & \;\;\alzer{\ttnepsk \times \ttnepsk}
    \end{array}\right]\left[\begin{array}{cc}\mathsf{P}_{\eps,k}^{\mbox{\scriptsize CW}}(\tau_0) & \mathsf{P}_{\eps,k}^{0}(\tau_0) \end{array}\right]^T\!\!\!,
\end{equation}
where $\left[\begin{array}{cc}\mathsf{P}_{\eps,k}^{\mbox{\scriptsize CW}}(\tau_0) & \mathsf{P}_{\eps,k}^{0}(\tau_0) \end{array}\right]$ is an orthogonal $2k \times 2k$ matrix, $\Cmat_{\tilde{X}^{(k)}_{\opt}\Wepsk}(\tau_0)\in\mR^{(2k-\ttnepsk)\times(2k-\ttnepsk)}$ is a symmetric positive-definite matrix, $\Cmat_{\tilde{X}^{(k)}_{\opt}\Wepsk}(\tau_0)\succ 0$, the columns of $\mathsf{P}_{\eps,k}^{\mbox{\scriptsize CW}}(\tau_0)\in\mR^{2k\times(2k-\ttnepsk)}$ form an orthonormal basis for $\mathrm{range}\Big(\Cmat_{\Xinkopt\Wepsk}(\tau_0)\Big)$ and the columns of $\mathsf{P}_{\eps,k}^{0}(\tau_0)\in\mR^{2k\times \ttnepsk}$ form an orthonormal basis for the null space of $ \Cmat_{\Xinkopt\Wepsk}(\tau_0)$. Then,
		  \begin{equation*}
		  \left[\begin{array}{cc}
		        \mathsf{P}_{\eps,k}^{\mbox{\scriptsize CW}}(\tau_0) & \mathsf{P}_{\eps,k}^{0}(\tau_0)
		    \end{array}\right]^T
		    \left[\begin{array}{cc}
		        \Xinkopt \\
		        \Wepsk 
		    \end{array}\right]\EqDist{}\left[\begin{array}{cc}
		        \mathrm{B}_{\eps}^{ (2k-\ttnepsk)} \\
		       \alzer{\ttnepsk\times 1}
		    \end{array}\right], \quad \mathrm{B}_{\eps}^{ (2k-\ttnepsk)}\sim \dsN\left(\alzer{(2k-\ttnepsk    )\times 1}, \Cmat_{\tilde{X}^{(k)}_{\opt}\Wepsk}(\tau_0)\right),
		  \end{equation*}
		  and we obtain that
		  \begin{eqnarray*}
		      & & \hspace{-1.5cm} \left[\begin{array}{cc}
		        \Xinkopt \\
		        \Wepsk 
		    \end{array}\right]^T  \tilde{\Cmat}_{\eps}^{(k)}(\tau_0) \left[\begin{array}{cc}
		        \Xinkopt \\
		        \Wepsk 
		    \end{array}\right]\notag \\
		    &=& \left[\begin{array}{cc}
		        \Xinkopt \\
		        \Wepsk 
		    \end{array}\right]^T\cdot\left[\begin{array}{cc} 		  
		    \mathsf{P}_{\eps,k}^{\mbox{\scriptsize CW}}(\tau_0) & \mathsf{P}_{\eps,k}^{0}(\tau_0) \end{array}\right]\cdot \left[\begin{array}{cc}\mathsf{P}_{\eps,k}^{\mbox{\scriptsize CW}}(\tau_0) & \mathsf{P}_{\eps,k}^{0}(\tau_0) \end{array}\right]^T\cdot \tilde{\Cmat}_{\eps}^{(k)}(\tau_0)\cdot\\
		    &  & \qquad\qquad\qquad\qquad \left[\begin{array}{cc}\mathsf{P}_{\eps,k}^{\mbox{\scriptsize CW}}(\tau_0) & \mathsf{P}_{\eps,k}^{0}(\tau_0) \end{array}\right]\cdot \left[\begin{array}{cc}\mathsf{P}_{\eps,k}^{\mbox{\scriptsize CW}}(\tau_0) & \mathsf{P}_{\eps,k}^{0}(\tau_0) \end{array}\right]^T\cdot\left[\begin{array}{cc}
		        \Xinkopt \\
		        \Wepsk 
		    \end{array}\right]\notag\\
		    \ifFullVersion
		    &\EqDist{}& \left[\begin{array}{cc}
		        \mathrm{B}_{\eps}^{ (2k-\ttnepsk)} \\
		       \alzer{\ttnepsk\times 1}
		    \end{array}\right]^T\cdot\left[\begin{array}{cc}\mathsf{P}_{\eps,k}^{\mbox{\scriptsize CW}}(\tau_0) & \mathsf{P}_{\eps,k}^{0}(\tau_0) \end{array}\right]^T\cdot \tilde{\Cmat}_{\eps}^{(k)}(\tau_0)\cdot \left[\begin{array}{cc}\mathsf{P}_{\eps,k}^{\mbox{\scriptsize CW}}(\tau_0) & \mathsf{P}_{\eps,k}^{0}(\tau_0) \end{array}\right]\cdot \left[\begin{array}{cc}
		        \mathrm{B}_{\eps}^{ (2k-\ttnepsk)} \\
		       \alzer{\ttnepsk\times 1}
		    \end{array}\right]\notag \\
		    &=&
		    \fi
		    \left(\mathrm{B}_{\eps}^{ (2k-\ttnepsk)} \right)^T\cdot \left(\mathsf{P}_{\eps,k}^{\mbox{\scriptsize CW}}(\tau_0)\right)^T\cdot \tilde{\Cmat}_{\eps}^{(k)}(\tau_0)\cdot \mathsf{P}_{\eps,k}^{\mbox{\scriptsize CW}}(\tau_0)\cdot \mathrm{B}_{\eps}^{ (2k-\ttnepsk)}.
		    \end{eqnarray*}
Observe that $\mathsf{P}_{\eps,k}^{\mbox{\scriptsize CW}}(\tau_0)$ is a full-rank matrix, and since  $\tilde{\Cmat}_{\eps}^{(k)}(\tau_0)$ is also a full-rank matrix, then  $\big(\mathsf{P}_{\eps,k}^{\mbox{\scriptsize CW}}(\tau_0)\big)^T\cdot \tilde{\Cmat}_{\eps}^{(k)}(\tau_0)\cdot \mathsf{P}_{\eps,k}^{\mbox{\scriptsize CW}}(\tau_0)$ is full-rank. Since $\Cmat_{\tilde{X}^{(k)}_{\opt}\Wepsk}(\tau_0)\succ 0$ and symmetric,  it can be expressed as  
\ifFullVersion
\cite[Thm. 13.11]{banerjee2014linear}\footnote{\cite[Thm. 13.11]{banerjee2014linear}: (Square root of a p.d. (non-negative definite (n.n.d.)) matrix): If $\mathsf A$ is an $n \times n$ p.d. (n.n.d.) 
matrix, then there exists an p.d. (n.n.d.) matrix $\mathsf B$ such that $\mathsf A = \mathsf B^2$.} \fi
$\Cmat_{\tilde{X}^{(k)}_{\opt}\Wepsk}(\tau_0)=\big(\matR{\eps,k}(\tau_0)\big)^2$; where $\matR{\eps,k}(\tau_0)\in \mR^{(2k-\ttnepsk)\times (2k-\ttnepsk)}$ is a  positive-definite symmetric matrix. Then, letting $\matR{\eps,k}^{-1}(\tau_0) \in \mR^{(2k-\ttnepsk)\times (2k-\ttnepsk)}$ denote the inverse of $\matR{\eps,k}(\tau_0)$, we can write
\begin{eqnarray*}
    \left[\begin{array}{cc}
        \Xinkopt \\
        \Wepsk 
    \end{array}\right]^T \tilde{\Cmat}_{\eps}^{(k)}(\tau_0)\left[\begin{array}{cc}
        \Xinkopt \\
        \Wepsk 
    \end{array}\right] &\EqDist{}&
    \ifFullVersion  
	    \left(\mathrm{B}_{\eps}^{ (2k-\ttnepsk)} \right)^T\cdot \big(\mathsf{P}_{\eps,k}^{\mbox{\scriptsize CW}}\big)^T\cdot \tilde{\Cmat}_{\eps}^{(k)}(\tau_0)\cdot \mathsf{P}_{\eps,k}^{\mbox{\scriptsize CW}}\cdot \mathrm{B}_{\eps}^{ (2k-\ttnepsk)}\notag\\
		&\EqDist{}&
    \fi  
    \Big(\mathrm{B}_{\eps}^{ (2k-\ttnepsk)} \Big)^T\cdot\matR{\eps,k}^{-1}(\tau_0)\cdot \matR{\eps,k}(\tau_0)\cdot \big(\mathsf{P}_{\eps,k}^{\mbox{\scriptsize CW}}(\tau_0)\big)^T\cdot \tilde{\Cmat}_{\eps}^{(k)}(\tau_0)\cdot \mathsf{P}_{\eps,k}^{\mbox{\scriptsize CW}}(\tau_0)\notag\\
    &  &\qquad \cdot\matR{\eps,k}(\tau_0)\cdot \matR{\eps,k}^{-1}(\tau_0)\cdot\mathrm{B}_{\eps}^{ (2k-\ttnepsk)}.\notag
\end{eqnarray*}
Next, observe that
\begin{equation*}
   \Gamma_{\eps}^{(2k-\ttnepsk)}\triangleq \matR{\eps,k}^{-1}(\tau_0)\cdot\mathrm{B}_{\eps}^{ (2k-\ttnepsk)}\sim \dsN\left(\alzer{(2k-\ttnepsk)\times 1},\iden{2k-\ttnepsk}\right),
\end{equation*}
and note that since $\matR{\eps,k}(\tau_0)$  and $\big(\mathsf{P}_{\eps,k}^{\mbox{\scriptsize CW}}(\tau_0)\big)^T\cdot \tilde{\Cmat}_{\eps}^{(k)}(\tau_0)\cdot \mathsf{P}_{\eps,k}^{\mbox{\scriptsize CW}}(\tau_0)$ are full-rank matrices, then also $\matR{\eps,k}(\tau_0)\cdot \big(\mathsf{P}_{\eps,k}^{\mbox{\scriptsize CW}}(\tau_0)\big)^T\cdot \tilde{\Cmat}_{\eps}^{(k)}(\tau_0)\cdot \mathsf{P}_{\eps,k}^{\mbox{\scriptsize CW}}(\tau_0)\cdot\matR{\eps,k}(\tau_0)$ is a full-rank, square, symmetric, real, indefinite matrix, whose rank is $2k-\ttnepsk$. Thus, we can write \cite[Thm. 11.27]{banerjee2014linear}:
\begin{eqnarray*}
   \ttCepsk(\tau_0)  & \triangleq & \matR{\eps,k}(\tau_0)\cdot \big(\mathsf{P}_{\eps,k}^{\mbox{\scriptsize CW}}(\tau_0)\big)^T\cdot \tilde{\Cmat}_{\eps}^{(k)}(\tau_0)\cdot \mathsf{P}_{\eps,k}^{\mbox{\scriptsize CW}}(\tau_0)\cdot\matR{\eps,k}(\tau_0)\\
   & = &\big(\myMatrix{P}{\eps,k}(\tau_0)\big)^T\cdot\myMatrix{D}{\eps,k}(\tau_0)\cdot \myMatrix{P}{\eps,k}(\tau_0), \qquad \big(\myMatrix{P}{\eps,k}(\tau_0)\big)^T\cdot \myMatrix{P}{\eps,k}(\tau_0)=\iden{2k-\ttnepsk},
\end{eqnarray*}
where $\myMatrix{D}{\eps,k}(\tau_0)\in \mR^{(2k-\ttnepsk)\times (2k-\ttnepsk)}$ is a diagonal matrix whose diagonal elements are the eigenvalues of  $\ttCepsk(\tau_0)\in \mR^{(2k-\ttnepsk)\times (2k-\ttnepsk)}$. Let $d_{\eps,ii,k}^{\{\tau_0\}}$ denote the $i$-th eigenvalue of the matrix $\ttCepsk(\tau_0)$.  Since $\ttCepsk(\tau_0)$ is full-rank it follows that $d_{\eps,ii,k}^{\{\tau_0\}}\neq 0$ for $0\leq i\leq 2k-\ttnepsk-1$. Using this representation, we write
\begin{eqnarray}
    \left[\begin{array}{cc}
        \Xinkopt \\
        \Wepsk 
    \end{array}\right]^T \tilde{\Cmat}_{\eps}^{(k)}(\tau_0)\left[\begin{array}{cc}
        \Xinkopt \\
        \Wepsk 
    \end{array}\right]&\EqDist{}&\Big(\Gamma_{\eps}^{(2k-\ttnepsk)}\Big)^T\cdot\big(\myMatrix{P}{\eps,k}(\tau_0)\big)^T\cdot\myMatrix{D}{\eps,k}(\tau_0)\cdot \myMatrix{P}{\eps,k}(\tau_0)\cdot\Gamma_{\eps}^{(2k-\ttnepsk)}\notag\\
    &\EqDist{}&\Big(\tilde{\Gamma}_{\eps}^{(2k-\ttnepsk)}\Big)^T\cdot \myMatrix{D}{\eps,k}(\tau_0) \cdot\tilde{\Gamma}_{\eps}^{(2k-\ttnepsk)},
    \label{eqn:eigendecompositionCtilde}
\end{eqnarray}
where we define
\begin{equation}
   \label{eqn:chisquare_dist}
   \tilde{\Gamma}_{\eps}^{(2k-\ttnepsk)}\triangleq \myMatrix{P}{\eps,k}(\tau_0)\cdot\Gamma_{\eps}^{(2k-\ttnepsk)}\sim\dsN\left(\alzer{(2k-\ttnepsk)\times 1},\iden{2k-\ttnepsk}\right).
\end{equation}
Eventually, we obtain
\begin{equation*}
    \tilde{V}_{k,\eps}(\tau_0)\EqDist{}-\Big(\tilde{\Gamma}_{\eps}^{(2k-\ttnepsk)}\Big)^T\cdot\myMatrix{D}{\eps,k}(\tau_0)\cdot \tilde{\Gamma}_{\eps}^{(2k-\ttnepsk)}=\mathop{\sum}\limits_{i=0}^{2k-\ttnepsk-1} \left(-d_{\eps,ii,k}^{\{\tau_0\}}\right)\cdot\tGmmiepsksqr, \quad d_{\eps,ii,k}^{\{\tau_0\}}\in \mR,
\end{equation*}
where $\tilde{\Gamma}_{\eps,i,k}$, $0\leq i\leq 2k-\ttnepsk-1$ denotes the $i$-th element of the vector $\tilde{\Gamma}_{\eps}^{(2k-\ttnepsk)}$.  Observe from \eqref{eqn:chisquare_dist} that the elements $\tilde{\Gamma}_{\eps,i,k}$ are \ac{iid} Gaussian \acp{rv}, hence, $\tGmmiepsksqr$ is a central chi-square random variable with a single degree of freedom \cite[Example 5.2]{Papoulis2002}, which is denoted as $\tGmmiepsksqr\sim \mathcal{X}^2(1)$, and $\tGmmiepsksqr$, $0\leq i\leq 2k-\ttnepsk-1$, are mutually independent  $\mathcal{X}^2(1)$ \acp{rv}. We can now express $\zkeps{k}{\eps}\Big( \cdf{\Xinkopt|\tau_0}|\tau_0\Big)$ as:
\begin{equation*}
    \zkeps{k}{\eps}\Big( \cdf{\Xinkopt|\tau_0}|\tau_0\Big)\EqDist{}\frac{1}{2k}\cdot \log\left(\frac{\Det\big(\corrmatye(\tau_0)\big)}{\Det\big(\corrmatwe(\tau_0)\big)}\right)+ \frac{\log(e)}{2k} \mathop{\sum}\limits_{i=0}^{2k-\ttnepsk-1} \left(-d_{\eps,ii,k}^{\{\tau_0\}}\right)\cdot\tGmmiepsksqr.
\end{equation*}

Examining $\zkeps{k}{\eps}\Big( \cdf{\Xinkopt|\tau_0}|\tau_0\Big)$, we note that since $\E\Big\{\zkeps{k}{\eps}\Big( \cdf{\Xinkopt|\tau_0}|\tau_0\Big)\Big\}=\frac{1}{2k}\cdot \log\left(\frac{\Det\big(\corrmatye(\tau_0)\big)}{\Det\big(\corrmatwe(\tau_0)\big)}\right)$ \cite[Eqn. (7.31)]{GraybookEntropy}, then it necessarily  should hold that $\mathop{\sum}\limits_{i=0}^{2k-\ttnepsk-1}\!\!\! \left(-d_{\eps,ii,k}^{\{\tau_0\}}\right)=0$. This can be verified via a direct derivation:
\begin{subequations}
    \begin{eqnarray}
       & &\hspace{-0.5cm}   \mathop{\sum}\limits_{i=0}^{2k-\ttnepsk-1} \!\!\!\!\! d_{\eps,ii,k}^{\{\tau_0\}} \notag\\
       & = &  \Tr\big\{\ttCepsk(\tau_0)\big\}\notag\\
       &=&\Tr\left\{\matR{\eps,k}(\tau_0)\cdot \big(\mathsf{P}_{\eps,k}^{\mbox{\scriptsize CW}}(\tau_0)\big)^T\cdot \tilde{\Cmat}_{\eps}^{(k)}(\tau_0)\cdot \mathsf{P}_{\eps,k}^{\mbox{\scriptsize CW}}(\tau_0)\cdot\matR{\eps,k}(\tau_0)\right\}\notag\\
        &=&\Tr\left\{ \tilde{\Cmat}_{\eps}^{(k)}(\tau_0)\cdot \mathsf{P}_{\eps,k}^{\mbox{\scriptsize CW}}(\tau_0)\cdot\matR{\eps,k}(\tau_0)\cdot \matR{\eps,k}(\tau_0)\cdot \big(\mathsf{P}_{\eps,k}^{\mbox{\scriptsize CW}}(\tau_0)\big)^T\right\}\notag\\
        &=&\Tr\left\{\tilde{\Cmat}_{\eps}^{(k)}(\tau_0)\cdot \mathsf{P}_{\eps,k}^{\mbox{\scriptsize CW}}(\tau_0)\cdot\Cmat_{\tilde{X}^{(k)}_{\opt}\Wepsk}(\tau_0)\cdot \big(\mathsf{P}_{\eps,k}^{\mbox{\scriptsize CW}}(\tau_0)\big)^T\right\}\notag\\
        &\stackrel{(a)}{=}&\Tr\left\{\tilde{\Cmat}_{\eps}^{(k)}(\tau_0)\cdot \left[\begin{array}{cc}\mathsf{P}_{\eps,k}^{\mbox{\scriptsize CW}}(\tau_0) & \mathsf{P}_{\eps,k}^{0}(\tau_0) \end{array}\right]\left[\begin{array}{cc}
        \Cmat_{\tilde{X}^{(k)}_{\opt}\Wepsk}(\tau_0) & \;\;\alzer{(2k-\ttnepsk)\times \ttnepsk} \\
        \alzer{\ttnepsk\times (2k-\ttnepsk)} & \;\;\alzer{\ttnepsk\times \ttnepsk}
        \end{array}\right]\left[\begin{array}{cc}\mathsf{P}_{\eps,k}^{\mbox{\scriptsize CW}}(\tau_0) & \mathsf{P}_{\eps,k}^{0}(\tau_0) \end{array}\right]^T\right\}\notag\\
         &\stackrel{(b)}{=}&\Tr\left\{\tilde{\Cmat}_{\eps}^{(k)}(\tau_0)\cdot \Cmat_{\Xinkopt\Wepsk}(\tau_0)\right\}\label{eqn:Sum_dn_stp1}\\
        &=&-\Tr\left\{\!\left[\begin{array}{cc}
        \left(\corrmatxeopt(\tau_0)+\corrmatwe(\tau_0)\right)^{-1}\corrmatxeopt(\tau_0) & \left(\corrmatxeopt(\tau_0)+\corrmatwe(\tau_0)\right)^{-1}\corrmatwe(\tau_0)\\
		\left(\corrmatxeopt(\tau_0)+\corrmatwe(\tau_0)\right)^{-1}\corrmatxeopt(\tau_0) & \quad \left(\corrmatxeopt(\tau_0)+\corrmatwe(\tau_0)\right)^{-1}\corrmatwe(\tau_0) -\iden{k}
		\end{array}\right]\!\right\}\notag\\
		&=&\Tr\Big\{\!-\left(\corrmatxeopt(\tau_0)+\corrmatwe(\tau_0)\right)^{-1}\!\!\corrmatxeopt(\tau_0)\notag\\
		& & \qquad \qquad \qquad\qquad\qquad -\left(\corrmatxeopt(\tau_0)+\corrmatwe(\tau_0)\right)^{-1}\!\!\corrmatwe(\tau_0) +\iden{k}\Big\}\\
		&= & 0,
		\label{eqn:sum_d}
	\end{eqnarray}
\end{subequations}
where (a) follows since  
\ifFullVersion   
    \begin{eqnarray}
        & & \hspace{-2cm}\left[\begin{array}{cc}\mathsf{P}_{\eps,k}^{\mbox{\scriptsize CW}}(\tau_0) & \quad \mathsf{P}_{\eps,k}^{0} (\tau_0)\end{array}\right]\left[\begin{array}{cc}
	    \Cmat_{\tilde{X}^{(k)}_{\opt}\Wepsk}(\tau_0) & \qquad\alzer{(2k-\ttnepsk)\times \ttnepsk} \\
	    \alzer{\ttnepsk\times (2k-\ttnepsk)} & \alzer{\ttnepsk\times \ttnepsk}
		\end{array}\right]\left[\begin{array}{cc}\mathsf{P}_{\eps,k}^{\mbox{\scriptsize CW}}(\tau_0) & \quad\mathsf{P}_{\eps,k}^{0}(\tau_0) \end{array}\right]^T\notag\\
		&=& \left[\begin{array}{cc}\mathsf{P}_{\eps,k}^{\mbox{\scriptsize CW}}(\tau_0)\Cmat_{\tilde{X}^{(k)}_{\opt}\Wepsk}(\tau_0) & \quad \alzer{2k \times \ttnepsk} \end{array}\right]\left[\begin{array}{cc}\big(\mathsf{P}_{\eps,k}^{\mbox{\scriptsize CW}}(\tau_0)\big)^T \\ \big(\mathsf{P}_{\eps,k}^{0}(\tau_0)\big)^T \end{array}\right]\notag\\
		&=& \mathsf{P}_{\eps,k}^{\mbox{\scriptsize CW}}(\tau_0)\cdot\Cmat_{\tilde{X}^{(k)}_{\opt}\Wepsk}(\tau_0)\cdot \big(\mathsf{P}_{\eps,k}^{\mbox{\scriptsize CW}}(\tau_0)\big)^T; \notag
	\end{eqnarray}
\fi     
and (b) follows from \eqref{eqn:Cxw_decomp}.
		   
\begin{sloppypar}
We now compute $\Tr\Big\{\left(\ttCepsk(\tau_0)\right)^2\Big\}$: Begin by using Eqn. \eqref{eqn:Sum_dn_stp1} and write  
\begin{align*}
   &\Tr\Big\{\left(\ttCepsk(\tau_0)\right)^2\Big\}=\Tr\left\{\tilde{\Cmat}_{\eps}^{(k)}(\tau_0)\cdot
   \Cmat_{\Xinkopt\Wepsk}(\tau_0)
   \cdot \tilde{\Cmat}_{\eps}^{(k)}(\tau_0)\cdot 
   \Cmat_{\Xinkopt\Wepsk}(\tau_0)
    \right\}.
\end{align*} 
\ifFullVersion  
    Note that 
    \begin{align}
	    \label{eqn:matrx_square_for_var}
        & \tilde{\Cmat}_{\eps}^{(k)}(\tau_0)\cdot \Cmat_{\Xinkopt\Wepsk}(\tau_0)\cdot\tilde{\Cmat}_{\eps}^{(k)}(\tau_0)\cdot \Cmat_{\Xinkopt\Wepsk}(\tau_0)\notag\\
        & =\left[\begin{array}{cc}
		-\left(\corrmatxeopt(\tau_0)+\corrmatwe(\tau_0)\right)^{-1}\!\!\corrmatxeopt(\tau_0) & -\left(\corrmatxeopt(\tau_0)+\corrmatwe(\tau_0)\right)^{-1}\!\!\corrmatwe(\tau_0)\\
         -\left(\corrmatxeopt(\tau_0)+\corrmatwe(\tau_0)\right)^{-1}\!\!\corrmatxeopt(\tau_0) & \quad -\left(\corrmatxeopt(\tau_0)+\corrmatwe(\tau_0)\right)^{-1}\!\!\corrmatwe(\tau_0) +\iden{k}
         \end{array}\right]^2
    \end{align}
    We now evaluate explicitly the elements of the matrix product\footnote{see { \tt https://mathworld.wolfram.com/BlockMatrix.html} for the product of block matrices.} in Eqn. \eqref{eqn:matrx_square_for_var}:\\
	Element (1,1):
	\begin{align}
	     &\left(\corrmatxeopt(\tau_0)+\corrmatwe(\tau_0)\right)^{-1}\corrmatxeopt(\tau_0)\left(\corrmatxeopt(\tau_0)+\corrmatwe(\tau_0)\right)^{-1}\corrmatxeopt(\tau_0) \notag\\
	     &\qquad + \left(\corrmatxeopt(\tau_0)+\corrmatwe(\tau_0)\right)^{-1}\corrmatwe(\tau_0) \left(\corrmatxeopt(\tau_0)+\corrmatwe(\tau_0)\right)^{-1}\corrmatxeopt(\tau_0)\notag\\
	     &=\left(\corrmatxeopt(\tau_0)+\corrmatwe(\tau_0)\right)^{-1}\left(\corrmatxeopt(\tau_0)+\corrmatwe(\tau_0)\right)\left(\corrmatxeopt(\tau_0)+\corrmatwe(\tau_0)\right)^{-1}\corrmatxeopt(\tau_0)\notag\\
         &=\left(\corrmatxeopt(\tau_0)+\corrmatwe(\tau_0)\right)^{-1}\corrmatxeopt(\tau_0);\notag
    \end{align}

    \noindent   Element (1,2):
    \begin{align}
         &\left(\corrmatxeopt(\tau_0)+\corrmatwe(\tau_0)\right)^{-1}\corrmatxeopt(\tau_0)\left(\corrmatxeopt(\tau_0)+\corrmatwe(\tau_0)\right)^{-1}\corrmatwe(\tau_0)\notag \\
         & \qquad + \left(\corrmatxeopt(\tau_0)+\corrmatwe(\tau_0)\right)^{-1}\corrmatwe(\tau_0) \left(\corrmatxeopt(\tau_0)+\corrmatwe(\tau_0)\right)^{-1}\corrmatwe(\tau_0)\notag\\
         & \qquad \qquad
	     -\left(\corrmatxeopt(\tau_0)+\corrmatwe(\tau_0)\right)^{-1}\corrmatwe(\tau_0)\notag\\
	     &=\left(\corrmatxeopt(\tau_0)+\corrmatwe(\tau_0)\right)^{-1}\left(\corrmatxeopt(\tau_0) \left(\corrmatxeopt(\tau_0)+\corrmatwe(\tau_0)\right)^{-1}\right.\notag\\
	     &\qquad \left.+\corrmatwe(\tau_0) \left(\corrmatxeopt(\tau_0)+\corrmatwe(\tau_0)\right)^{-1}-\iden{k}\right)\corrmatwe(\tau_0)\notag\\
	     &=\left(\corrmatxeopt(\tau_0)+\corrmatwe(\tau_0)\right)^{-1}\left(\left(\corrmatxeopt(\tau_0) +\corrmatwe(\tau_0) \right)\cdot\phantom{\bigg)} \right.\notag\\
	     & \qquad\qquad\qquad\qquad\qquad\qquad\qquad \left.\left(\corrmatxeopt(\tau_0)+\corrmatwe(\tau_0)\right)^{-1}-\iden{k}\right)\corrmatwe = \alzer{k\times k};\notag
	\end{align}
		    
	\noindent Element (2,1):
	\begin{align}
	     &\left(\corrmatxeopt(\tau_0)+\corrmatwe(\tau_0)\right)^{-1}\corrmatxeopt(\tau_0)\left(\corrmatxeopt(\tau_0)+\corrmatwe(\tau_0)\right)^{-1}\corrmatxeopt(\tau_0)\notag\\
	     & \qquad + \left(\corrmatxeopt(\tau_0)+\corrmatwe(\tau_0)\right)^{-1}\corrmatwe(\tau_0) \left(\corrmatxeopt(\tau_0)+\corrmatwe(\tau_0)\right)^{-1}\corrmatxeopt(\tau_0)\notag\\
        &\qquad \qquad -\left(\corrmatxeopt(\tau_0)+\corrmatwe(\tau_0)\right)^{-1}\corrmatxeopt(\tau_0)\notag\\
        &=\left(\left(\corrmatxeopt(\tau_0)+\corrmatwe(\tau_0)\right)^{-1}\corrmatxeopt(\tau_0)+\left(\corrmatxeopt(\tau_0)+\corrmatwe(\tau_0)\right)^{-1}\corrmatwe(\tau_0)-\iden{k}\right)\cdot\notag\\
        & \qquad \qquad \qquad \left(\corrmatxeopt(\tau_0)+\corrmatwe(\tau_0)\right)^{-1}\corrmatxeopt(\tau_0)\notag\\
	     &=\left(\left(\corrmatxeopt(\tau_0)+\corrmatwe(\tau_0)\right)^{-1}\left(\corrmatxeopt(\tau_0)+\corrmatwe(\tau_0)\right)-\iden{k}\right)\cdot\notag\\
	     & \qquad \qquad \left(\corrmatxeopt(\tau_0)+\corrmatwe(\tau_0)\right)^{-1}\corrmatxeopt(\tau_0)=\alzer{k\times k};\notag
	\end{align}
		    
    \noindent  Element (2,2):
	\begin{align}
	     &\left(\corrmatxeopt(\tau_0)+\corrmatwe(\tau_0)\right)^{-1}\corrmatxeopt(\tau_0)\left(\corrmatxeopt(\tau_0)+\corrmatwe(\tau_0)\right)^{-1}\corrmatwe(\tau_0) \notag\\
	     & \qquad + \left(\corrmatxeopt(\tau_0)+\corrmatwe(\tau_0)\right)^{-1}\corrmatwe(\tau_0) \left(\corrmatxeopt(\tau_0)+\corrmatwe(\tau_0)\right)^{-1}\corrmatwe(\tau_0)\notag\\
	     & \qquad \qquad \qquad   -2\left(\corrmatxeopt(\tau_0)+\corrmatwe(\tau_0)\right)^{-1}\corrmatwe(\tau_0) +\iden{k}\notag\\
	     &=\left(\corrmatxeopt(\tau_0)+\corrmatwe(\tau_0)\right)^{-1}\left(\corrmatxeopt(\tau_0)\left(\corrmatxeopt(\tau_0)+\corrmatwe(\tau_0)\right)^{-1}\right.\notag\\
	     & \qquad \qquad \qquad \qquad  \left.+\corrmatwe(\tau_0)\left(\corrmatxeopt(\tau_0)+\corrmatwe(\tau_0)\right)^{-1}-2\iden{k}\right)\corrmatwe(\tau_0)+\iden{k}\notag\\
        &=\left(\corrmatxeopt(\tau_0)+\corrmatwe(\tau_0)\right)^{-1}\left(\left(\corrmatxeopt(\tau_0)+\corrmatwe(\tau_0)\right)\left(\corrmatxeopt(\tau_0)+\corrmatwe(\tau_0)\right)^{-1}\right.\notag\\
        &\qquad \qquad \qquad \qquad \qquad \qquad \qquad \qquad \left.\phantom{\bigg(}-2\iden{k}\right)\corrmatwe(\tau_0)+\iden{k}\notag\\ 
	     &=\iden{k}-\left(\corrmatxeopt(\tau_0)+\corrmatwe(\tau_0)\right)^{-1}\corrmatwe(\tau_0)\notag\\
	     &=\left(\corrmatxeopt(\tau_0)+\corrmatwe(\tau_0)\right)^{-1}\!\!\!\!\cdot \left(\corrmatxeopt(\tau_0)+\corrmatwe(\tau_0)\right)-\left(\corrmatxeopt(\tau_0)+\corrmatwe(\tau_0)\right)^{-1}\corrmatwe(\tau_0)\notag\\
	      &=\left(\corrmatxeopt(\tau_0)+\corrmatwe(\tau_0)\right)^{-1}\cdot \corrmatxeopt(\tau_0).\notag
	\end{align}
	Thus,
	\begin{align}
	\label{eqn:matrixprid_squared}
	     &\tilde{\Cmat}_{\eps}^{(k)}(\tau_0)\cdot \Cmat_{\Xinkopt\Wepsk}(\tau_0)\cdot\tilde{\Cmat}_{\eps}^{(k)}(\tau_0)\cdot \Cmat_{\Xinkopt\Wepsk}(\tau_0)\notag\\
	     & \qquad = \left[\begin{array}{cc}
	     \left(\corrmatxeopt(\tau_0)+\corrmatwe(\tau_0)\right)^{-1}\corrmatxeopt(\tau_0) & \alzer{k\times k}\\
	     \alzer{k\times k} & \quad \left(\corrmatxeopt(\tau_0)+\corrmatwe(\tau_0)\right)^{-1}\corrmatxeopt(\tau_0)
	     \end{array}\right].
    \end{align}
\fi 
hence, $\Tr\Big\{\left(\ttCepsk(\tau_0)\right)^2\Big\}= 2\cdot \Tr\Big\{ \left(\corrmatxeopt(\tau_0)+\corrmatwe(\tau_0)\right)^{-1}\!\!\corrmatxeopt(\tau_0)\Big\}$.
With this result we can compute the variance of $\tilde{V}_{k,\eps}(\tau_0)$, denoted by $\tvarepsk(\tau_0)$, as follows:
\begin{eqnarray}
    \tvarepsk(\tau_0)&=&\mathop{\sum}\limits_{i=0}^{2k-\ttnepsk-1} \!\!\! \mathrm{var}\left(d_{\eps,ii,k}^{\{\tau_0\}}\cdot\tGmmiepsksqr\right)\notag\\
	 &=&2\cdot\mathop{\sum}\limits_{i=0}^{2k-\ttnepsk-1} \!\!\! \left(d_{\eps,ii,k}^{\{\tau_0\}}\right)^2\notag\\
	 &=&2\cdot\Tr\Big\{\left(\ttCepsk(\tau_0)\right)^2\Big\} \notag\\
	 \ifFullVersion 
		    &\stackrel{(a)}{=}& 4\cdot\Tr\left\{\left(\corrmatxeopt(\tau_0)+\corrmatwe(\tau_0)\right)^{-1}\corrmatxeopt(\tau_0)\right\}\notag\\
		    &=& 4\cdot\Tr\left\{\iden{k}-\left(\corrmatxeopt(\tau_0)+\corrmatwe\right)^{-1}\corrmatwe\right\}\notag\\
    \fi  
    &=& 4\cdot\bigg(k-\Tr\Big\{\Big(\iden{k}+\big(\corrmatwe(\tau_0)\big)^{-1}\corrmatxeopt(\tau_0)\Big)^{-1}\Big\}\bigg).\label{eqn:Expression_tvarnk}
\end{eqnarray}
\ifFullVersion
    where (a) follows from \eqref{eqn:matrixprid_squared}.
\fi
\end{sloppypar}


\begin{sloppypar}
From the results of \eqref{eqn:sum_d} and \eqref{eqn:Expression_tvarnk} it follows that
    $\E \left\{\zkeps{k}{\eps} \Big(\cdf{\Xinkopt|\tau_0}|\tau_0 \Big)  \right\}  =  \frac{1}{k}I\big(\Xinkopt; \Yepsk|\tau_0\big)$, and 
    $\mbox{var}\Big(\zkeps{k}{\eps} \Big(\cdf{\Xinkopt|\tau_0}|\tau_0 \Big) \Big)  \le  \frac{3}{k}$.
Since the variance of $\zkeps{k}{\eps} \Big(\cdf{\Xinkopt|\tau_0}|\tau_0 \Big)$ decreases as $k$ increases, then, by Chebyshev's inequality \cite[Eqn. (5-88)]{Papoulis2002}, we obtain
$\Pr\Big(\Big|\zkeps{k}{\eps} \Big(\cdf{\Xinkopt|\tau_0}|\tau_0 \Big)-\frac{1}{k}I\big(\Xinkopt; \Yepsk|\tau_0\big)\Big|>\frac{1}{k^{1/3}} \Big) <\frac{3}{k^{1/3}}$, and we conclude that 
$\forall\delta > 0$, $\exists k_0(\delta)\in\mNp$ s.t. $\forall k>k_0(\delta)$, it follows that 
\begin{equation}
    \label{eqn:Bounding_PrZepsk}
    \Pr\Big(\zkeps{k}{\eps} \Big(\cdf{\Xinkopt|\tau_0}|\tau_0 \Big) <\frac{1}{k}I\big(\Xinkopt; \Yepsk|\tau_0\big)-\delta\Big) <3\delta.
\end{equation}
Note that by definition of the limit-inferior, $\forall \delta>0$, $\exists k_1(\delta)\in\mNp$ s.t. $\forall k>k_1(\delta)$ (recall that $\Xinkopt$ maximizes $\frac{1}{k}I\big(\Xepsk; \Yepsk|\tau_0\big)$)
\[
    \frac{1}{k}I\big(\Xinkopt; \Yepsk|\tau_0\big) > \mathop{\liminf}\limits_{k \rightarrow \infty}\frac{1}{k}I\big(\Xinkopt; \Yepsk|\tau_0\big) -\delta,
\]
hence, for all $k>\max\big\{k_0(\delta), k_1(\delta) \big\}$.
\begin{equation*}
    \Pr\Big(\zkeps{k}{\eps} \Big(\cdf{\Xinkopt|\tau_0}|\tau_0 \Big) < \mathop{\liminf}\limits_{k \rightarrow \infty}\frac{1}{k}I\big(\Xinkopt; \Yepsk|\tau_0\big)-2\delta\Big) <3\delta,
\end{equation*}
and we conclude that 
\begin{eqnarray}
    {\rm p-}\mathop{\lim\!\inf}\limits_{k \rightarrow \infty} \zkeps{k}{\eps}\Big( \cdf{\Xinkopt|\tau_0}|\tau_0\Big) & \triangleq & \sup\left\{\alpha \in \mR\; \big| \mathop{\lim}\limits_{k \rightarrow \infty}\Pr \Big(\zkeps{k}{\eps}\Big( \cdf{\Xinkopt|\tau_0}|\tau_0\Big) < \alpha \Big) 
    =  0   \right\}\notag\\
    \label{eqn:Relationship_PliminfZkeps_and_MIeps}
    &\ge &  \mathop{\liminf}\limits_{k \rightarrow \infty}\frac{1}{k}I\big(\Xinkopt; \Yepsk|\tau_0\big).
\end{eqnarray}

Next, we consider the power constraint. The following lemma asserts that a Gaussian codebook generated according to a distribution which satisfies the trace constraints $\frac{1}{k}\Tr\Big\{\corrmatxeopt(\tau_0)\Big\}\le P$ 
and $\frac{1}{k^2}\Tr\Big\{\big(\corrmatxeopt(\tau_0)\big)^2\Big\}\mathop{\longrightarrow}\limits_{k\rightarrow\infty}0$, satisfies the per-codeword power constraint \eqref{eqn:AsyncConst1} asymptotically as $k\rightarrow\infty$ with a probability which is arbitrarily close to $1$:

\begin{lemma}
    \label{lemma:PowerConstrSatisfied}
    Let $\Xinkopt\sim\dsN\big(\alzer{k},\corrmatxeopt(\tau_0)\big)$, with $\frac{1}{k}\Tr\Big\{\corrmatxeopt(\tau_0)\Big\}\le P$, and assume that  $\frac{1}{k^2}\Tr\Big\{\big(\corrmatxeopt\big)\Big\}\mathop{\longrightarrow}\limits_{k\rightarrow\infty}0$. Then, $\forall \delta >0$, there exists $k_{\delta}\in\mNp$ such that $\forall k\in\mNp$,  $k>k_{\delta}$ it holds that  $\Pr\Big(\frac{1}{k}\sum_{i=0}^{k-1} \big(X_{\rm opt}[i]\big)^2 \le P\Big)>1-\delta$.
\end{lemma}
\begin{IEEEproof}
    We consider the distribution of $\big(\Xinkopt\big)^T\cdot \Xinkopt$. First note that $\corrmatxeopt(\tau_0)$ is in general positive semidefinite. Let $\tkx$ denote the number of zero eigenvalue of $\corrmatxeopt(\tau_0)$. Then, as in \eqref{eqn:Cxw_decomp}, $\corrmatxeopt(\tau_0)$ can be decomposed as 
    \begin{equation}
    \label{eqn:CxOnly_decomp}
        \corrmatxeopt(\tau_0) =
    \left[\begin{array}{cc}\mathsf{P}_{X,k}^{\mbox{\scriptsize CW}}(\tau_0) & \mathsf{P}_{X,k}^{0}(\tau_0) \end{array}\right]\left[\begin{array}{cc}
    \Cmat_{\tilde{X}^{(k)}_{\opt}}(\tau_0) & \;\;\alzer{(k-\tkx)\times \tkx} \\
        \alzer{\tkx\times (k-\tkx)} & \;\;\alzer{\tkx \times \tkx}
    \end{array}\right]\left[\begin{array}{cc}\mathsf{P}_{X,k}^{\mbox{\scriptsize CW}}(\tau_0) & \mathsf{P}_{X,k}^{0}(\tau_0) \end{array}\right]^T\!\!\!,
\end{equation}
where $\left[\begin{array}{cc}\mathsf{P}_{X,k}^{\mbox{\scriptsize CW}}(\tau_0) & \mathsf{P}_{X,k}^{0}(\tau_0) \end{array}\right]$ is an orthogonal $k \times k$ matrix, $\Cmat_{\tilde{X}^{(k)}_{\opt}}(\tau_0)\in\mR^{(k-\tkx)\times(k-\tkx)}$ is a symmetric positive-definite matrix.
\ifFullVersion
 Then 
 \[
    \left[\begin{array}{cc}\mathsf{P}_{X,k}^{\mbox{\scriptsize CW}}(\tau_0) & \mathsf{P}_{X,k}^{0}(\tau_0) \end{array}\right]^T \corrmatxeopt(\tau_0) \left[\begin{array}{cc}\mathsf{P}_{X,k}^{\mbox{\scriptsize CW}}(\tau_0) & \mathsf{P}_{X,k}^{0}(\tau_0) \end{array}\right]= \left[\begin{array}{cc}
    \Cmat_{\tilde{X}^{(k)}_{\opt}}(\tau_0) & \;\;\alzer{(k-\tkx)\times \tkx} \\
        \alzer{\tkx\times (k-\tkx)} & \;\;\alzer{\tkx \times \tkx}
    \end{array}\right]
 \]
 and we obtain 
 \[
    \left[\begin{array}{cc}\mathsf{P}_{X,k}^{\mbox{\scriptsize CW}}(\tau_0)  & \mathsf{P}_{X,k}^{0}(\tau_0) \end{array}\right]^T\cdot \Xinkopt \sim \dsN\left(\alzer{k},\left[\begin{array}{cc}
    \Cmat_{\tilde{X}^{(k)}_{\opt}}(\tau_0) & \;\;\alzer{(k-\tkx)\times \tkx} \\
        \alzer{\tkx\times (k-\tkx)} & \;\;\alzer{\tkx \times \tkx}
    \end{array}\right]\right).
 \]
 Hence, repeating the steps leading to the derivation of \eqref{eqn:chisquare_dist} we obtain
 \begin{align}
  \left(\mathsf{P}_{X,k}^{\mbox{\scriptsize CW}}(\tau_0)\right)^T\cdot \Xinkopt & \sim \dsN\left(\alzer{k-\tkx}, \Cmat_{\tilde{X}^{(k)}_{\opt}}(\tau_0)\right)\notag\\
    \Rightarrow \Gamma_X^{(k)}\triangleq\Big(\Cmat_{\tilde{X}^{(k)}_{\opt}}(\tau_0)\Big)^{-\frac{1}{2}}\cdot \left(\mathsf{P}_{X,k}^{\mbox{\scriptsize CW}}(\tau_0)\right)^T\cdot \Xinkopt & \sim \dsN\left(\alzer{k},\iden{k-\tkx}\right). \notag
 \end{align}
 \fi
 It follows that the \ac{rv} $\Gamma_X^{(k)}$ is a vector of $k-\tkx$ i.i.d. Gaussian \acp{rv}, $\big\{\Gamma_{X,i}\big\}_{i=0}^{k-\tkx-1}$, each has a zero mean an unit variance. Therefore, we can write
 \begin{align}
     \big(\Xinkopt\big)^T\cdot \Xinkopt & \EqDist{} 
     \big(\Xinkopt\big)^T\cdot \left[\begin{array}{cc}\mathsf{P}_{X,k}^{\mbox{\scriptsize CW}}(\tau_0) & \mathsf{P}_{X,k}^{0}(\tau_0) \end{array}\right] \cdot  \left[\begin{array}{cc}\mathsf{P}_{X,k}^{\mbox{\scriptsize CW}}(\tau_0) & \mathsf{P}_{X,k}^{0}(\tau_0) \end{array}\right]^T \cdot \Xinkopt\notag\\
     & \EqDist{} 
     \big(\Xinkopt\big)^T\cdot \left(\mathsf{P}_{X,k}^{\mbox{\scriptsize CW}}(\tau_0)\right) \cdot  \left(\mathsf{P}_{X,k}^{\mbox{\scriptsize CW}}(\tau_0)\right)^T \cdot \Xinkopt\notag\\
      & \EqDist{} \!
     \big(\Xinkopt\big)^T\!\!\cdot\! \left(\mathsf{P}_{X,k}^{\mbox{\scriptsize CW}}(\tau_0)\right)\! \cdot\! \Big(\!\Cmat_{\tilde{X}^{(k)}_{\opt}}(\tau_0)\!\Big)^{-\frac{1}{2}} \!\!\!\cdot\! \Cmat_{\tilde{X}^{(k)}_{\opt}}(\tau_0)\!\cdot\! \Big(\!\Cmat_{\tilde{X}^{(k)}_{\opt}}(\tau_0)\!\Big)^{-\frac{1}{2}}\!\!\!\cdot \! \left(\mathsf{P}_{X,k}^{\mbox{\scriptsize CW}}(\tau_0)\right)^T\!\! \cdot\! \Xinkopt\notag\\
      & \EqDist{} 
      \Big(\Gamma_X^{(k)}\Big)^T\cdot \Cmat_{\tilde{X}^{(k)}_{\opt}}(\tau_0)\cdot \Gamma_X^{(k)}.\notag
 \end{align}
 As $\Cmat_{\tilde{X}^{(k)}_{\opt}}(\tau_0)\succ 0$, and symmetric we can write its eigenvalue decomposition as  $\Cmat_{\tilde{X}^{(k)}_{\opt}}(\tau_0) = \big(\Pmat_X(\tau_0)\big)^T \cdot \Dmat_X(\tau_0) \cdot\Pmat_X(\tau_0)$, where $\Pmat_X(\tau_0)$ is an orthogonal matrix and $\Dmat_X(\tau_0)$ a diagonal matrix with $k-\tkx$ positive elements $\big\{d_{X,i}^{\{\tau_0\}}\big\}_{i=0}^{k-\tkx}$. Finally we conclude that 
 \[
    \frac{1}{k}\big(\Xinkopt\big)^T\cdot \Xinkopt  \EqDist{} \frac{1}{k}\sum_{i=0}^{k-\tkx-1}d_{X,i}^{\{\tau_0\}}\cdot \Gamma_{X,i}^2,
 \]
 where $\Gamma_{X,i}^2\sim\chi^2(1)$, chi-square \acp{rv}, mutually independent over the index $i$. Then
 \begin{subequations}
\label{eqn:xsqr_mean}
 \begin{align}
   \dsE\Big\{\frac{1}{k}\big(\Xinkopt\big)^T\cdot \Xinkopt\Big\} & = \frac{1}{k}\sum_{i=0}^{k-\tkx-1}d^{\{\tau_0\}}_{X,i} = \frac{1}{k}\Tr\big\{\Cmat_{\tilde{X}^{(k)}_{\opt}}(\tau_0)\}\notag\\
   & = \frac{1}{k}\Tr\big\{\corrmatxeopt(\tau_0)\big\} \notag\\
   &\stackrel{(a)}{\le} P\notag\\
   \var\Big(\frac{1}{k}\big(\Xinkopt\big)^T\cdot \Xinkopt\Big) & = \sum_{i=0}^{k-\tkx-1}\frac{1}{k^2}\Big(d^{\{\tau_0\}}_{X,i}\Big)^2 \cdot  \var\big(\Gamma_{X,i}^2\big)\notag\\
   & = \frac{2}{k^2} \cdot \Tr\Big(\big(\corrmatxeopt(\tau_0)\big)^2\Big),\notag
 \end{align}
 \end{subequations}
 where (a) follows by choice of the statistics used for generating the channel input $\Xinkopt$.
 As by assumption $\frac{1}{k^2}\Tr\Big\{\big(\corrmatxeopt(\tau_0)\big)^2\Big\}\mathop{\longrightarrow}\limits_{k\rightarrow\infty}0$, then repeating  the argument in the discussion after \eqref{eqn:Expression_tvarnk}, we can apply 
 Chebyshev's inequality and conclude that for any arbitrary $\delta$, taking $k$ sufficiently large we obtain
$\Pr\Big(\Big|\frac{1}{k}\big(\Xinkopt\big)^T\cdot \Xinkopt- (P-\delta)\Big|>\delta \Big) <\delta$.
 \end{IEEEproof}

\label{pg:ExplanationLowerBound}
As $\Xinkopt$ has a specific distribution, and it satisfies the per-codeword power constraint \eqref{eqn:AsyncConst1} with a probability arbitrarily close to $1$, as $k$ increases, then, from the general capacity formula \cite[Thm. 3.6.1]{han2003information} it follows that ${\rm p-}\mathop{\lim\!\inf}\limits_{k \rightarrow \infty} \zkeps{k}{\eps}\Big( \cdf{\Xinkopt|\tau_0}|\tau_0\Big)$ is a lower bound on capacity $\Capacity_\myEps(\tau_0)$. Hence, we obtain the following lower bound on capacity: 
\begin{align}
    \Capacity_\myEps(\tau_0) & \ge {\rm p-}\mathop{\lim\!\inf}\limits_{k \rightarrow \infty} \zkeps{k}{\eps}\Big( \cdf{\Xinkopt|\tau_0}|\tau_0\Big)\notag\\
    &\ge  \mathop{\liminf}\limits_{k \rightarrow \infty}\frac{1}{k}I\big(\Xinkopt;\Yepsk|\tau_0\big).
    \label{eqn:lower_bound_capacity_given_tau0}
\end{align}


Next, recall that from Fano's inequality, for any $[R,k]$ code, designed for delay $\tau_0$, having an average  probability of error $P_e^{l}(\tau_0)\le\rho$, $\rho\in[0,1)$, we obtain (see, e.g., \cite[Thm. 3]{verdu1994general}):
    \begin{eqnarray}
    R &\stackrel{(a)}{\leq}& \frac{1}{1-\rho}\cdot\frac{1}{k}I\left(\bX^{(k)}; \bYepsk|\tau_0\right) + \frac{h(\rho)}{k} \notag\\
    &  \stackrel{(b)}{\leq}& \frac{1}{1-\rho}\cdot\frac{1}{k}\mathop{\sup}\limits_{\left\{ \dsE_U\{\frac{1}{l}\sum_{i=0}^{l-1}\big(\xin_{U}\left[i\right]\big)^2\}\le P\right\}_{k\in\mNp}} I\left(\bX^{(k)}; \bYepsk|\tau_0\right)  + \frac{h(\rho)}{k}  \notag\\
    &  \stackrel{(c)}{\leq}& \frac{1}{1-\rho}\cdot\frac{1}{k}\mathop{\sup}\limits_{\left\{\cdf{\bX^{(k)}|\tau_0}:\; \dsE\{\frac{1}{l}\sum_{i=0}^{l-1}\big(\bX\left[i\right]\big)^2\}\le P\right\}_{k\in\mNp}} I\left(\bX^{(k)}; \Yepsk|\tau_0\right) + \frac{h(\rho)}{k} \notag
\end{eqnarray}
where in (a) $\bX^{(k)}$ is an \ac{rv} which places probability mass of $\frac{1}{|\mU|}$ on each codeword $\xin_{u}^{(l)}$, $u\in\mU$, and $\bYepsk = \bXk + \Wepsk$; (b) follows as when each codeword $u\in\mU$ satisfies $\frac{1}{l}\sum_{i=0}^{l-1}\big(\xin_{u}\left[i\right]\big)^2\le P$, then the average over all codewords in the codebook satisfies the same constraint;  (c) follows as we define $\bX[i] = x_U[i]$ and then relax the restrictions on the input codebook by directly maximizing 
over the \ac{rv} $\bX^{(k)}$.
Hence, we obtain the upper bound on capacity as 
 \begin{equation}
    \label{eqn:cap_eps_uppr_bnd}
    \Capacity_\myEps(\tau_0)\leq\mathop{\lim\!\inf}\limits_{k\rightarrow \infty}\mathop{\sup}\limits_{\left\{\cdf{\bX {}^{(k)}|\tau_0}:\; \dsE\{\frac{1}{l}\sum_{i=0}^{l-1}\big(\bX\left[i\right]\big)^2\}\le P\right\}_{k\in\mNp}} \frac{1}{k}I\left(\bX^{(k)}; \Yepsk|\tau_0\right).
\end{equation}

By \cite{cover1989gaussian}, for every given $k\in\mNp$, the supremum in Eqn. \eqref{eqn:cap_eps_uppr_bnd} is achieved 
by Gaussian random vector.
If the optimal distribution in \eqref{eqn:cap_eps_uppr_bnd} satisfies the conditions of Lemma \ref{lemma:PowerConstrSatisfied}, then 
\eqref{eqn:cap_eps_uppr_bnd} is maximized by the distribution 
$\Xinkopt\sim\cdf{\Xinkopt|\tau_0}$, which is used in the derivation of the lower bound \eqref{eqn:lower_bound_capacity_given_tau0}, i.e.,
\begin{equation}
    \label{eqn:upper_bound_capacity}
    \Capacity_\myEps(\tau_0)\leq\mathop{\lim\!\inf}\limits_{k\rightarrow \infty} \frac{1}{k}I\left(\Xinkopt; \Yepsk|\tau_0\right),
\end{equation}
which, combined with the lower bound of \eqref{eqn:lower_bound_capacity_given_tau0}, results in $\Capacity_\myEps(\tau_0)=\mathop{\lim\!\inf}\limits_{k\rightarrow \infty} \frac{1}{k}I\big(\Xepsk; \Yepsk|\tau_0\big)$.

Finally, as the sampling interval is incommensurate with $\Tc$, then for the transmission of asymptotically long sequence of {\em messages}, the sequence of sampling phases $\tau_0$ is a uniformly distributed sequence over the interval $[0,\Tc)$, \cite[Example 2.1]{KuipersNiederreiter:UniformDistributionBook}. As transmitter's and receiver's knowledge of $\tau_0\in [0,\Tc)$ allows both units to select the appropriate codebook with a rate of 
	$\mathop{\lim\!\inf}\limits_{k\rightarrow \infty} \frac{1}{k}I\big(\Xinkopt; \Yepsk|\tau_0\big)$, it follows that when rate adaptation is allowed, capacity can be expressed as the average rate
\[
    C_\eps = \frac{1}{\Tc}\int_{\tau_0=0}^{\Tc} C_\eps(\tau_0)\mbox{d}\tau_0.
\]
	\end{sloppypar}

\bigskip

\section{Proof of Thm. \ref{thm:AsycCap}}
\label{app:Proof2}

\subsection{Convergence of the Noise Correlation Matrices and Their Inverses}
\label{Appndx:NoiseCorrelationConvergence}
Define the set $\mK \triangleq \left\{0,1, 2,..., k-1\right\}$, and
consider the $k$-dimensional, zero-mean, real random vectors $W_{n}^{(k)}$ and $W_{\myEps}^{(k)}$. 
Recall the definition of the correlation matrix 
$\corrmatwe(\tau_0)$ in Eqn. \eqref{eqn:corrmatdefe} and define the correlation matrix
$\corrmatwn(\tau_0)$ in a similar manner:
\begin{equation}
\label{eqn:corrmatdefn}
        \left(\corrmatwn(\tau_0)\right)_{u,v}  \triangleq  \dsE\big\{W_{n}[u]\cdot W_{n}[v]\big|\tau_0\big\} \equiv \disccorrn^{\{\tau_0\}}[v,u-v],
\end{equation}
for $(u,v) \in \mK \times \mK$.
Note that since $\dsE\big\{W_{n}[u]\cdot W_{n}[v]\big|\tau_0\big\} = \dsE\big\{W_{n}[v]\cdot W_{n}[u]|\big|\tau_0\big\}$, then
$\disccorrn^{\{\tau_0\}}[v,u-v] = \disccorrn^{\{\tau_0\}}[u,v-u]$, and $ \Big(\corrmatwn(\tau_0)\Big)_{u,v} =  \Big(\corrmatwn(\tau_0)\Big)_{v,u}$. Similarly, $\disccorr^{\{\tau_0\}}[v,u-v] = \disccorr^{\{\tau_0\}}[u,v-u]$, and $ \Big(\corrmatwe(\tau_0)\Big)_{u,v} =  \Big(\corrmatwe(\tau_0)\Big)_{v,u}$.

Next, we note that by the definition of $\eps_n\triangleq \frac{\lfloor n\cdot \eps\rfloor}{n}$ it directly follows that $\frac{n\eps-1}{n}\le \myEps_n \le \frac{n\eps}{n}$, hence,
\begin{equation}
\label{eqn:limeps}
    \lim_{n\rightarrow \infty}\eps_n = \myEps.
\end{equation}
Define $\disccorrn^{\{\tau_0\}}[i,\Delta] \triangleq \contcorr\left(i\cdot \frac{\Tc}{p+\eps_n}+\tau_0, \Delta\cdot \frac{\Tc}{p+\eps_n}\right)$. Then, by the definition of a continuous 
\ifFullVersion
function\footnote{
From \cite[Def. 11.1.5]{Rankin07}: A function $f$, defined on a general interval $\mA$, is said to be continuous on $\mA$ if it is continuous at every point $c$ in $\mA$.\\
From \cite[Def. 11.1.1]{Rankin07}: The function $f$ is said to be continuous at the point $c$ if $f(x) \rightarrow f(c)$ as $x \rightarrow c$. Equivalently, let $f(x)$ be defined on some interval $(c-\delta_0, c+\delta_0)$. Then, for every $\myEps > 0$ there exists a positive $\delta(\myEps,c)$ such that
\[
    \big| f(x)-f(c) \big| < \myEps, \qquad \forall x\in \big(c-\delta(\myEps,c), c+\delta(\myEps,c)\big).
\]},
\fi
    we obtain that continuity of $\contcorr(t,\lambda)$ in $t$ and in $\lambda$, combined with \eqref{eqn:limeps}, implies that $\forall i, \Delta \in \mZ$,
\begin{eqnarray}
\label{eqn:limcorr}
    \lim_{n\rightarrow \infty} \disccorrn^{\{\tau_0\}}[i,\Delta] & = &  \lim_{n\rightarrow \infty} \contcorr\left(i\cdot \frac{\Tc}{p+\eps_n}+\tau_0, \Delta\cdot \frac{\Tc}{p+\eps_n}\right)\notag \\
    &= &  \contcorr\left(i\cdot \frac{\Tc}{p+\eps}+\tau_0, \Delta\cdot \frac{\Tc}{p+\eps}\right) \equiv \disccorr^{\{\tau_0\}}[i,\Delta].
\end{eqnarray}
Recall that as the autocorrelation function $\Cwc(t,\lambda)$ is bounded and continuous in $(t, \lambda)\in\mR^2$, periodic in $t\in\mR$, and is zero $\forall |\lambda|\ge \lambda_m$, then it is 
bounded and uniformly continuous with respect to time $t \in [0,\Tc]$ and lag $\lambda\in\mR$ [23,
Ch. III, Thm. 3.13]. Also recall that the correlation matrices $\corrmatwe(\tau_0)$ and  $\corrmatwn(\tau_0)$ are non-singular (the rationale for this assumption is given in Comment \ref{rem:CovMatIsPosDef}). Combining these properties with the definitions 
of the correlation matrices $\corrmatwe(\tau_0)$ and $\corrmatwn(\tau_0)$ 
in \eqref{eqn:corrmatdefe} and \eqref{eqn:corrmatdefn}, respectively, and with the limit in \eqref{eqn:limcorr} we obtain that
\begin{equation}
\label{eqn:Matn2e}
    \lim_{n\rightarrow \infty} \max_{\substack{\tau_0\in [0,\Tc],\\(u,v)\in \mK \times \mK}} \left\{ \left| \left(\corrmatwn(\tau_0)\right)_{u,v}  - \left(\corrmatwe(\tau_0)\right)_{u,v}\right| \right\} = 0.
\end{equation}

\begin{sloppypar}
Next, consider the mapping $m_k: \mR^{k^2} \mapsto \mR^{k^2}$, defined via
\[
    m_k\big(\Cmat^{(k)}\big) = \big(\Cmat^{(k)}\big)^{-1}, \qquad \Cmat^{(k)} \in \mR^{k^2}.
\]
This is a continuous mapping over the set of positive-definite $k\times k$ matrices $\Cmat^{(k)}\succ 0$, see, e.g., \cite[Eqns. (1.5)-(1.6)]{stewart1969continuity}. 
Consider now the positive-definite matrix $\big(\corrmatwe(\tau_0)\big)^{-1}$. By the strict diagonal dominance of $\corrmatwe(\tau_0)$ (see condition \eqref{eqn:ConditionThmUnifConv_2})  we obtain that 
\begin{eqnarray}
	 \maxEig\Big\{\big(\corrmatwe(\tau_0)\big)^{-1}\Big\}
	 & \stackrel{(a)}{\le} & \norm{\big(\corrmatwe(\tau_0)\big)^{-1}}_1\notag\\
	 %
	 & \stackrel{(b)}{=}& \norm{\big(\corrmatwe(\tau_0)\big)^{-1}}_{\infty}\notag\\
	 &\stackrel{(c)}{\le}& \bigg(\mathop{\min}\limits_{0 \leq u \leq  k-1}\bigg\{\Big|\big(\corrmatwe(\tau_0)\big)_{u,u}\Big|-\mathop{\sum}\limits_{v=0, v\neq u}^{k-1}\Big|\big(\corrmatwe(\tau_0)\big)_{u,v}\Big|\bigg\}\bigg)^{-1}\notag\\
	 \label{eqn:UpperBndMaxEig}
	 & \stackrel{(d)}{\le} & \bigg(\mathop{\min}\limits_{0 \leq t \leq \Tc}\bigg\{\Cwc(t,0)-2\Memd\cdot\mathop{\max}\limits_{|\lambda|>\frac{\Tc}{p+1}}\big\{|\Cwc(t, \lambda)|\big\}\bigg\}\bigg)^{-1},
\end{eqnarray}
where (a) follows from the upper bound of \cite[Thm. 5]{dembo1988bounds}; (b) follows from the symmetry of 
$\big(\corrmatwe(\tau_0)\big)^{-1}$, due to which we can obtain explicitly 
$\norm{\big(\corrmatwe(\tau_0)\big)^{-1}}_1\triangleq \mathop{\max}\limits_{0\leq v \leq k-1}\Big\{\mathop{\sum}\limits_{u=0}^{k-1}\big|\big(\big(\corrmatwe(\tau_0)\big)^{-1}\big)_{u,v}\Big|\Big\}=\mathop{\max}\limits_{0\leq v \leq k-1}\Big\{\mathop{\sum}\limits_{u=0}^{k-1}\big|\big(\big(\corrmatwe(\tau_0)\big)^{-1}\big)_{v,u}\big|\Big\}=\mathop{\max}\limits_{0\leq u \leq k-1}\Big\{\mathop{\sum}\limits_{v=0}^{k-1}\big|\big(\big(\corrmatwe(\tau_0)\big)^{-1}\big)_{u,v}\big|\Big\}=\norm{\big(\corrmatwe(\tau_0)\big)^{-1}}_{\infty}$; (c) follows from the bound in \cite[Eqn. (4)]{ahlberg1963convergence} 
(see also \cite[Eqn. (3)]{moravca2008bounds}), as, by condition \eqref{eqn:ConditionThmUnifConv_2}, the matrix $\corrmatwe(\tau_0)$ is \ac{sdd}, see Comment \ref{rem:cond_strict_diag}; and lastly,	 (d) follows from Eqn. \eqref{eqn:diagonal_dominance}. Evidently, the same  bound applies to $\maxEig\Big\{\big(\corrmatwn(\tau_0)\big)^{-1}\Big\}$ and we note that it is independent of $\tau_0$. Then, since the magnitudes of the elements of a positive-definite real, symmetric matrix are upper-bounded by its largest   eigenvalue\footnote{\label{footnote:boundMatElement}
	 for a positive-definite real, symmetric matrix $\Cmat$, and two vectors $\evec_i$ and $\evec_j$, we have
	 \[
	     0<(\evec_i-\evec_j)^T\cdot\Cmat\cdot (\evec_i-\evec_j)=\evec_i^T\cdot\Cmat\cdot \evec_i+\evec_j^T\cdot\Cmat\cdot \evec_j-\evec_i^T\cdot\Cmat\cdot \evec_j-\evec_j^T\cdot\Cmat\cdot \evec_i \Rightarrow \frac{1}{2}\cdot\big(\evec_i^T\cdot\Cmat\cdot \evec_i+\evec_j^T\cdot\Cmat\cdot \evec_j\big)  > \evec_i^T\cdot\Cmat\cdot \evec_j
	 \]
	 \[
	     0<(\evec_i+\evec_j)^T\cdot\Cmat\cdot (\evec_i+\evec_j)=\evec_i^T\cdot\Cmat\cdot \evec_i+\evec_j^T\cdot\Cmat\cdot \evec_j+\evec_i^T\cdot\Cmat\cdot \evec_j+\evec_j^T\cdot\Cmat\cdot \evec_i \Rightarrow -\frac{1}{2}\cdot\big(\evec_i^T\cdot\Cmat\cdot \evec_i+\evec_j^T\cdot\Cmat\cdot \evec_j\big)  < \evec_i^T\cdot\Cmat\cdot \evec_j
	\]
	Hence, $0\le\left|\evec_i^T\cdot\Cmat\cdot \evec_j \right|<\frac{1}{2}\cdot\big(\evec_i^T\cdot\Cmat\cdot \evec_i+\evec_j^T\cdot\Cmat\cdot \evec_j\big)$. Now, letting $\evec_i$ denote the all-zero vector except for $1$ at the $i$-th element ($i\in\mN$), we note that the $(i,j)$-th element of $\Cmat$ is given by
	\[
        \left|\left(\Cmat\right)_{i,j}\right|=\left|\evec_i^T\cdot\Cmat\cdot \evec_j\right| < \frac{1}{2}\cdot\big(\evec_i^T\cdot\Cmat\cdot \evec_i+\evec_j^T\cdot\Cmat\cdot \evec_j\big)
		\le \mathop{\max \mathrm{Eig}}\left\{\Cmat\right\},
    \]
    where the last inequality follows from \cite[Thm. 4.2.2]{horn2012matrix}.
    \ifFullVersion   
	   See also:\\
	  {\rm https://mathoverflow.net/questions/235861/largest-element-in-inverse-of-a-positive-definite-symmetric-matrix}\\
	  {\rm https://math.stackexchange.com/questions/29787/bounds-on-inverse-elements-of-hermitian-matrices}
	 \fi   
	} 
we conclude that the elements of the matrices $\big(\corrmatwe(\tau_0)\big)^{-1}$ and  $\big(\corrmatwn(\tau_0)\big)^{-1}$ belong to a finite interval whose end points are independent of $k$, $\tau_0$ and $n$.

Next, note that boundedness of $\Cwc(t,\lambda)$ (see Section \ref{subsec:problem_formulation}) implies that the elements of $\corrmatwn(\tau_0)$ and of $\corrmatwe(\tau_0)$ are all bounded.
It therefore follows from  \eqref{eqn:Matn2e}, the boundedness of the elements of  $\corrmatwe(\tau_0)$,  $\corrmatwn(\tau_0)$, $\big(\corrmatwe(\tau_0)\big)^{-1}$ and  $\big(\corrmatwn(\tau_0)\big)^{-1}$, boundedness and uniform continuity of the \ac{ct} correlation function $\contcorr\left(t, \lambda \right)$ in $t$ and in $\lambda$,  continuity of the mapping $m_k: \mR^{k^2} \mapsto \mR^{k^2}$,\footnote{Since the inverse is a continuous mapping from a compact and finite set to a compact and finite set.} and from \cite[Thm. 11.2.3]{Rankin07} that
\begin{equation}
\label{Eqn:convergence_of_inverse_correlation_matrix}
     \lim_{n\rightarrow \infty} \max_{\substack{\tau_0\in [0,\Tc],\\(u,v)\in \mK \times \mK}} \left\{ \left| \Big(\big(\corrmatwn(\tau_0)\big)^{-1}\Big)_{u,v}  - \Big(\big(\corrmatwe(\tau_0)\big)^{-1}\Big)_{u,v}\right| \right\} = 0.
\end{equation}
\end{sloppypar}




\subsection[Showing convergence of the mutual information for finite k]{Showing that $\limn \frac{1}{k}I(\Xnkopt;\Ynk|\taunk\opt)=\frac{1}{k} I(\Xinkopt;\Yepsk|\tauepsk\opt)$ for the Optimal Sampling Phases and Inputs Distributions $\big(\taunk\opt,$ $\cdf{\Xnkopt|\taunk\opt}\big)$ and $\big(\tauepsk\opt, \cdf{\Xinkopt|\tauepsk\opt}\big)$}
\label{appndx:convergence of mutual informations}

\label{pg:Def_mCxk}
Consider a fixed $k\in\mNp$, let 
\begin{equation}
\label{eqn:Def_setCx}
 \mCxk\triangleq \bigg\{\tau\in [0,\Tc), \Cmatxk\in \mR^{k\times k}\Big|\sum_{i=0}^{k-1}\left(\Cmatxk\right)_{ii}\leq k\cdot P, \; \Cmatxk=\left(\Cmatxk\right)^T, \Cmatxk\succcurlyeq 0 \bigg\}, 
 \end{equation}
 and denote
\begin{subequations}
    \label{eqn:defOptProbs}
    \begin{eqnarray}
		\label{eqn:defOptSolCorr_Xeps}
		(\taunk\opt,\Cmatxnk\opt) & = & \mathop{\argmax}\limits_{(\tau_0,\Cmatxk)\in \mathcal{C}_{X^{(k)}}} \frac{1}{2k} \log \Big(\Det\big({\Cmat}_{X^{(k)}}+\corrmatwn(\tau_0)\big)/\Det\big(\corrmatwn(\tau_0)\big)\Big)\\
		\label{eqn:defOptSolCorr_Xn}
		(\tauepsk\opt,\Cmatxepsk\opt) & = & \mathop{\arg\!\max}\limits_{(\tau_0,\Cmatxk)\in \mathcal{C}_{X^{(k)}}} \frac{1}{2k} \log\Big( \Det\big({\Cmat}_{X^{(k)}}+\corrmatwe(\tau_0)\big)/\Det\big(\corrmatwe(\tau_0)\big)\Big).
	\end{eqnarray} 
\end{subequations}
Then, the zero-mean Gaussian random vectors $\Xnkopt$ and $\Xinkopt$, with covariance matrices $\Cmatxnk\opt$ and $\Cmatxepsk\opt$, respectively, at the respective sampling phases $\taunk\opt$ and $\tauepsk\opt$, maximize the mutual information expressions $\frac{1}{k}I(\Xnk;\Ynk|\tau_0)$ and $\frac{1}{k} I(\Xepsk;\Yepsk|\tau_0)$, respectively, when maximization is over all sampling phases and associated input distributions which satisfy the respective  trace  constraint, $\frac{1}{k}\Tr\big\{\corrmatxn\big\}\le P$, $\frac{1}{k}\Tr\big\{\corrmatxe\big\}\le P$, see, e.g., \cite[Eqn. (6)]{cover1989gaussian}. We now have the following lemma:
\begin{lemma}  
	\label{lem:tightness}
	When $k\in\mNp$ is fixed, 		$\Xnkopt$ and $\Xinkopt$ are zero-mean Gaussian random vectors with covariance matrices $\Cmatxnk\opt$ and $\Cmatxepsk\opt$, respectively, where $(\taunk\opt,\Cmatxnk\opt)$ and $(\tauepsk\opt,\Cmatxepsk\opt)$ satisfy \eqref{eqn:defOptProbs}, then 
	\begin{equation}
	\label{eqn:Limit_of_mutual_information_vs_n}
	   \limn \frac{1}{k}I(\Xnkopt;\Ynk|\taunk\opt)=\frac{1}{k} I(\Xinkopt;\Yepsk|\tauepsk\opt).
	\end{equation}
\end{lemma}
		
\begin{IEEEproof}
	First, recall that by \cite{cover1989gaussian},   when the covariance matrix and sampling phase pairs are given as  $(\taunk\opt,\Cmatxnk\opt)$ and $(\tauepsk\opt,\Cmatxepsk\opt)$ satisfying Eqns. \eqref{eqn:defOptProbs}, then the mutual information expressions in \eqref{eqn:Limit_of_mutual_information_vs_n} 
	evaluated for $\Xnkopt$ and $\Xinkopt$  Gaussian inputs processes corresponding to the sampling phase-correlation matrix pairs of Eqns. \eqref{eqn:defOptProbs},  
	are equal to the maximal values of the objective functions in \eqref{eqn:defOptProbs}:
	\begin{eqnarray*}
	    \frac{1}{k}I(\Xnkopt;\Ynk|\taunk\opt) & = &  \frac{1}{2k} \log \Big(\Det \big(\Cmatxnk\opt+\corrmatwn(\taunk\opt)\big)/\Det\big(\corrmatwn(\taunk\opt)\big)\Big)\\
		\frac{1}{k} I(\Xinkopt;\Yepsk|\tauepsk\opt) & = & \frac{1}{2k} \log\Big( \Det\big(\Cmatxepsk\opt+\corrmatwe(\tauepsk\opt)\big)/\Det\big(\corrmatwe(\tauepsk\opt)\big)\Big).
	\end{eqnarray*}
	Therefore, convergence of the limit in \eqref{eqn:Limit_of_mutual_information_vs_n} corresponds to having that the optimal values of the objective function in the optimization problem  \eqref{eqn:defOptSolCorr_Xeps} converge, as $n\rightarrow\infty$, to the optimal value of the objective function in \eqref{eqn:defOptSolCorr_Xn}. To prove this convergence we employ \cite[Thm. 2.1]{kanniappan1983uniform}\footnote{\cite[Thm. 2.1]{kanniappan1983uniform}: Let $f_n \rightarrow f$ uniformly as $n\rightarrow\infty$. Then, the sequence of problems $P_n: a_n=\mathop{\inf}\limits_{x\in \mX}f_n(x)$ converges to the problem $P: a=\mathop{\inf}\limits_{x\in \mX}f(x)$ as $n\rightarrow\infty$.\\
	See additional conditions regarding the application of \cite[Thm. 2.1]{kanniappan1983uniform} in Footnotes \ref{footnote:conditions} and \ref{footnote:convex}.}. The main requirement for the application of \cite[Thm. 2.1]{kanniappan1983uniform} is that $\mathop{\lim}\limits_{n\rightarrow \infty} \frac{1}{2k} \log \Big( \Det\big({\Cmat}_{X^{(k)}}+\corrmatwn(\tau_0)\big)/\Det\big(\corrmatwn(\tau_0)\big)\Big)=\frac{1}{2k} \log \Big(\Det\big({\Cmat}_{X^{(k)}}+\corrmatwe(\tau_0)\big)/\Det\big(\corrmatwe(\tau_0)\big)\Big)$ {\em uniformly} over $\mathcal{C}_{X^{(k)}}$. In the following we show that such a uniform convergence holds.

    \label{pg:modify2_due_to_power}
    It follows from the limit in \eqref{Eqn:convergence_of_inverse_correlation_matrix} that for any $\delta_1>0$ there exists $n_0(\delta_1) \in \mNp$ sufficiently large such that for all $0\leq l,q \leq k-1$, $\tau_0\in[0,\Tc)$, and for all $n>n_0(\delta_1)$, it holds that  $\Big|\Big(\big(\corrmatwn(\tau_0)\big)^{-1}-\big(\corrmatwe(\tau_0)\big)^{-1}\Big)_{l,q}\Big|\leq \delta_1$. Now, we note that since ${\Cmat}_{X^{(k)}}$ is a positive semi-definite matrix which satisfies a constraint on the sum of its diagonal elements \eqref{eqn:Def_setCx}, then from the Cauchy-Schwartz inequality \cite[Eqn. (9-176)]{Papoulis2002}, \cite[Sec. 3.6]{grimmett2001probability} we have that
	\begin{equation}
	     \label{eqn:Bounding the magnitude of Cx}
      \left|\big({\Cmat}_{X^{(k)}}\big)_{l,q}\right|\!=\!\big|\E\{X[l]X[q]\}\big|\!\leq\!\sqrt{\E\left\{(X[l])^2\right\}\E\left\{(X[q])^2\right\}}\!\leq\! \sqrt{(k\cdot P)^2}=k\cdot P,\;\forall\; 0\leq l,q\leq k-1.
	\end{equation}
    It thus follows that all the matrices $\Cmatxk \in \mCxk$ have bounded elements. We now bound \label{pg:modify3_due_to_power} $\bigg|\Big(\big(\corrmatwn(\tau_0)\big)^{-1}{\Cmat}_{X^{(k)}}\Big)_{l,q}-\Big(\big(\corrmatwe(\tau_0)\big)^{-1}{\Cmat}_{X^{(k)}}\Big)_{l,q}\bigg|$ for all $n > n_0(\delta_1)$ as follows:
	\begin{align*}
	     & \Big|\Big(\big(\corrmatwn(\tau_0)\big)^{-1}{\Cmat}_{X^{(k)}}\Big)_{l,q}-\Big(\big(\corrmatwe(\tau_0)\big)^{-1}{\Cmat}_{X^{(k)}}\Big)_{l,q}\Big|\\
	     &=\bigg|\mathop{\sum}\limits_{m=0}^{k-1}\Big(\big(\corrmatwn(\tau_0)\big)^{-1}-\big(\corrmatwe(\tau_0)\big)^{-1}\Big)_{l,m}\left({\Cmat}_{X^{(k)}}\right)_{m,q}\bigg|\notag \\
	     &\leq\delta_1 \cdot \mathop{\sum}\limits_{m=0}^{k-1}\big|\left({\Cmat}_{X^{(k)}}\right)_{m,q}\big|\leq \delta_1 \cdot P \cdot k^2,
	\end{align*}
	where we recall that $P$ and $k$ are finite and given. Thus, for any $\delta>0, \exists n_0(\delta) \in \mNp$ sufficiently large such that for all $0\leq l,q\leq k-1$, $n>n_0(\delta)$,   and for all $(\tau_0, \Cmatxk ) \in \mCxk$
	\begin{equation}
	\label{eqn:Convergence_of_CWminus1Cx}
		\bigg|\Big(\big(\corrmatwn(\tau_0)\big)^{-1}{\Cmat}_{X^{(k)}}\Big)_{l,q}-\Big(\big(\corrmatwe(\tau_0)\big)^{-1}{\Cmat}_{X^{(k)}}\Big)_{l,q}\bigg|\leq \delta.
	\end{equation}


\begin{sloppypar}			
    Observe that the product $\big(\corrmatwe(\tau_0)\big)^{-1}\corrmatxe$ has non-negative eigenvalues, \cite[Thm. 7.5]{Zhang:MatrixTheoryBasicResults}, and the maximal eigenvalue, denoted $\mathop{\max\! \mathrm{Eig}}\Big\{\big(\corrmatwe(\tau_0)\big)^{-1}\corrmatxe\Big\}$, can be upper bounded as follows:  
    \begin{eqnarray}
    \label{eqn:max_eig_val_eps}
        & &\mathop{\max\!\mathrm{Eig}}\Big\{\big(\corrmatwe(\tau_0)\big)^{-1}\corrmatxe\Big\}\notag\\
        \ifFullVersion  
            & & \qquad \stackrel{(a)}{\leq} \mathop{\max\! \mathrm{Eig}}\left\{\left(\corrmatwe(\tau_0)\right)^{-1}\right\}\cdot\mathop{\max\! \mathrm{Eig}}\left\{\corrmatxe\right\}\notag\\
        \fi 
        &   & \qquad \stackrel{(b)}{\leq}  \mathop{\max\! \mathrm{Eig}}\left\{\left(\corrmatwe(\tau_0)\right)^{-1}\right\} \cdot \norm{\corrmatxe}_1\notag\\
        \ifFullVersion   
            &   & \qquad \stackrel{(c)}{\leq}  \mathop{\max\! \mathrm{Eig}}\left\{\left(\corrmatwe(\tau_0)\right)^{-1}\right\} \cdot k^2\cdot P,\notag\\
           &   & \qquad \stackrel{(d)}{\leq} \underbrace{\frac{k^2\cdot P}{\mathop{\min}\limits_{0 \leq t \leq \Tc}\bigg\{\Cwc(t,0)- 2\Memd\cdot\mathop{\max}\limits_{|\lambda|>\frac{\Tc}{p+1}} \big\{|\Cwc(t, \lambda)|\big\}\bigg\}}}_{\triangleq \beta_0(k)}, \\
        \fi 
        & & \notag
    \end{eqnarray}
    where (a) follows from \cite[Eqn. (9)]{zhang2006eigenvalue}, see also \cite[Thm. 8.12]{Zhang:MatrixTheoryBasicResults}; in (b) we use the upper bound from \cite[Thm. 5]{dembo1988bounds};
    \ifFullVersion
        step (c) follows 
        since using the bound in \eqref{eqn:Bounding the magnitude of Cx} we obtain
        $\norm{\corrmatxe}_1\triangleq \mathop{\max}\limits_{0\leq v \leq k-1}\Big\{\mathop{\sum}\limits_{u=0}^{k-1}\big|\big({\corrmatxe}\big)_{u,v}\Big|\Big\}
        \leq k^2\cdot P$. Lastly, step (d) follows
        from the bound in \eqref{eqn:UpperBndMaxEig}.
    \fi   
    
    Since the magnitudes of the elements of $\Big(\corrmatwe(\tau_0)\Big)^{-1}$ are upper bounded by 
    \eqref{eqn:UpperBndMaxEig} (see Footnote \ref{footnote:boundMatElement}) and the magnitudes of the elements of $\corrmatxe$ are upper bounded by $k\cdot P$, we obtain that the magnitudes of the elements of the matrix product $\corrmatwe(\tau_0)\big)^{-1}\corrmatxe$ are upper bounded by 
    $k^2\cdot \bigg(\mathop{\min}\limits_{0 \leq t \leq \Tc}\bigg\{\Cwc(t,0)- 2\Memd\cdot\mathop{\max}\limits_{|\lambda|>\frac{\Tc}{p+1}}\big\{|\Cwc(t, \lambda)|\big\}\bigg\}\bigg)^{-1} \cdot P \equiv \beta_0(k)$. As the upper bound $\beta_0(k)$  is finite and independent of $\eps$, then the same upper bound applies also to the magnitudes of the elements of $\big(\corrmatwn(\tau_0)\big)^{-1}{\Cmat}_{X^{(k)}}$, and we conclude that magnitudes of the elements of $\big(\corrmatwn(\tau_0)\big)^{-1}{\Cmat}_{X^{(k)}}$ and of  $\big(\corrmatwe(\tau_0)\big)^{-1}{\Cmat}_{X^{(k)}}$ are all finite and upper bounded by a bound which increases as $k^2$, and is independent of $n$ and $\tau_0$.
	
	Consider next the ordered sets of eigenvalues of $\big(\corrmatwn(\tau_0)\big)^{-1}\corrmatxe$ and of $\big(\corrmatwe(\tau_0)\big)^{-1}\corrmatxe$  arranged in descending order: Let $\Lambda_{i,n}^{(k)}\left({\Cmat}_{X^{(k)}};\tau_0\right)\equiv \Lambda_i^{(k)} \Big\{\big(\corrmatwn(\tau_0)\big)^{-1}{\Cmat}_{X^{(k)}}\Big\}$ and $\Lambda_{i,\eps}^{(k)}\left({\Cmat}_{X^{(k)}};\tau_0\right)\equiv \Lambda_i^{(k)} \Big\{\big(\corrmatwe(\tau_0)\big)^{-1}{\Cmat}_{X^{(k)}}\Big\}$, $0\le i \le k-1$.
	Then, continuity of the eigenvalues of square real matrices \cite[Sec. 2.4.9, Thm. 2.4.9.2]{horn2012matrix}, combined with the boundedness of the elements of $\big(\corrmatwe(\tau_0)\big)^{-1}{\Cmat}_{X^{(k)}}$  and of $\big(\corrmatwn(\tau_0)\big)^{-1}{\Cmat}_{X^{(k)}}$, boundedness of their corresponding eigenvalues,  and the fact that convergence of $\big(\corrmatwn(\tau_0)\big)^{-1}{\Cmat}_{X^{(k)}}$ to $\big(\corrmatwe(\tau_0)\big)^{-1}{\Cmat}_{X^{(k)}}$ is uniform over  $\mathcal{C}_{X^{(k)}}$, see Eqn. \eqref{eqn:Convergence_of_CWminus1Cx}, imply that the ordered sets of eigenvalues of $\big(\corrmatwn(\tau_0)\big)^{-1}{\Cmat}_{X^{(k)}}$ converge to the ordered set of eigenvalues of $\big(\corrmatwe(\tau_0)\big)^{-1}{\Cmat}_{X^{(k)}}$ uniformly in $(\tau_0, \Cmatxk)\in \mathcal{C}_{X^{(k)}}$,\ifFullVersion\footnote{https://math.stackexchange.com/questions/110573/continuous-mapping-on-a-compact-metric-space-is-uniformly-continuous}
	\fi
	namely, $\forall \delta>0$, $\exists \tilde{n}_0(\delta) \in \mNp$ sufficiently large such that for all $n>\tilde{n}_0(\delta)$,   and for all $(\tau_0,\Cmatxk)\in \mathcal{C}_{X^{(k)}}$
	\begin{equation}
        \label{eqn:unif_converge_eigenvalues}
		\left| \Lambda_{i,n}^{(k)}\left({\Cmat}_{X^{(k)}};\tau_0\right)- \Lambda_{i,\eps}^{(k)}\left({\Cmat}_{X^{(k)}};\tau_0\right)\right|\leq \delta, \quad 0\leq i \leq k-1.
	\end{equation}
\end{sloppypar}		
			
			\label{pg:Taylor_Start}
	Lastly, consider the distance between the objective functions in \eqref{eqn:defOptProbs}: For any $(\tau_0, \Cmatxk)\in \mathcal{C}_{X^{(k)}}$, the distance between the objective functions can now be expressed as\footnote{Note that since $\corrmatwe(\tau_0)\big)^{-1}\succ 0$, symmetric, then $\corrmatwe(\tau_0)\big)^{-1}\cdot{\Cmat}_{X^{(k)}}+\iden{k}=\corrmatwe(\tau_0)\big)^{-\frac{1}{2}}\cdot\Big(\corrmatwe(\tau_0)\big)^{-\frac{1}{2}}{\Cmat}_{X^{(k)}}\cdot\corrmatwe(\tau_0)\big)^{-\frac{1}{2}}+\iden{k}\Big)\cdot\corrmatwe(\tau_0)\big)^{\frac{1}{2}}$ }:
	\begin{align*}
	    &\Big|\frac{1}{2k}\log\!\det \Big(\big(\corrmatwn(\tau_0)\big)^{-1}{\Cmat}_{X^{(k)}}+\iden{k}\Big)-\frac{1}{2k}\log\!\det \Big(\big(\corrmatwe(\tau_0)\big)^{-1}{\Cmat}_{X^{(k)}}+\iden{k}\Big)\Big| \notag\\
	    &\qquad =\frac{1}{2k}\bigg|\mathop{\sum}\limits_{i=0}^{k-1}\log \Big(1+\Lambda_{i,n}^{(k)}\big({\Cmat}_{X^{(k)}};\tau_0\big)\Big)-\mathop{\sum}\limits_{i=0}^{k-1}\log \Big(1+\Lambda_{i,\eps}^{(k)}\big({\Cmat}_{X^{(k)}};\tau_0\big)\Big)\bigg|.
	\end{align*}
	Using the first order Taylor expansion 
	\ifFullVersion
		\cite[Pg. 415-418]{spivak1994calculus}\footnote{ $\log_2 \left(1+x\right)=\log_2 \left(1+a\right) +\frac{\mathrm{d}}{\mathrm{d}x}\log_2(1+x)|_{x=a}\left(x-a\right)+ \frac{1}{2!}\frac{\mathrm{d^2}}{\mathrm{d}x^2}\log_2(1+x)|_{x=t}\left(x-a\right)^2=\log_2 \left(1+a\right) +\frac{1}{\ln{2}}\frac{1}{1+a}\left(x-a\right)-\frac{1}{2} \frac{1}{\ln{2}}\frac{1}{(1+t)^2}\left(x-a\right)^2$, where $t\in\big[\min\{a,x\}, \max\{a,x\}\big]$}
	\fi
	we can write 
	\begin{align*}
	    &\log \left(1+\Lambda_{i,n}^{(k)}\left({\Cmat}_{X^{(k)}};\tau_0\right)\right)=\log \left(1+\Lambda_{i,\eps}^{(k)}\left({\Cmat}_{X^{(k)}};\tau_0\right)\right)\notag \\ & \hspace{2cm}+\frac{1}{\ln{2}}\frac{1}{1+\Lambda_{i,\eps}^{(k)}\left({\Cmat}_{X^{(k)}};\tau_0\right)}\left(\Lambda_{i,n}^{(k)}\left({\Cmat}_{X^{(k)}};\tau_0\right)- \Lambda_{i,\eps}^{(k)}\left({\Cmat}_{X^{(k)}};\tau_0\right)\right)\notag \\ 
	    & \hspace{2cm} + \xi_i\left(\Lambda_{i,n}^{(k)}\left({\Cmat}_{X^{(k)}};\tau_0\right)\right),
	\end{align*}
	where $\xi_i\left(\Lambda_{i,n}^{(k)}\left({\Cmat}_{X^{(k)}};\tau_0\right)\right)$ is a reminder term.
	Thus, $\forall n>\tilde{n}_0(\delta)$  such that \eqref{eqn:unif_converge_eigenvalues} is satisfied, we obtain $\forall (\tau_0,\Cmatxk)\in \mathcal{C}_{X^{(k)}}$  that 
	\begin{eqnarray}
	    & &\hspace{-1cm}\frac{1}{2k}\left|\log\!\det \left(\left(\corrmatwn(\tau_0)\right)^{-1}{\Cmat}_{X^{(k)}}+\iden{k}\right)-\log\!\det \left(\left(\corrmatwe(\tau_0)\right)^{-1}{\Cmat}_{X^{(k)}}+\iden{k}\right)\right|\notag\\
		\ifFullVersion
		    &=& \frac{1}{2k}\Bigg|\mathop{\sum}\limits_{i=0}^{k-1}\Bigg(\log \left(1+\Lambda_{i,\eps}^{(k)}\left({\Cmat}_{X^{(k)}};\tau_0\right)\right)\notag \\  
		    & &  \hspace{2cm}+\frac{1}{\ln{2}}\frac{1}{1+\Lambda_{i,\eps}^{(k)}\left({\Cmat}_{X^{(k)}};\tau_0\right)}\left(\Lambda_{i,n}^{(k)}\left({\Cmat}_{X^{(k)}};\tau_0\right)- \Lambda_{i,\eps}^{(k)}\left({\Cmat}_{X^{(k)}};\tau_0\right)\right)\notag \\ 
		    & & \hspace{2cm} + \xi_i\Big(\Lambda_{i,n}^{(k)}\big({\Cmat}_{X^{(k)}};\tau_0\big)\Big)\Bigg)\notag\\
		    & & \hspace{2cm}-\mathop{\sum}\limits_{i=0}^{k-1}\log \left(1+\Lambda_{i,\eps}^{(k)}\left({\Cmat}_{X^{(k)}};\tau_0\right)\right)\Bigg| \notag \\
		\fi
		&=&\frac{1}{2k}\bigg|\mathop{\sum}\limits_{i=0}^{k-1}\bigg(\frac{1}{\ln{2}}\frac{1}{1+\Lambda_{i,\eps}^{(k)}\big({\Cmat}_{X^{(k)}};\tau_0\big)}\Big(\Lambda_{i,n}^{(k)}\big({\Cmat}_{X^{(k)}};\tau_0\big)- \Lambda_{i,\eps}^{(k)}\big({\Cmat}_{X^{(k)}};\tau_0\big)\Big)\notag \\ 
		& &\hspace{2cm} + \xi_i\Big(\Lambda_{i,n}^{(k)}\big({\Cmat}_{X^{(k)}};\tau_0\big)\Big)
		\bigg)\bigg| \notag \\
		& \leq & 2\frac{1}{2k} \mathop{\sum}\limits_{i=0}^{k-1}\left|\Lambda_{i,n}^{(k)}\left({\Cmat}_{X^{(k)}};\tau_0\right)- \Lambda_{i,\eps}^{(k)}\left({\Cmat}_{X^{(k)}};\tau_0\right)\right|
		+\frac{1}{2k} \mathop{\sum}\limits_{i=0}^{k-1} \left|\xi_i\left(\Lambda_{i,n}^{(k)}\left({\Cmat}_{X^{(k)}};\tau_0\right)\right)\right|\notag \\
		&\leq& \delta + \frac{1}{2}
		\mathop{\max}\limits_{0\leq i \leq k-1}\left|\xi_i\left(\Lambda_{i,n}^{(k)}\left({\Cmat}_{X^{(k)}};\tau_0\right)\right)\right|\notag\\
		& \stackrel{(a)}{\leq} & \delta + \frac{1}{2}\delta^2.\notag
	\end{eqnarray}
	Note that uniformity in $\mathcal{C}_{X^{(k)}}$ of the inequality in step (a)  follows from the uniform boundedness of $\left|\Lambda_{i,n}^{(k)}\left({\Cmat}_{X^{(k)}};\tau_0\right)- \Lambda_{i,\eps}^{(k)}\left({\Cmat}_{X^{(k)}};\tau_0\right)\right|$ over $\mathcal{C}_{X^{(k)}}$, as established in \eqref{eqn:unif_converge_eigenvalues}, combined with the following bound, derived using \cite[Pg. 418]{spivak1994calculus}:
	\begin{eqnarray}
	    & & \hspace{-1cm} \Big|\xi_i\left(\Lambda_{i,n}^{(k)}\left({\Cmat}_{X^{(k)}};\tau_0\right)\right)\Big| 	=\bigg|\frac{1}{2}\cdot\frac{1}{\ln{2}}\cdot\frac{1}{\big(1+\zeta\big)^2} \Big(\Lambda_{i,n}^{(k)}\big({\Cmat}_{X^{(k)}};\tau_0\big)- \Lambda_{i,\eps}^{(k)}\big({\Cmat}_{X^{(k)}};\tau_0\big)\Big)^2\bigg|, 	\notag\\ 
	    & & \;\min\big\{\Lambda_{i,\eps}^{(k)}\big({\Cmat}_{X^{(k)}};\tau_0\big),\Lambda_{i,n}^{(k)}\big({\Cmat}_{X^{(k)}};\tau_0\big)\big\}\le \zeta \le 		\max\big\{\Lambda_{i,\eps}^{(k)}\big({\Cmat}_{X^{(k)}};\tau_0\big),\Lambda_{i,n}^{(k)}\big({\Cmat}_{X^{(k)}};\tau_0\big)\big\} \notag\\
		& \stackrel{(a')}{\Rightarrow} & 		
		\Big|\xi_i\left(\Lambda_{i,n}^{(k)}\left({\Cmat}_{X^{(k)}};\tau_0\right)\right)\Big| \le
		\frac{1}{2}\cdot\frac{1}{\ln{2}}\Big(\Lambda_{i,n}^{(k)}\big({\Cmat}_{X^{(k)}};\tau_0\big)- \Lambda_{i,\eps}^{(k)}\big({\Cmat}_{X^{(k)}};\tau_0\big)\Big)^2\le\delta^2, 	\notag
	\end{eqnarray}
	uniformly over $\mathcal{C}_{X^{(k)}}$, where (a$'$) follows since $\zeta\ge 0$, by the non-negativity of the eigenvalues of $\big(\corrmatwe(\tau_0)\big)^{-1}\corrmatxe$.
\label{pg:Taylor_End}

	\begin{sloppypar}
		It follows that the distance between the $\log\!\det$ functions is uniformly upper bounded for all $(\tau_0,\Cmatxk) \in \mathcal{C}_{X^{(k)}}$, hence, convergence of $\frac{1}{2k}\log\!\det \Big(\big(\corrmatwn(\tau_0)\big)^{-1}{\Cmat}_{X^{(k)}}+\iden{k}\Big)$ to $\frac{1}{2k}\log\!\det \Big(\big(\corrmatwe(\tau_0)\big)^{-1}{\Cmat}_{X^{(k)}}+\iden{k}\Big)$ as $n\rightarrow\infty$ is uniform over the feasible set $\mathcal{C}_{X^{(k)}}$. We conclude  that for a sequence of  optimization problems \eqref{eqn:defOptSolCorr_Xeps},  the objective functions  converge uniformly to a limiting objective function \eqref{eqn:defOptSolCorr_Xn}. Thus, it follows from the proof of \cite[Thm. 2.1]{kanniappan1983uniform}\footnote{\label{footnote:conditions}For the application of \cite[Thm. 2.1]{kanniappan1983uniform}: $X$ in the theorem corresponds to the union of the interval $[0,\Tc)$ and the space of real symmetric, positive semidefinite matrices, subject to a  constraint on their trace: $\left\{{\Cmat}_{X^{(k)}}\in \mR^{k\times k}\bigg|\frac{1}{k}\sum_{i=0}^{k-1}\left({\Cmat}_{X^{(k)}}\right)_{ii}\leq P,\;  {\Cmat}_{X^{(k)}}=\left({\Cmat}_{X^{(k)}}\right)^T, {\Cmat}_{X^{(k)}}\succcurlyeq 0 \right\}$, which is a convex space. $Y$ in the theorem corresponds to the set of real numbers and the positive cone corresponds to the set of non-negative real numbers, thus, this cone is clearly normal \cite[Example 6.3.5]{DrabekMilota:MethodsNonlinearAnalysis}. }\textsuperscript{,}\footnote{\label{footnote:convex}Note that while  \cite[Thm. 2.1]{kanniappan1983uniform} is stated for convex objectives, convexity of the objective is not required for the convergence of the optimal objective values, only for the convergence of the optimal solutions. As we are not interested in the convergence of the optimal solutions (i.e., not interested in the convergence of $\big(\taunk\opt,\Cmatxnk\opt\big) $), then we can apply the steps in proof of   \cite[Thm. 2.1]{kanniappan1983uniform} to conclude that the optimal objective {\em values} converge also for non-convex objectives, without considering convergence of the solutions.} that the sequence of  optimal objective values of the sequence of problems \eqref{eqn:defOptSolCorr_Xeps} converges to the  optimal objective value of the limiting problem \eqref{eqn:defOptSolCorr_Xn}:
		\begin{equation*}
		     \mathop{\lim}\limits_{n\rightarrow \infty}\frac{1}{2k}\log\!\det \Big(\big(\corrmatwn(\taunk\opt)\big)^{-1}\corrmatxn\opt+\iden{k}\Big) = \frac{1}{2k}\log\!\det \Big(\big(\corrmatwe(\tauepsk\opt)\big)^{-1}\corrmatxe\opt+\iden{k}\Big).
		 \end{equation*}
	\end{sloppypar}
\end{IEEEproof}
	

\subsection{Equivalence Between and \texorpdfstring{$C_{\eps}$}{Ceps} and \texorpdfstring{$\liminfcn$}{liminfCn}
for the Setup of Theorem \ref{thm:AsycCap}}
Let $\cdf{\Xinkopt|\tauepsk\opt}$ denote a $k$-dimensional Gaussian \ac{cdf} with a correlation matrix denoted by $\corrmatxe\opt$,  such that $\big(\tauepsk\opt,\corrmatxe\opt\big)$ maximizes $\frac{1}{k}I(\Xink;\Yepsk|\tauepsk)$. Let $ \zkeps{k}{\eps}\Big( \cdf{\Xinkopt|\tauepsk\opt}|\tauepsk\opt\Big)$ denote the corresponding mutual information density rate. We now have the following Lemma:
\begin{lemma}     
			\label{lem:Pliminf}
			For the setup of Theorem \ref{thm:AsycCap} it holds that
			\begin{eqnarray}			
			\label{eqn:pliminfk_to_liminfn}
			C_{\eps}={\rm p-}\mathop{\lim\!\inf}\limits_{k \rightarrow \infty} \zkeps{k}{\eps}\Big( \cdf{\Xinkopt|\tauepsk\opt}|\tauepsk\opt\Big)&=&\liminfcn,
			\end{eqnarray}
            where $C_n$	is defined in Eqn. \eqref{eqn:DefCn}.

		\end{lemma}

\begin{IEEEproof}		
First, consider the upper bound on capacity: 
Recall that  by Lemma \ref{lem:tightness}, for every finite $k\in\mNp$, letting 
$\taunk\opt$ and $\tauepsk\opt$ denote the optimal sampling phases within the noise period, and letting 
$\Xnkopt$ and $\Xinkopt$ denote the corresponding Gaussian inputs with the optimal correlation matrices,  it holds that 
$\limn \frac{1}{k}I(\Xnkopt;\Ynk|\taunk\opt)=\frac{1}{k} I(\Xinkopt;\Yepsk|\tauepsk\opt)$. 
Then, we can write
\begin{eqnarray}
\Capacity_\myEps 
& \stackrel{(a)}{\le} & \liminfk \frac{1}{k}I\Big(\Xinkopt; \Yepsk|\tauepsk\opt\Big)\notag\\
& \stackrel{(b)}{=} &  \mathop{\lim\!\inf}\limits_{k\rightarrow \infty} \limn\frac{1}{k}I\left(\Xnkopt; \Ynk|\taunk\opt\right)\notag\\
& \stackrel{(c)}{=}  &  \liminfk \liminfn\frac{1}{k}I\left(\Xnkopt; \Ynk|\taunk\opt\right)\notag\\
& \stackrel{(d)}{\le} & \liminfk \liminfcn(\taunk\opt)\notag\\
& \stackrel{(e)}{\le} & \liminfk \liminfcn\notag\\
\label{appndxEqn:UpperBound}
& = & \liminfcn,
\end{eqnarray}
where 
(a) follows from   similar arguments as in the derivation of \eqref{eqn:cap_eps_uppr_bnd}:
\begin{eqnarray*}
    \Capacity_{\eps} & \le &
    \mathop{\liminfk}\mathop{\sup}\limits_{\left\{\substack{\big( \cdf{\bXk|\btauepsk}, \btauepsk\big):\; \frac{1}{l}\sum_{i=0}^{l-1}(\barx_u\left[i\right])^2\le P,\; u\in\mU,\\ \btauepsk\in[0,\Tc]} \right\}_{k\in\mNp}} \frac{1}{k}I\left(\bXk; \bYepsk|\btauepsk\right)\\
    & \leq & \mathop{\liminfk}\mathop{\sup}\limits_{\left\{\substack{\big( \cdf{\bXk|\btauepsk}, \btauepsk\big):\; \dsE\big\{\frac{1}{l}\sum_{i=0}^{l-1}(\bX\left[i\right])^2\big\}\le P,\\ \btauepsk\in[0,\Tc]}\right\}_{k\in\mNp}} \frac{1}{k}I\left(\bXk; \bYepsk|\btauepsk\right),
\end{eqnarray*}
where $\bYepsk$ is the channel output for input $\bXk$, and $\btauepsk$ is the sampling phase, and in the first inequality we maximize over all input distributions which facilitate selection of a codebook which satisfies the per-codeword power constraint. 
As the channel \eqref{eqn:AsnycModel1} is an additive Gaussian noise channel, then, subject to a trace constraint on the input correlation matrix, it follows that for every given $k\in\mNp$, the mutual information expression between the channel input and its output is maximized by Gaussian inputs \cite[Eqns. (4), (30)]{cover1989gaussian}, which satisfy the trace constraint, and we  use the maximizing sampling phase according to \eqref{eqn:defOptSolCorr_Xn}. 
Step (b) follows from  Lemma  \ref{lem:tightness}; and 
step (c) follows since the limit in $n$ exists and is finite \cite[Thm. 33.1.1]{Rankin07}. 
Step (d) follows as the channel \eqref{eqn:AsnycModel2} 
is a \ac{dt} \ac{acgn} channel, then by the discussion in \cite[Eqn. (14)-(16)]{shlezinger2016capacity} it is equivalent to a finite-memory stationary multivariate Gaussian channel. 
Hence, Step (d) follows directly from the converse in \cite[Eqn. (21)]{brandenburg1974capacity}:
For every finite $k\in\mNp$ and a given $\taunk\opt$, the achievable rate is not greater than the capacity:
$C_n\big(\taunk\opt\big) \ge \frac{1}{k}I\big(\Xnk; \Ynk|\taunk\big)$. 
Lastly,  step (e) follows as by \eqref{eqn:DefCn}, $C_n\ge C_n\big(\taunk\opt\big)$.

To show achievability of $\liminfcn$, we consider a transmission scheme which partitions the transmitted sequence into finite-length blocks, all of length $k\in\mNp$, and appends each $k$-block with a guard interval, sufficient to facilitate statistical independence between the noise process samples belonging to the different $k$-blocks as well as to facilitate synchronization of the transmission start times of all $k$-blocks to begin at the initial sampling phase $\tauepsk\opt$, which is selected to maximize the mutual information of the $k$-block subject to a trace constraint, as in \eqref{eqn:defOptSolCorr_Xn}. Note that such synchronization is permissible as the setup of Thm. \ref{thm:AsycCap} allows for transmission delay. We now detail the operations at the transmitter and at the receiver, recalling that $\Weps[i]$ has a {\em finite memory} $\Memd<\infty$. In the following description we use the tilde symbol to denote the channel input and output with the additional guard intervals, as well as the appropriate noise sequence. The channel inputs which carry information and the processed outputs used for decoding are denoted without the tilde symbol, and so does the corresponding noise sequence.

\subsubsection{Transmitter's Operations} 
\label{par:Tx-Scheme}
Let $k\in\mNp$ denote the blocklength, and set the duration of the guard interval between subsequent blocks to $\Memd\cdot \Tsamp(\eps)+\Dg$ time units (in \ac{ct}), $0\le\Dg\le\Tc$ will be explicitly defined later in the proof. Due to the finite memory and Gaussianity of the noise process it follows that noise sequences belonging to  different $k$-blocks are statistically independent. 
Moreover, the additional guard time of $\Dg$ beyond $\Memd$ allows the transmitter to synchronize the sampling phase of the $i$-th $k$-block, denoted $\tauepsk^{(i)}$, to the sampling phase which maximizes the $k$-block mutual information, i.e., $\tauepsk^{(i)}=\tauepsk\opt$. The codewords are generated according to a Gaussian distribution $\cdf{\Xinkopt|\tauepsk\opt}$ with a correlation matrix $\corrmatxe\opt$, s.t. $\big(\tauepsk\opt, \corrmatxe\opt \big)$ are selected according to \eqref{eqn:defOptSolCorr_Xn}. Hence, a codeword  consisting of $l\cdot k$ code symbols is transmitted over a time interval corresponding to $l\cdot (k+\Memd+\Dg/\Tsamp(\eps))$ channel symbols. Note that as the transmitter knows the correlation function of the noise, and, naturally, knows its own symbol interval $\Tsamp(\eps)$, it can deterministically compute the delay $\Dg$ needed to arrive again at $\tauepsk\opt$, which is given by  
\[
    \Dg\!=\!\left\{\begin{array}{cl}
         \tauepsk\opt\! -\!\Big(\big(\tauepsk\opt + (k+\Memd)\cdot\Tsamp(\eps)\big)\!\!\!\!\!\mod \Tc\Big) & ,\Big(\!\big(\tauepsk\opt + (k+\Memd)\!\cdot\!\Tsamp(\eps)\big)\!\!\!\!\!\mod\! \Tc\!\Big)\!\! < \! \tauepsk\opt\\
         \!\!\tauepsk\opt \!+ \!\Tc \!- \! \Big(\!\big(\tauepsk\opt \!+\! (k\!+\!\Memd)\!\cdot\!\Tsamp(\eps)\big)\!\!\!\!\!\mod \!\Tc\!\Big) & ,\Big(\!\big(\tauepsk\opt + (k+\Memd)\!\cdot\!\Tsamp(\eps)\big)\!\!\!\!\!\mod \!\Tc\!\Big)\!\! > \!\tauepsk\opt
    \end{array}\right. \!.
\]
A codebook $\mC\mB_k^{(i)}(\tauepsk\opt)$ of rate $R$ for the $i$-th $k$-block is generated by selecting $2^{kR}$ codewords   randomly and independently according to $\cdf{\Xinkopt|\tauepsk\opt}$. The codeword  for the $i$-th $k$-block, $1\le i \le l$, denoted $X_{(i-1)\cdot k}^{i\cdot k-1}$,  is 
independent of the codewords selected for the  other $k$-blocks.  

Note that with this construction, the codebook satisfies that for each message $u\in\mU$,
\begin{eqnarray}
    \frac{1}{l\cdot k}\sum_{i=0}^{l\cdot k -1}\big(x_u[i]\big)^2  & = & \frac{1}{l}\sum_{l'=0}^{l-1}\Big(\frac{1}{k}\sum_{k'=0}^{k-1} \big(x_u[l' \cdot k + k']\big)^2\Big)\notag\\
    & = & \frac{1}{k}\sum_{k'=0}^{k-1}\Big(\frac{1}{l}\sum_{l'=0}^{l-1} \big(x_u[l' \cdot k + k']\big)^2\Big)
    \stackrel{(\mbox{\scriptsize in prob.})}{\mathop{\longrightarrow}\limits_{l\rightarrow \infty}}  \frac{1}{k}\sum_{k'=0}^{k-1} \dsE\Big\{\big(X[k']\big)^2\Big\} \le P,\qquad
    \label{eqn:power_constr_check_proof_thm2}
\end{eqnarray}
where the limit as $l\rightarrow\infty$ follows from the weak law of large numbers, see e.g., \cite[Sec. 9.1]{cover2006elements}, and the inequality follows from trace constraint in the definition of $\mC_{X^{(k)}}$. It follows that for every $\delta>0$, we can select $l\in\mNp$ sufficiently large s.t. $\forall u \in \mU$, 
\[
    \Pr\Big(\frac{1}{l\cdot k}\sum_{i=0}^{l\cdot k -1}\big(X_u[i]\big)^2  > P\Big) \le \delta.
\]

A message $u$ of rate $R$ is transmitted via a codeword $X_{u}^{(l\cdot k)}\equiv \Big\{X_{u, (i-1)\cdot k}^{i\cdot k-1}\Big\}_{i=1}^l$, by splitting the information bit sequence of length $l\cdot k \cdot R$ into $l$ blocks, each contains $k\cdot R$ bits, where each block of $k \cdot R$ bits is mapped into a codeword of length $k$, e.g., the $i$-th block of the message $u$ is mapped into $X_{u,(i-1)\cdot k}^{i\cdot k-1}\in \mC\mB_k^{(i)}(\tauepsk\opt)$. Lastly, the transmitted $X_{u}^{(l\cdot k)}$ is transmitted as a sequence  $\tX^{(l\cdot (k+\Memd))}$ of $l\cdot(k+\Memd)$ samples, which is sent over $l\cdot(k+\Memd)\cdot \Tsamp(\eps) +l\cdot \Dg$ time units. The rate of this scheme is then $R\cdot\frac{k}{k+\Memd+\Dg/\Tsamp(\eps)}=R\cdot \big(1-\frac{\Memd+\Dg/\Tsamp(\eps)}{k+\Memd+\Dg/\Tsamp(\eps)}\big)$.

\subsubsection{Receiver's Operations}
\label{par:Rx-scheme}
At the beginning of reception, the receiver identifies the start time of the received sequence. From that point, as the symbol rate at the receiver is synchronized with $\Tsamp(\eps)$, the receiver can maintain $k$-block synchronization as applied at the transmitter: Let $\tY_{\eps}^{(l\cdot (k+\Memd))}=\tX^{(l\cdot (k+\Memd))}+W_{\eps}^{(l\cdot (k+\Memd))}$ denote the received samples, observed over $l\cdot(k+\Memd)\cdot \Tsamp(\eps) + l\cdot \Dg$ time units, obtained by receiving a block of $k+\Memd$ samples, and then waiting for $\Delta_g$ times units to process the next block of $k+\Memd$ samples. The receiver then keeps only the first $k$ samples of each block, discarding the last 
$\Memd$ samples. This processing results in a received sequence of $l$ $k$-blocks denoted
$\Big\{Y_{\eps, (i-1)\cdot k}^{i\cdot k-1}\Big\}_{i=1}^{l} \equiv Y_{\eps}^{(l\cdot k)}$, 
which are used by the decoder to receover the message $u\in\mU$.

In the following we denote the mutual information density rate for the $i$-th $k$-block, with initial sampling time $\tauepsk^{(i)}$ and input distribution $\cdf{ X_{(i-1)\cdot k}^{i\cdot k-1}|\tauepsk^{(i)}}$, with
\[
    \zkeps{k}{\eps}^{(i)}\big( \cdf{ X_{(i-1)\cdot k}^{i\cdot k-1}|\tauepsk^{(i)}}|\tauepsk^{(i)}\big) \triangleq 
    \frac{1}{k}\log\Bigg(\frac{\pdf{Y_{\eps, (i-1)\cdot k}^{i\cdot k-1} | X_{(i-1)\cdot k}^{i\cdot k-1},\tauepsk^{(i)}}\big( Y_{\eps, (i-1)\cdot k}^{i\cdot k-1} \big| X_{(i-1)\cdot k}^{i\cdot k-1}, \tauepsk^{(i)}\big) }{\pdf{Y_{\eps, (i-1)\cdot k}^{i\cdot k-1}|\tauepsk^{(i)}}\big( Y_{\eps, (i-1)\cdot k}^{i\cdot k-1}|\tauepsk^{(i)}\big) }\Bigg).
\]
Note that the guard interval also facilitates statistical independence between channel outputs at the decoder corresponding to different transmitted {\em messages}. 
The addition of the guard interval effectively decreases the information rate, however, as the blocklength $k$ increases, the impact of such fixed-length guard interval on the information rate becomes asymptotically negligible, and hence, it does not impact capacity. Recall that $k\in\mNp$  denotes the length of a block of symbols without guard interval.
%

Using the above scheme it is shown that when $\Big\{X_{(i-1)\cdot k,\; {\rm opt}}^{i\cdot k-1}\Big\}_{i=1}^l$  are Gaussian with the optimal covariance matrix, $\corrmatxe\opt$, designed for the optimal sampling phase $\tauepsk\opt$, and $X^{(l \cdot k)}_{\rm opt}$ is the corresponding input sequence, then 
\begin{equation}
\label{eqn:lower_bound_Pliming_Proof_Thm2}
    C_{\eps}\ge {\rm p-}\mathop{\lim\!\inf}\limits_{k \rightarrow \infty} \zkeps{k\cdot l}{\eps}\left( \cdf{X^{(l\cdot k)}_{\rm opt}|\tauepsk\opt}|\tauepsk\opt\right) \ge \liminfcn,
\end{equation}	
where the first inequality follows from  \cite[Thm. 3.6.1]{han2003information}, as 
by \eqref{eqn:power_constr_check_proof_thm2} the constructed codebook satisfies the per-codeword power constraint \eqref{eqn:AsyncConst1} with a probability which is arbitrarily close to $1$, and the limit-inferior in probability of any given input process which satisfies the per-codeword power constraint \eqref{eqn:AsyncConst1}, clearly does not exceed capacity. 
Hence, it remains to show the inequality on the \ac{rhs}. \label{pg:liminf_lower_bounds_capacity}

Letting $\tauepsk\opt$ denote the sampling phase at the start of the transmitted message, then as the guard interval facilitates statistical independence between the $k$-blocks we obtain that 
\begin{eqnarray}
	& & \hspace{-1cm}\zkeps{l\cdot k}{\eps}\big( \cdf{X^{(l\cdot k)}|\tauepsk\opt}|\tauepsk\opt\big)\notag\\
	& \triangleq & \frac{1}{l\cdot k}\log \Bigg( \frac{\pdf{Y_{\eps}^{(l \cdot k )} | X^{(l\cdot k)},\tauepsk}\big( Y_{\eps}^{(l\cdot k )} \big| X^{(l\cdot k )}, \tauepsk\opt\big) }{\pdf{Y_{\eps}^{(l\cdot k )}|\tauepsk}\big(Y_{\eps}^{(l\cdot k )}|\tauepsk\opt\big) }\Bigg)\notag\\
	& = & \frac{1}{l\cdot k}\log \Bigg(\frac{\pdf{\big\{Y_{\eps, (i-1)\cdot k}^{i\cdot k-1}\big\}_{i=1}^{l} \big| \big\{X_{(i-1)\cdot k}^{i\cdot k-1}\big\}_{i=1}^{l},\tauepsk}\Big( \big\{Y_{\eps, (i-1)\cdot k}^{i\cdot k-1}\big\}_{i=1}^{l} \Big| \big\{X_{(i-1)\cdot k}^{i\cdot k-1}\big\}_{i=1}^{l}, \tauepsk\opt\Big) }{\pdf{\big\{Y_{\eps, (i-1)\cdot k}^{i\cdot k-1}\big\}_{i=1}^{l}\big|\tauepsk}\big( \big\{Y_{\eps, (i-1)\cdot k}^{i\cdot k-1}\big\}_{i=1}^{l}\big|\tauepsk\opt\big) }\Bigg)\notag\\
	& = & \frac{1}{l\cdot k}\log\Bigg( \frac{\pdf{\big\{W_{\eps, (i-1)\cdot k}^{i\cdot k-1}\big\}_{i=1}^{l} \big| \tauepsk}\Big( \big\{Y_{\eps, (i-1)\cdot k}^{i\cdot k-1} - X_{ (i-1)\cdot k}^{i\cdot k-1}\big\}_{i=1}^{l} \Big|  \tauepsk\opt\Big) }{\pdf{\big\{Y_{\eps, (i-1)\cdot k}^{i\cdot k-1}\big\}_{i=1}^{l}\big|\tauepsk}\big( \big\{Y_{\eps, (i-1)\cdot k}^{i\cdot k-1}\big\}_{i=1}^{l}\big|\tauepsk\opt\big) }\Bigg)\notag\\
	& \stackrel{(a)}{=} & \frac{1}{l\cdot k}\log \Bigg(\prod_{i=1}^l\frac{\pdf{W_{\eps, (i-1)\cdot k}^{i\cdot k-1} | \tauepsk^{(i)}}\big( Y_{\eps, (i-1)\cdot k}^{i\cdot k-1} - X_{ (i-1)\cdot k}^{i\cdot k-1}\big| \tauepsk^{(i)}\big) }{\pdf{Y_{\eps, (i-1)\cdot k}^{i\cdot k-1}|\tauepsk^{(i)}}\big( Y_{\eps, (i-1)\cdot k}^{i\cdot k-1}|\tauepsk^{(i)}\big) }\Bigg)\notag\\
	& \stackrel{(b)}{=}  & \frac{1}{l\cdot k}\log \Bigg(\prod_{i=1}^l\frac{\pdf{Y_{\eps, (i-1)\cdot k}^{i\cdot k-1} | X_{(i-1)\cdot k}^{i\cdot k-1},\tauepsk}\big( Y_{\eps, (i-1)\cdot k}^{i\cdot k-1} \big| X_{(i-1)\cdot k}^{i\cdot k-1}, \tauepsk\opt\big) }{\pdf{Y_{\eps, (i-1)\cdot k}^{i\cdot k-1}|\tauepsk}\big( Y_{\eps, (i-1)\cdot k}^{i\cdot k-1}|\tauepsk\opt\big) }\Bigg)\notag\\
	& = & \frac{1}{l}\sum_{i=1}^l  \frac{1}{k}\log\Bigg(\frac{\pdf{Y_{\eps, (i-1)\cdot k}^{i\cdot k-1} | X_{(i-1)\cdot k}^{i\cdot k-1},\tauepsk}\big( Y_{\eps, (i-1)\cdot k}^{i\cdot k-1} \big| X_{(i-1)\cdot k}^{i\cdot k-1}, \tauepsk\opt\big) }{\pdf{Y_{\eps, (i-1)\cdot k}^{i\cdot k-1}|\tauepsk}\big( Y_{\eps, (i-1)\cdot k}^{i\cdot k-1}|\tauepsk\opt\big) }\Bigg)\notag\\
	\label{eqn:Zeps_as_a_sum}
	& = & \frac{1}{l}\sum_{i=1}^l \zkeps{k}{\eps}^{(i)}\big( \cdf{ X_{(i-1)\cdot k}^{i\cdot k-1}|\tauepsk\opt}|\tauepsk\opt\big),
\end{eqnarray}
	where in (a), $\tauepsk^{(i)}$ denotes the  sampling phase of the $i$-th $k$-block which is generated by the transmission scheme, and the equality follows  since both the input $k$-blocks 
	$\big\{X_{(i-1)\cdot k}^{i\cdot k-1}\big\}_{i=1}^{l}$ and the noise $k$-blocks 
	$\big\{W_{\eps,(i-1)\cdot k}^{i\cdot k-1}\big\}_{i=1}^{l}$
	are mutually independent (over $1\le i \le l$): For the noise, $\big\{W_{\eps,(i-1)\cdot k}^{i\cdot k-1}\big\}_{i=1}^{l}$, independence among the $k$-blocks follows as the $k$-blocks are separated more than $\Memd$ samples apart, while the noise memory is $\Memd$; for the channel output, $\big\{Y_{\eps,(i-1)\cdot k}^{i\cdot k-1}\big\}_{i=1}^{l}$, independence follows as both the noise $k$-blocks are independent (as explained above) and the input $k$-blocks are independent by the assumption of uniformity and independence of the messages, as well as the codebook generation process;
	step (b) follows since at each block, sampling phase synchronization is applied, which results in  $\tauepsk^{(i)}=\tauepsk\opt$ for all $k$-blocks $1\le i \le l$.
	
\begin{sloppypar}

Note that as the sampling phase (within a period of the noise correlation function), $\tauepsk\opt$, is identical for all $k$-blocks, it follows that the noise process has the {\em same correlation matrix} for every $k$-block. Therefore, the \ac{cdf} $\cdf{ X_{(i-1)\cdot k, {\rm opt}}^{i\cdot k-1}|\tauepsk\opt}$ which maximizes the mutual information for the $i$-th $k$-block, $\frac{1}{k}I( X_{(i-1)\cdot k}^{i\cdot k-1};Y_{\eps,(i-1)\cdot k}^{i\cdot k-1}|\tauepsk\opt)$, is identical for all the $k$-blocks 
\label{pg:Max_Dist_is_Identical_for_All}
(i.e., for all of the $l$ blocks, each of length $k$): $\cdf{X_{(i-1)\cdot k,{\rm opt}}^{i\cdot k-1}|\tauepsk\opt}= \cdf{X^{( k)}_{\rm opt}|\tauepsk\opt}$, $i= 1, 2, ..., l$. 
Letting $\cdf{X_{\rm opt}^{(l\cdot k)}|\tauepsk\opt}=\prod_{i=1}^l \cdf{X_{(i-1)\cdot k, {\rm opt}}^{i\cdot k -1}|\tauepsk\opt}$, we conclude that the \acp{rv} $\Big\{\zkeps{k}{\eps}^{(i)}\big( \cdf{ X_{(i-1)\cdot k, {\rm opt}}^{i\cdot k-1}|\tauepsk\opt}|\tauepsk\opt\big)\Big\}_{i=1}^l$ all have identical Gaussian distributions. Thus,
\begin{subequations}
	\begin{eqnarray*}
	\dsE\Big\{\zkeps{l\cdot k}{\eps}\big( \cdf{X_{\rm opt}^{(l\cdot k)}|\tauepsk\opt}|\tauepsk\opt\big) \Big\}
	& = &  \dsE\bigg\{ \frac{1}{l}\sum_{i=1}^l \zkeps{k}{\eps}^{(i)}\big( \cdf{ X_{(i-1)\cdot k, {\rm opt}}^{i\cdot k-1}|\tauepsk\opt}|\tauepsk\opt\big)\bigg\}\notag\\
	& = &  \frac{1}{l}\sum_{i=1}^l \dsE\bigg\{\zkeps{k}{\eps}^{(i)}\big( \cdf{ X_{(i-1)\cdot k, {\rm opt}}^{i\cdot k-1}|\tauepsk\opt}|\tauepsk\opt\big)\bigg\}\notag\\
    & \stackrel{(a)}{=} & \frac{1}{k}I\big(\Xinkopt; \Yepsk|\tauepsk\opt\big)\\
	\mbox{var}\Big(\zkeps{l\cdot k}{\eps}\big( \cdf{X_{\rm opt}^{(l\cdot k)}|\tauepsk\opt}|\tauepsk\opt\big) \Big)& \stackrel{(b)}{=} &
	\frac{1}{l^2}\sum_{i=1}^l
	\mbox{var}\Big(\zkeps{k}{\eps}^{(i)}\big( \cdf{ X_{(i-1)\cdot k, {\rm opt}}^{i\cdot k-1}|\tauepsk\opt}|\tauepsk\opt\big) \Big)\notag\\
	& \stackrel{(c)}{\le} & \frac{3}{l\cdot k},
	\end{eqnarray*}
\end{subequations}
where steps (a) and (c) follow from the derivation in Appendix \ref{Appndx:Capacity_General}, after Eqn. \eqref{eqn:Expression_tvarnk}:
$\E \left\{\zkeps{k}{\eps} \left(\cdf{X^{(k)}_{{\rm opt}}|\tauepsk}|\tauepsk \right)  \right\}  =  \frac{1}{k}I\big(\Xepsk_{\opt}; \Yepsk|\tauepsk\big)$ and $\mbox{var}\Big(\zkeps{k}{\eps}\big( \cdf{ X^{(k)}_{{\rm opt}}|\tauepsk}|\tauepsk\big) \Big)  \le  \frac{3}{k}$;  step (b) follows since the mutual information density rates of the different $k$-blocks are mutually independent.
    
Since the variance of $\zkeps{l\cdot k}{\eps}\big( \cdf{X_{\rm opt}^{(l\cdot k)}|\tauepsk\opt}|\tauepsk\opt\big)$ decreases as $l\cdot k$ increases, then, by Chebyshev's inequality \cite[Eqn. (5-88)]{Papoulis2002}, we obtain
$\Pr\Big(\Big|\zkeps{l\cdot k}{\eps}\big( \cdf{X_{\rm opt}^{(l\cdot k)}|\tauepsk\opt}|\tauepsk\opt\big)-\frac{1}{k}I\big(\Xinkopt; \Yepsk|\tauepsk\opt\big)\Big|>\frac{1}{(l\cdot k)^{1/3}} \Big) <\frac{3}{(l\cdot k)^{1/3}}$, and we conclude that $\forall k\in\mNp$ and $\forall\delta > 0$, $\exists l_0(k,\delta)\in\mNp$ s.t. $\forall l>l_0(k,\delta)$, it follows that 
\begin{equation}
\label{eqn:Bounding_PrZepsk_LemmaB}
    \Pr\Big(\zkeps{l\cdot k}{\eps}\big( \cdf{X_{\rm opt}^{(l\cdot k)}|\tauepsk\opt}|\tauepsk\opt\big) <\frac{1}{k}I\big(\Xinkopt; \Yepsk|\tauepsk\opt\big)-\delta\Big) <3\delta.
\end{equation}

\label{pg:modify1_due_to_power}
To show achievability of $\liminfcn$ we will 
show the \ac{rhs} of \eqref{eqn:lower_bound_Pliming_Proof_Thm2}.
First, recall that 
\[
    {\rm p-}\mathop{\lim\!\inf}\limits_{k \rightarrow \infty} \zkeps{k}{\eps}\big( \cdf{\Xink|\tauepsk}|\tauepsk\big)  = \sup\left\{\alpha \in \mR\; \big| \mathop{\lim}\limits_{k \rightarrow \infty}\Pr \big(\zkeps{k}{\eps}\left( \cdf{\Xink|\tauepsk}|\tauepsk\right) < \alpha \big) = 0   \right\}, 
\]
hence,  $\liminfcn$ is achievable if, 
for a given finite and bounded constant $\xi$ (that will be defined later),
$\forall \delta>0$, $\Pr \Big(\zkeps{l\cdot k}{\eps}\big( \cdf{X^{(l\cdot k)}|\tauepsk}|\tauepsk\big) < \liminfcn-\delta\cdot(5+2\cdot\xi) \Big)$ can be made arbitrarily small by properly selecting  $k\in\mNp$, $\tauepsk$ and   $\cdf{\Xink|\tauepsk}$, and taking
$l$ sufficiently large.
Next, define
\begin{eqnarray}
    \gamma(k) &\triangleq & \log\Big(P\cdot k + \mathop{\max}\limits_{0\le t \le \Tc}\big\{\Cwc(t, 0) \big\}\Big)\notag\\
    \label{eqn:gammadef}
    & & \qquad +\log(e)\cdot \frac{1}{\mathop{\min}\limits_{0 \leq t \leq \Tc}\bigg\{\Cwc(t,0)-2\Memd\cdot\mathop{\max}\limits_{|\lambda|>\frac{\Tc}{p+1}}\big\{ |\Cwc(t, \lambda)|\big\}\bigg\}}, \phantom{xxxxx}
\end{eqnarray}
and pick $k\in \mNp$ s.t. $\frac{ \Memd\cdot \gamma(k)}{k+\Memd} < \delta$. 
%
In addition, $k$ is selected sufficiently large such that the values of the sampled variance $\disccorr^{\{\tau_0\}}[i,\Delta]$ over a $k$-block starting at sampling phase $\tau_0$ satisfy
\begin{equation}
    \label{eqn:UD_conv}
    \Big|\frac{1}{k}\sum_{i=0}^{k-1} \disccorr^{\{\tau_0\}}[i,0]-\frac{1}{\Tc}\int_{t=0}^{\Tc}\Cwc(t+\tau_0,0){\rm d}t\Big|<\frac{\delta}{2}, \qquad \forall \tau_0 \in[0,\Tc).
\end{equation}
Such a selection is possible since the sampling interval $\Tsamp(\eps)$ is incommensurate with the noise period $\Tc$. Then, for a sufficiently large $k$, the sampling points will be nearly uniformly distributed over $[0,\Tc)$, see also 
discussion after Eqn. \eqref{eqn:upper_bound_capacity}. 
{
By definition of a uniformly distributed modulo $1$ sequence\footnote{\color{black} Note that switching the modulo from $1$ to modulo $\Tc$ amounts to scaling of the time axis, which can be incorporated into the definition in a straightforward manner.}, \cite[Def. 1.1]{KuipersNiederreiter:UniformDistributionBook}, it follows that the empirical distribution of the sampling instances approaches a uniform distribution on $[0,\Tc]$. Then, by \cite[Thm. 1.1]{KuipersNiederreiter:UniformDistributionBook}, as the correlation function (at any given lag, and hence also at lag $\lambda=0$) is a continuous mapping of the time, then \eqref{eqn:UD_conv} follows.
}

When $k$ is fixed, we pick $n\in\mNp$ s.t. $\liminf\limits_{n_0\rightarrow\infty}C_{n_0}  <  C_n + \delta$, and also 
\begin{equation}
\label{eqn:mutual_information_bounded_distance}
    \Big|\frac{1}{k}I(\Xinkopt;\Yepsk|\tauepsk\opt) - \frac{1}{k}I(\Xnkopt; \Ynk |\taunk\opt)\Big| < \delta, 
\end{equation}
where we recall that $\cdf{\Xinkopt|\tauepsk\opt}$ and $\cdf{\Xnkopt|\taunk\opt}$ are Gaussian \acp{cdf}, $\big(\taunk\opt, \Xnkopt\big)$ maximize $\frac{1}{k}I(\Xnk; \Ynk |\taunk)$, and $\big(\tauepsk\opt, \Xinkopt\big)$ maximize $\frac{1}{k}I(\Xink; \Yepsk |\tauepsk)$. Such $n\in\mNp$ exists by the convergence in  Lemma \ref{lem:tightness}.	
Lastly, the selected $n$ is increase to guarantee that 
\[
    \Big|\frac{1}{k}\sum_{i=0}^{k-1} \disccorr^{\{\tau_0\}}[i,0]-\frac{1}{k}\sum_{i=0}^{k-1} \disccorrn^{\{\tau_0\}}[i,0]\Big|<\frac{\delta}{2}, \qquad \forall \tau_0 \in[0,\Tc).
\]
which implies that 
\begin{equation}
    \label{eqn:n-selec-cond-3}
    \Big|\frac{1}{k}\sum_{i=0}^{k-1} \disccorrn^{\{\tau_0\}}[i,0]-\frac{1}{\Tc}\int_{t=0}^{\Tc}\Cwc(t+\tau_0,0){\rm d}t\Big|<\delta, \qquad \forall \tau_0 \in[0,\Tc).
\end{equation}

After picking $n$, we pick $l\in\mNp$ such that 
\[
    C_n \le \frac{1}{l\cdot (k+\Memd)}I\big(\tX_n^{l\cdot (k+\Memd)}; \tY_n^{l\cdot (k+\Memd)}|\tau_{n,l}\big)  +\delta,\qquad \tau_{n,l}\in [0,\Tc],
\]
where $\tX_n[i]$ is the capacity-achieving input process for the channel \eqref{eqn:AsnycModel2}, see  \cite[Thm. 1]{shlezinger2016capacity}, and $\tY_n[i]$ is the corresponding output process:  $\tY_n[i]=  \tX_n[i]+ W_n[i]$. Such $\tau_{n,l}$ exists since we can set $\tau_{n,l} = \mathop{\argmax}\limits_{\tau_0\in[0,\Tc]}C_n(\tau_0)\triangleq\tau_n\opt$. Then, for  $\tau_n\opt$, $k$ and $n$, as $\tX_n[i]$ is the capacity-achieving input process, there is a codeword length beyond which the mutual information between the channel input and output is less than $\delta$ apart from capacity.
Next, define \acp{rv} the $\Xvec_1$, $\Xvec_2$, $\Yvec_1$, and $\Yvec_2$ as follows:
\begin{subequations}
\begin{eqnarray}
    \Xvec_1 & = & \big\{ \tX_{n,i\cdot\Memd + (i-1)\cdot k }^{i\cdot(k + \Memd)-1} \big\}_{i=1}^l \qquad \Xvec_2 = \big\{ \tX_{n, (i-1)\cdot (k+\Memd) }^{(i-1)\cdot(k + \Memd)+\Memd-1} \big\}_{i=1}^l\\
    \Yvec_1 & = & \big\{ \tY_{n,i\cdot\Memd + (i-1)\cdot k }^{i\cdot(k + \Memd)-1} \big\}_{i=1}^l \qquad \Yvec_2 = \big\{ \tY_{n, (i-1)\cdot (k+\Memd) }^{(i-1)\cdot(k + \Memd)+\Memd-1} \big\}_{i=1}^l.
\end{eqnarray}
\end{subequations}
We note that $(\Xvec_2^T,\Xvec_1^T)^T$ is a {\em permutation} of the vector $\tX_n^{l\cdot (k+\Memd)}$. Let us denote this permutation with the matrix $\Pmat$, i.e., $(\Xvec_2^T,\Xvec_1^T)^T = \Pmat\cdot \tX_n^{l\cdot (k+\Memd)}$. Similarly, we write $(\Wvec_2^T,\Wvec_1^T)^T=(\Yvec_2^T,\Yvec_1^T)^T-(\Xvec_2^T,\Xvec_1^T)^T = \Pmat\cdot W_n^{l\cdot (k+\Memd)}$. Note that $\cov\big((\Wvec_2^T, \Wvec_1^T)^T\big|\tau_{n,l}\big) = \dsE\big\{\Pmat\cdot W_n^{l\cdot (k+\Memd)} \cdot \big(W_n^{l\cdot (k+\Memd)}\big)^T \cdot \Pmat^T\big|\tau_{n,l}\big\}= \Pmat\cdot \Cmat_{W_n^{l\cdot (k+\Memd)}}(\tau_{n,l}) \cdot \Pmat^T$. With these definitions, applying the chain rules for differential entropy and for mutual information,  we write 
\begin{eqnarray*}
    & & \hspace{-1.5cm}I\big(\tX_n^{l\cdot (k+\Memd)}; \tY_n^{l\cdot (k+\Memd)}|\tau_{n,l}\big) \\
    & \equiv & I(\Xvec_1, \Xvec_2;\Yvec_1,\Yvec_2|\tau_{n,l})\\
     & = & I(\Xvec_1;\Yvec_1,\Yvec_2|\tau_{n,l}) + I(\Xvec_2;\Yvec_1,\Yvec_2|\Xvec_1, \tau_{n,l})\\
    & = & I(\Xvec_1;\Yvec_1|\tau_{n,l}) + I(\Xvec_1;\Yvec_2|\Yvec_1, \tau_{n,l}) + I(\Xvec_2;\Yvec_1,\Yvec_2|\Xvec_1, \tau_{n,l})\\
    & = &  I(\Xvec_1;\Yvec_1|\tau_{n,l}) + h(\Yvec_2|\Yvec_1, \tau_{n,l}) - h(\Yvec_2|\Xvec_1, \Yvec_1, \tau_{n,l}) + h(\Yvec_1,\Yvec_2|\Xvec_1, \tau_{n,l})\\
    &   &  \qquad \qquad - h(\Yvec_1,\Yvec_2|\Xvec_1, \Xvec_2, \tau_{n,l})\\
   & = &  I(\Xvec_1;\Yvec_1|\tau_{n,l}) + h(\Yvec_2|\Yvec_1, \tau_{n,l}) - h(\Yvec_2|\Xvec_1, \Yvec_1, \tau_{n,l}) + h(\Yvec_1|\Xvec_1, \tau_{n,l})\\
   &   & \qquad \qquad + h(\Yvec_2|\Xvec_1, \Yvec_1, \tau_{n,l}) - h(\Yvec_1,\Yvec_2|\Xvec_1, \Xvec_2, \tau_{n,l})\\
   & = &  I(\Xvec_1;\Yvec_1|\tau_{n,l}) + h(\Yvec_2|\Yvec_1, \tau_{n,l})  + h(\Yvec_1|\Xvec_1, \tau_{n,l})  - h(\Yvec_1,\Yvec_2|\Xvec_1, \Xvec_2, \tau_{n,l})\\
   & \stackrel{(a)}{=} &  I(\Xvec_1;\Yvec_1|\tau_{n,l}) + h(\Yvec_2|\Yvec_1, \tau_{n,l})  + h(\Wvec_1|\Xvec_1, \tau_{n,l})  - h(\Wvec_1,\Wvec_2|\Xvec_1, \Xvec_2, \tau_{n,l})\\
   & \stackrel{(b)}{=} &  I(\Xvec_1;\Yvec_1|\tau_{n,l}) + h(\Yvec_2|\Yvec_1, \tau_{n,l})  + h(\Wvec_1| \tau_{n,l})  - h(\Wvec_1,\Wvec_2| \tau_{n,l})\\
   & = &  I(\Xvec_1;\Yvec_1|\tau_{n,l}) + h(\Yvec_2|\Yvec_1, \tau_{n,l})  - h(\Wvec_2|\Wvec_1, \tau_{n,l}),
\end{eqnarray*}
where (a) follows as $\Yvec_1=\Xvec_1+\Wvec_1$ and $\Yvec_2=\Xvec_2+\Wvec_2$; and (b) follows as $(\Wvec_1, \Wvec_2) \independent (\Xvec_1, \Xvec_2)$. 	Next, denoting the maximal diagonal element of the matrix $\Amat$ with $\maxDiag\big\{\Amat\big\}$, we can write
\label{pg:Bounding the difference of entropies}
\begin{eqnarray*}
   &  & \hspace{-1cm} h(\Yvec_2|\Yvec_1, \tau_{n,l})  - h(\Wvec_2|\Wvec_1, \tau_{n,l}) \\
   & \stackrel{(a)}{\le} & 	    h(\Yvec_2| \tau_{n,l})  - h(\Wvec_2|\Wvec_1, \tau_{n,l})\\
   & \stackrel{(b)}{=} & \logdet\big((2\pi e)\cdot\cov(\Yvec_2|\tau_{n,l})\big) \\
   &  & \qquad - \logdet\Big((2\pi e)\cdot\big(\cov(\Wvec_2|\tau_{n,l}) -\cov(\Wvec_2,\Wvec_1|\tau_{n,l})\big(\cov(\Wvec_1|\tau_{n,l})\big)^{-1}\cov(\Wvec_1,\Wvec_2|\tau_{n,l})\big) \Big)\\
   & = &  \logdet\big( \cov(\Yvec_2|\tau_{n,l})\big)  - \logdet\Big(\cov(\Wvec_2|\tau_{n,l})\\
   &  & \qquad \qquad \qquad \qquad 
   -\cov(\Wvec_2,\Wvec_1|\tau_{n,l})\big(\cov(\Wvec_1|\tau_{n,l})\big)^{-1}\cov(\Wvec_1,\Wvec_2|\tau_{n,l})\Big)\\
   & \stackrel{(c)}{\le} & (l\cdot \Memd)\cdot\log\big(P\cdot k + \mathop{\max}\limits_{0\le t \le \Tc}\big\{\Cwc(t, 0) \big\}\big)\\
   &  & \qquad   - \logdet\Big(\cov(\Wvec_2|\tau_{n,l})-\cov(\Wvec_2,\Wvec_1|\tau_{n,l})\big(\cov(\Wvec_1|\tau_{n,l})\big)^{-1}\cov(\Wvec_1,\Wvec_2|\tau_{n,l}) \Big)\\
   & \stackrel{(d)}{\le} &  (l\cdot \Memd)\cdot\log\big(P\cdot k + \mathop{\max}\limits_{0\le t \le \Tc}\big\{\Cwc(t, 0) \big\}\big)\\
   &  & \qquad   +\log(e)\cdot\bigg( \Tr\Big\{\Big(\cov(\Wvec_2|\tau_{n,l})-\cov(\Wvec_2,\Wvec_1|\tau_{n,l})\big(\cov(\Wvec_1|\tau_{n,l})\big)^{-1}\cov(\Wvec_1,\Wvec_2|\tau_{n,l})\Big)^{-1} \Big\}\\
   & & 
   \qquad \qquad \qquad -l\cdot \Memd \bigg)\\
   & \le &  (l\cdot \Memd)\cdot\log\big(P \cdot k + \mathop{\max}\limits_{0\le t \le \Tc}\big\{\Cwc(t, 0) \big\}\big)\\
   &  & \;   + (l\cdot \Memd)\cdot\log(e)\cdot\maxDiag\Big\{\Big(\cov(\Wvec_2|\tau_{n,l})\\
   &  & \qquad\qquad\qquad\qquad\qquad\qquad -\cov(\Wvec_2,\Wvec_1|\tau_{n,l})\big(\cov(\Wvec_1|\tau_{n,l})\big)^{-1}\cov(\Wvec_1,\Wvec_2|\tau_{n,l})\Big)^{-1} \Big\}\\
   & \stackrel{(e)}{\le} &  (l\cdot \Memd)\cdot\log\big(P \cdot k + \mathop{\max}\limits_{0\le t \le \Tc}\big\{\Cwc(t, 0) \big\}\big)\\
   &  & \qquad   + \log(e)\cdot(l\cdot \Memd)\cdot\maxDiag\Big\{\Big(\cov\big((\Wvec_2^T,\Wvec_1^T)^T|\tau_{n,l}\big)\Big)^{-1} \Big\}\\
   & \stackrel{(f)}{\le} &  (l\cdot \Memd)\cdot\log\big(P  \cdot k + \mathop{\max}\limits_{0\le t \le \Tc}\big\{\Cwc(t, 0) \big\}\big)   + (l\cdot \Memd)\cdot\log(e)\cdot\norm{\big(\Pmat\cdot \Cmat_{W_n^{l\cdot (k+\Memd)}}(\tau_{n,l}) \cdot \Pmat^T\big)^{-1}}_1\\
   & \stackrel{(g)}{\le} &  (l\cdot \Memd)\cdot\log\big(P \cdot k + \mathop{\max}\limits_{0\le t \le \Tc}\big\{\Cwc(t, 0) \big\}\big)   + (l\cdot \Memd)\cdot\log(e)\cdot\norm{\big( \Cmat_{W_n^{l\cdot (k+\Memd)}}(\tau_{n,l}) \big)^{-1}}_1\\
   & \stackrel{(h)}{\le} & (l\cdot \Memd) \cdot \log\big(P \cdot k + \mathop{\max}\limits_{0\le t \le \Tc}\big\{\Cwc(t, 0) \big\}\big)\\
   &  & \qquad + (l\cdot \Memd) \cdot \log(e)\cdot\frac{1}{\mathop{\min}\limits_{0 \leq t \leq \Tc}\bigg\{\Cwc(t,0)- 2\Memd\cdot\mathop{\max}\limits_{|\lambda|>\frac{\Tc}{p+1}}\Big\{|\Cwc(t, \lambda)|\Big\}\bigg\}},
\end{eqnarray*} 
where (a) follows from \cite[Corollary on Pg. 253]{cover2006elements}; (b) follows since $\Yvec_2|\tau_{n,l}$ and $(\Wvec_2^T,\Wvec_1^T)^T|\tau_{n,l}$ are Gaussian vectors, and since the conditional covariance for jointly Gaussian \acp{rv} is independent of the conditioning value, see, e.g., \cite[Ch. 21.6]{Fristedt:ModernApproachProbabilityTheoryBook}:
\begin{eqnarray*}
    h(\Wvec_2|\Wvec_1,\tau_{n,l}) & = & \int_{\wvec_1\in \mR^{l\cdot k}} f_{\Wvec_1|\tau_{n,l}}(\wvec_1|\tau_{n,l})  h(\Wvec_2|\Wvec_1=\wvec_1,\tau_{n,l}) \mbox{d}\wvec_1\\
     & \stackrel{(a')}{=} &  \int_{\wvec_1\in \mR^{l\cdot k}} f_{\Wvec_1|\tau_{n,l}}(\wvec_1|\tau_{n,l}) \cdot \logdet\Big((2\pi e)\cdot\big(\cov(\Wvec_2|\tau_{n,l})\\  &  & \qquad\qquad\qquad -\cov(\Wvec_2,\Wvec_1|\tau_{n,l})\big(\cov(\Wvec_1|\tau_{n,l})\big)^{-1}\cov(\Wvec_1,\Wvec_2|\tau_{n,l})\big)\Big)\mbox{d}\wvec_1\\
     & = &   \logdet\Big((2\pi e)\cdot\big(\cov(\Wvec_2|\tau_{n,l})\\ 
     &  & \qquad\qquad\qquad -\cov(\Wvec_2,\Wvec_1|\tau_{n,l})\big(\cov(\Wvec_1|\tau_{n,l})\big)^{-1}\cov(\Wvec_1,\Wvec_2|\tau_{n,l})\big)\Big),
\end{eqnarray*}
where the expression for the conditional correlation matrix is step (a$'$) is given in \cite[Ch. 21.6]{Fristedt:ModernApproachProbabilityTheoryBook}.
Step (c) follows from Hadamard's inequality \cite[Eqn. (8.64)]{cover2006elements}, 	as the determinant of a symmetric positive semidefinite matrix is upper bounded by the product of its diagonal elements, and we take $l\cdot\Memd$ multiples of  the largest possible diagonal element to further upper bound this product. In step (d) we used \cite[Lemma 11.6]{GolubMeurant:MatricesMomentsandQuadrature} and the fact that for any positive $x$ it holds that $\log(x)=\ln(x)\cdot\log(e)\le x\cdot \log(e)$; step (e) follows since $\Big(\cov(\Wvec_2|\tau_{n,l})-\cov(\Wvec_2,\Wvec_1|\tau_{n,l})\big(\cov(\Wvec_1|\tau_{n,l})\big)^{-1}\cov(\Wvec_1,\Wvec_2|\tau_{n,l})\Big)^{-1}$ is the  upper-left block of the inverse covariance matrix of the vector $(\Wvec_2^T,\Wvec_1^T)^T$, namely, the upper-left block of   $\Big(\cov\big((\Wvec_2^T,\Wvec_1^T)^T|\tau_{n,l}\big)\Big)^{-1}$, \cite[Eqn.  (0.7.3.1)]{horn2012matrix}. Then, having more elements can only increase the maximum. In step (f) we upper bound the largest diagonal element by the matrix $1$-norm, as in step (a) in the derivation of \eqref{eqn:UpperBndMaxEig}; in step (g) we use the fact that permutation matrices are orthogonal, hence,
\[
    \big(\Pmat\cdot \Cmat_{W_n^{l\cdot (k+\Memd)}}(\tau_{n,l}) \cdot \Pmat^T\big)^{-1}=\big(\Pmat^T\big)^{-1}\cdot \big(\Cmat_{W_n^{l\cdot (k+\Memd)}}(\tau_{n,l})\big)^{-1} \cdot \Pmat^{-1}=\Pmat\cdot \big(\Cmat_{W_n^{l\cdot (k+\Memd)}}(\tau_{n,l})\big)^{-1} \cdot \Pmat^T,
\]
and  the fact that induced matrix norms are sub-multiplicative \cite[Ch. 5.6 and Example 5.6.4]{horn2012matrix}, thereby $\norm{\Pmat\cdot \big(\Cmat_{W_n^{l\cdot (k+\Memd)}}(\tau_{n,l})\big)^{-1} \cdot \Pmat^T}_{1}\le \norm{\Pmat}_{1} \cdot \norm{\big(\Cmat_{W_n^{l\cdot (k+\Memd)}}(\tau_{n,l})\big)^{-1}}_{1}\cdot \norm{ \Pmat^T}_{1} = \norm{\big(\Cmat_{W_n^{l\cdot (k+\Memd)}}(\tau_{n,l})\big)^{-1}}_{1}$, where the last equality follows since for a permutation matrix $\Pmat$, $\norm{\Pmat}_{1}=\norm{\Pmat^T}_{1}=1$. Lastly, (h) follows similarly to step (c) in the derivation of \eqref{eqn:max_eig_val_eps}. Using the definition of $\gamma(k)$ in \eqref{eqn:gammadef} we obtain 
\[
    h(\Yvec_2|\Yvec_1, \tau_{n,l})  - h(\Wvec_2|\Wvec_1, \tau_{n,l}) \le l\cdot \Memd\cdot \gamma(k).
\]
\end{sloppypar}

With this bound, letting $E_W \triangleq \frac{1}{\Tc}\int_{t=0}^{\Tc}\Cwc(t+\tau_0,0){\rm d}t$ and setting  $\xi\triangleq \frac{3}{P+E_W}<\infty$,  
considering $\delta \le \frac{P+E_W}{2}$  we can write 
\begin{eqnarray}
    C_n & \le &  \frac{1}{l\cdot (k+\Memd)}\cdot I\big(\tX_n^{l\cdot (k+\Memd)}; \tY_n^{l\cdot (k+\Memd)}|\tau_{n,l}\big)  +\delta \notag \\
    & \le & \frac{1}{l\cdot (k+\Memd)}\cdot I\Big( \big\{ \tX_{n,i\cdot\Memd + (i-1)\cdot k }^{i\cdot(k + \Memd)-1} \big\}_{i=1}^l; \big\{ \tY_{n,i\cdot\Memd + (i-1)\cdot k }^{i\cdot(k + \Memd)-1} \big\}_{i=1}^l\Big|\tau_{n,l}\Big)  +\frac{l\cdot \Memd\cdot \gamma(k)}{l\cdot (k+\Memd)}  + \delta \notag\\
    & = & \frac{1}{l\cdot (k+\Memd)}\cdot\bigg( h\Big(\big\{ \tY_{n,i\cdot\Memd + (i-1)\cdot k }^{i\cdot(k + \Memd)-1} \big\}_{i=1}^l\Big|\tau_{n,l}\Big) \notag\\
    &  & \qquad\qquad    
    -h\Big( \big\{ \tY_{n,i\cdot\Memd + (i-1)\cdot k }^{i\cdot(k + \Memd)-1} \big\}_{i=1}^l\Big|\big\{ \tX_{n,i\cdot\Memd + (i-1)\cdot k }^{i\cdot(k + \Memd)-1} \big\}_{i=1}^l,\tau_{n,l}\Big) \bigg)
    +\frac{ \Memd\cdot \gamma(k)}{k+\Memd}  + \delta \notag\\
    & = & \frac{1}{l\cdot (k+\Memd)}\cdot \bigg( h\Big(\big\{ \tY_{n,i\cdot\Memd + (i-1)\cdot k }^{i\cdot(k + \Memd)-1} \big\}_{i=1}^l\Big|\tau_{n,l}\Big) 
    -h\Big( \big\{ W_{n,i\cdot\Memd + (i-1)\cdot k }^{i\cdot(k + \Memd)-1} \big\}_{i=1}^l\Big|\tau_{n,l}\Big) \bigg)
    +\frac{ \Memd\cdot \gamma(k)}{k+\Memd}  + \delta \notag\\
    & \stackrel{(a)}{\le} & \frac{1}{l\cdot (k+\Memd)}\cdot \sum_{i=1}^l\bigg( h\Big(\tY_{n,i\cdot\Memd + (i-1)\cdot k }^{i\cdot(k + \Memd)-1}\Big|\tau_i\Big) 
    -h\Big( W_{n,i\cdot\Memd + (i-1)\cdot k }^{i\cdot(k + \Memd)-1}\Big|\tau_i\Big) \bigg)
    +\frac{ \Memd\cdot \gamma(k)}{k+\Memd}  + \delta \notag\\
    & = & \frac{1}{l\cdot (k+\Memd)}\cdot \sum_{i=1}^l\bigg( h\Big(\tY_{n,i\cdot\Memd + (i-1)\cdot k }^{i\cdot(k + \Memd)-1}\Big|\tau_i\bigg) 
    -h\Big( \tY_{n,i\cdot\Memd + (i-1)\cdot k }^{i\cdot(k + \Memd)-1}\Big|
    \tX_{n,i\cdot\Memd + (i-1)\cdot k }^{i\cdot(k + \Memd)-1}, \tau_i\Big) \bigg)
     \notag\\
    & & \qquad\qquad\qquad\qquad\qquad\qquad\qquad\qquad\qquad\qquad\qquad\qquad\qquad\qquad
    +\frac{ \Memd\cdot \gamma(k)}{k+\Memd} + \delta \notag\\
    \label{eqn:upper_bound_Cn_Step1}
    & = & \frac{1}{l\cdot (k+\Memd)}\cdot \sum_{i=1}^l I\Big(\tX_{n,i\cdot\Memd + (i-1)\cdot k }^{i\cdot(k + \Memd)-1};\tY_{n,i\cdot\Memd + (i-1)\cdot k }^{i\cdot(k + \Memd)-1}\Big|\tau_i\Big) 
    +\frac{ \Memd\cdot \gamma(k)}{k+\Memd}  + \delta \\
    & \stackrel{(b)}{\le} & \frac{1}{k+\Memd}\cdot \max_{\Big\{\substack{\cdf{\bXnk}:\;\frac{1}{k}\sum_{i=0}^{k-1}\dsE\{(\bXn[i])^2\}\le P, \\ \bar{\tau}_0\in[0,\Tc]}\Big\}} I\big(\bXnk;\bYnk\big|\bar{\tau}_0\big) +\frac{ \Memd\cdot \gamma(k)}{k+\Memd}  + \delta\cdot(1+2\cdot \xi) \notag\\
    & \stackrel{(c)}{=} & \frac{k}{k+\Memd}\cdot \frac{1}{k}I(\Xnkopt;\Ynk|\taunk\opt) + \frac{ \Memd\cdot \gamma(k)}{k+\Memd}   + \delta\cdot(1+2\cdot \xi) \notag\\
    \label{eqn:upper bound on Cn}
    & \stackrel{(d)}{\le} &  \frac{1}{k}I(\Xnkopt;\Ynk|\taunk\opt) + 2\delta\cdot (1+ \xi),
\end{eqnarray}
where in (a) $\tau_i$ denotes the sampling phase of the $i$-th block, and we also use the fact that conditioning decreases the differential entropy \cite[Corollary on Pg. 253]{cover2006elements}, and the fact that the noise process $W_n[i]$ has a memory of $\Memd$; 
in (b)  $\bY_n[i] = \bX_n[i] +W_n[i]$.
The bound is obtained as follows: The mutual information in \eqref{eqn:upper_bound_Cn_Step1} corresponds to the sum of the mutual information of $l$ blocks, each consisting of $k$ symbols. Note that the period of the noise in this additive Gaussian \ac{wscs} noise channel is $p_n$, and each period of $p_n$ samples consists of $n_k\triangleq p_n/k$ independent\footnote{Note that if $p_n/k$ is not an integer we can discard the samples of the  $\lceil  p_n/k \rceil$-th $k$-block such that the overall number of $k$ blocks in every period is  $n_k\triangleq \lfloor  p_n/k \rfloor$, and the remaining samples required to achieve $p_n$ samples  are set to zero. Asymptotically, as $n$ increases then $p_n$ increases, and  such an omission does not affect the rate.} \ac{mimo} subchannels, each of size $k\times k$.
  Applying the \ac{dcd}, as in the proof of \cite[Thm. 1]{shlezinger2015capacity} we obtain an equivalent $(n_k\cdot k) \times (n_k\cdot k)$ \ac{mimo} channel:
\begin{equation}
    \label{eqn:upper_bound_Cn_Step2}
    \tY^{(n_k\cdot k)}[i] = \tX^{(n_k\cdot k)}[i] + \tW_n^{(n_k\cdot k)}[i],
\end{equation}
where $\tW_n^{(n_k\cdot k)}[i]$ is a memoryless stationary noise process.
Then, as argued in \cite[Sec, I.D]{hirt1988capacity}, capacity subject to the power constraint \eqref{eqn:AsyncConst1} is equivalent to capacity subject to a (\ac{mimo}) per-symbol average power constraint, i.e.,
\[  
    \Tr\Big\{\tX^{(k\cdot n_k)} \cdot \big(\tX^{(k\cdot n_k)}\big)^T\Big\} < n_k \cdot k\cdot  P.
\]

\begin{sloppypar}

Next, observe that the capacity of the channel \eqref{eqn:upper_bound_Cn_Step2} subject to the above power constraint is obtained by waterfilling over the eigenvalues, see e.g., \cite[Eqn. (15)-(16)]{cover1989gaussian}. As by condition \eqref{eqn:cond_P}, $P>  \mathop{\max}\limits_{t\in [0,\Tc]}\Big( \Cwc(t,0)+\Memd\cdot\mathop{\max}\limits_{|\lambda|>\frac{\Tc}{p}}\big\{|\Cwc(t, \lambda)|\big\}\Big)$,
then, by \cite[Corollary 6.1.5]{horn2012matrix} it follows that that $P>\maxEig\big\{\Cmat_{W_n^{(\tk)}}(\tau_0)\big\}$ for any $\tk\in\mNp$. This implies that the waterfilling solution will use all the eigenvalues of the noise correlation matrix. Recall that by the selection of $n$, see \eqref{eqn:n-selec-cond-3},
for each $k$-block the trace of the noise correlation matrix is within $\delta$ from $E_W$. Then, letting $\Cmat_{\tW_n^{(n_k\cdot k)}}(\tau_0) \triangleq \dsE\Big\{\tW_n^{(n_k\cdot k)}\cdot \big(\tW_n^{(n_k\cdot k)}\big)^T\Big|\tau_0\Big\}$ we obtain from \eqref{eqn:n-selec-cond-3} that 
\[
    \Big|\Tr\Big\{\Cmat_{\tW_n^{(n_k\cdot k)}}(\tau_0)\Big\} - n_k \cdot k \cdot E_W\Big|<n_k\cdot k \cdot \delta.
\] The  waterfilling rate for this case is then 
\begin{align}
    R_{tot} &=\frac{1}{k\cdot n_k}\log\left(\frac{\Big(P+\frac{1}{n_k\cdot k}\Tr\Big\{\Cmat_{\tW_n^{(n_k\cdot k)}}(\tau_0)\Big\}\Big)^{k\cdot n_k}}{\Det\Big(\Cmat_{\tW_n^{(n_k\cdot k)}}(\tau_0)\Big)}\right)
    \label{eqn:sum_capacity_waterfilling}\\
    &\le \frac{1}{k\cdot n_k}\log\left(\frac{\Big(P+E_W + \delta\Big)^{k\cdot n_k}}{\Det\Big(\Cmat_{\tW_n^{(n_k\cdot k)}}(\tau_0)\Big)}\right)\notag\\
    &=\log\left(\frac{P+E_W + \delta}{\sqrt[k\cdot n_k]{\Det\Big(\Cmat_{\tW_n^{(n_k\cdot k)}}(\tau_0)\Big)}}\right).\label{eqn:upper_bound_in_rate_step2}
\end{align}

\end{sloppypar}

Due to the discarding of $\Memd$ samples every $k$ samples, the $n_k$ $k$-blocks are independent, thus,
\eqref{eqn:upper_bound_Cn_Step2} consists of $n_k$ parallel \ac{mimo} subchannels. The waterfilling solution to channel $i_c$ within this set results in a rate of 
\[  
    R_{i_c} = \frac{1}{k}\log\left(\frac{\Big(P+\frac{1}{ k}\Tr\Big\{\Cmat_{\tW_n^{(k)}}(\tau_{i_c})\Big\}\Big)^{k}}{\Det\Big(\Cmat_{\tW_n^{(k)}}(\tau_{i_c})\Big)}\right) 
    \le \log\left(\frac{P+E_W +\delta}{\sqrt[k]{\Det\Big(\Cmat_{\tW_n^{(k)}}(\tau_{i_c})\Big)}}\right)
\]
Then, the sum-rate over the $n_k$ $k$-blocks can be upper bounded as
\[
    \frac{1}{n_{k}}\sum_{i_c=1}^{n_k}R_{i_c} 
    \le
    \log\left(\frac{P+E_W +\delta}{\sqrt[n_k]{\prod_{i_c=1}^{n_k}\sqrt[k]{\Det\Big(\Cmat_{\tW_n^{(k)}}(\tau_{i_c})\Big)}}}\right)
    =  \log\left(\frac{P+E_W + \delta}{\sqrt[k\cdot n_k]{\Det\Big(\Cmat_{\tW_n^{(n_k\cdot k)}}(\tau_0)\Big)}}\right),
\]
which coincides with the upper bound in \eqref{eqn:upper_bound_in_rate_step2}. 
Finally, recalling that $\delta\le\frac{P+E_W}{2}$, then, as we are interested in bounding the distance between the upper bound and  \eqref{eqn:sum_capacity_waterfilling}, we consider
\begin{align*}
    &\log\left(P+E_W + \delta\right) -\log\left(P+E_W - \delta\right)\notag\\
    &\qquad = \log\left((P+E_W)\left(1+\frac{\delta }{P+E_W}\right)\right) - \log\left((P+E_W)\left(1-\frac{\delta }{P+E_W}\right)\right)\notag\\ 
    &\qquad = \log(e)\cdot \ln\left(1+\frac{\delta }{P+E_W}\right) - \log(e)\cdot \ln\left(1-\frac{\delta }{P+E_W}\right)\notag\\
    & \qquad \stackrel{(a')}{\le} \log(e)\cdot \left(\frac{\delta }{P+E_W} - \frac{-\frac{\delta }{P+E_W}} {1-\frac{\delta }{P+E_W}}\right) \notag\\
    &\qquad = \delta\cdot \log(e)\cdot \left(\frac{1}{P+E_W}+\frac{1}{P+E_W-\delta}\right)\\
    & \qquad \stackrel{(b')}{\le}  \delta\cdot \log(e)\cdot \frac{3}{P+E_W} \\
    & \qquad <2\cdot \delta\cdot  \xi,
\end{align*}
where in (a$'$) we used 
$\frac{x}{1+x}<\ln(1+x)<x$, 
$\forall x>-1$, which holds for $\delta \le \frac{P + E_W}{2}$; and (b$'$) follows as $\delta \le \frac{P + E_W}{2}$.
Thus, Step (b) in the derivation of \eqref{eqn:upper bound on Cn} follows as we can upper bound the capacity of the channel in \eqref{eqn:upper_bound_Cn_Step1} with the capacity of the best $k$-block subchannel, with a maximum error of $2\cdot\delta\cdot\xi$, and we further maximize the mutual information over the initial phase; 
\label{pg:modify4_due_to_power}
Step (c) follows as the maximization in Step (b) coincides with \eqref{eqn:defOptSolCorr_Xeps}.
As detailed in Section \ref{appndx:convergence of mutual informations}, we use $\taunk\opt$ to denote the optimal sampling phase, $\Xnkopt$ to denote the corresponding optimal input vector, and $\Ynk$ to denote the corresponding channel output, for the channel \eqref{eqn:AsnycModel2}. 
Step (d) follows  from our choice of $k\in\mNp$. 

Lastly, we analyze the mutual information density rate of the scheme considered in Subsections \ref{par:Tx-Scheme}-\ref{par:Rx-scheme}. For any $\delta\in\mR$ s.t. $0<\delta \le \frac{P+E_W}{2}$, we can select $k$, $n$ and $l$ s.t.
\begin{eqnarray*}
	& & \hspace{-1cm}\Pr\Big(\zkeps{l\cdot k}{\eps}\big( \cdf{X_{\rm opt}^{(l\cdot k)}|\tauepsk\opt}|\tauepsk\opt\big) \le \liminfcn - \delta\cdot(5+2\cdot\xi)\Big)\\
	& \stackrel{(a)}{\le} & \Pr\Big( \zkeps{l\cdot k}{\eps}\big( \cdf{X_{\rm opt}^{(l\cdot k)}|\tauepsk\opt}|\tauepsk\opt\big) \le C_n - \delta\cdot (4+2\cdot\xi)\Big)\\
	& \stackrel{(b)}{\le} & \Pr\Big( \zkeps{l\cdot k}{\eps}\big( \cdf{X_{\rm opt}^{(l\cdot k)}|\tauepsk\opt}|\tauepsk\opt\big) \le    \frac{1}{k}I(\Xnkopt;\Ynk|\taunk\opt) - 2\delta\Big)\\
	& \stackrel{(c)}{\le} & \Pr\Big( 
	\zkeps{l\cdot k}{\eps}\big( \cdf{X_{\rm opt}^{(l\cdot k)}|\tauepsk\opt}|\tauepsk\opt\big)
	\le  \frac{1}{k}I\big(\Xinkopt;\Yepsk\big|\tauepsk\opt\big) - \delta\Big)\\
	& \stackrel{(d)}{\le} & \delta,
\end{eqnarray*}
where (a) follows as $\liminf\limits_{n_0\rightarrow\infty}C_{n_0} < C_n + \delta$ by the selection of $n\in\mNp$; (b) follows from  \eqref{eqn:upper bound on Cn} and our selection of $k$, $n$ and $l$. Step (c) follows as  the mutual information expressions $\frac{1}{k}I(\Xnkopt;\Ynk|\taunk\opt)$ and $\frac{1}{k} I(\Xinkopt;\Yepsk|\tauepsk\opt)$ are maximized by Gaussian inputs \cite[Eqns. (4), (30)]{cover1989gaussian}. Then, due to Lemma \ref{lem:tightness}, for the fixed $k\in\mNp$, $n$ can be selected such that \eqref{eqn:mutual_information_bounded_distance} is satisfied. 	Lastly, step (d) follows from the bound in Eqn. \eqref{eqn:Bounding_PrZepsk_LemmaB}, as $l\in\mNp$ can be selected arbitrarily large. Consequently, we conclude that 
\[
    \lim_{l\rightarrow\infty}\Pr\Big(\zkeps{l\cdot k}{\eps}\big( \cdf{X_{\rm opt}^{(l\cdot k)}|\tauepsk\opt}|\tauepsk\opt\big) < \liminfcn \Big)=0.
\]
By the rate expression in Section  \ref{par:Tx-Scheme} we obtain that a rate of $\liminfcn\cdot \big(1-\frac{\Memd+\Dg/\Tsamp(\eps)}{k+\Memd+\Dg/\Tsamp(\eps)}\big)$ is achievable. Note that fixing $k$ sufficiently large, we can approach $\liminfcn$ arbitrarily close. As a final comment, we note that we considered blocklengths which are integer multiple of $k$. Since $k$ is fixed, then by taking $l$ sufficiently large, we can use zero-padding of up to $k-1$ zeros and obtain a code with any blocklength, causing only an arbitrarily small rate decrease.
	
It thus follows that $\liminfcn$ is both achievable and, by \eqref{appndxEqn:UpperBound}, it is the upper bound on the achievable rate, hence it is the capacity for the situation in which the transmitter can select the most appropriate sampling phase for transmission.
\end{IEEEproof}


\end{appendices}
\bibliographystyle{IEEEtran}
\bibliography{myreference.bib}

\end{document}

%% file: preamble.tex
\newcommand{\E}{\mathds{E}}

\newcommand{\myVec}[1]{{\boldsymbol{#1}}}
\newcommand{\myMat}[1]{\mathsf{#1}}
\newcommand{\myMatrix}[2]{\mathsf{#1}_{#2}}
\newcommand{\mySet}[1]{\mathcal{#1}}
\newcommand{\Cmat}{\mathsf{C}}
\newcommand{\Pmat}{\mathsf{P}}

\newcommand{\Amat}{\mathsf{A}}

\newcommand{\Bmat}{\mathsf{B}}
\newcommand{\Dmat}{\mathsf{D}}
\newcommand{\mB}{\mathcal{B}}
\newcommand{\mC}{\mathcal{C}}

\newcommand{\norm}[1]{\left\lVert#1\right\rVert}

\newcommand{\tGmmiepsksqr}{\big(\tilde{\Gamma}_{\eps,i,k}\big)^{2}}
\newcommand{\liminfcn}{\liminf\limits_{n\rightarrow\infty}C_n}

\newcommand{\limn}{\lim\limits_{n\rightarrow\infty}}
\newcommand{\Rs}{R_c}

\newcommand{\bX}{\bar{X}}

\newcommand{\Xink}{{X^{(k)}}}

\newcommand{\xin}{{x}}
\newcommand{\Yi}{{Y}}

\newcommand{\Xnkopt}{X_{n,\mbox{\scriptsize \rm opt}}^{(k)}}
\newcommand{\Xinkopt}{X_{\mbox{\scriptsize \rm opt}}^{(k)}}

\newcommand{\PCst}{P}
\newcommand{\opt}{^{\rm opt}}

\newcommand{\zkeps}[2]{Z_{#1,#2}}
\newcommand{\zk}[1]{Z_{#1}}

\newcommand{\iden}[1]{\mathsf{I}_{#1}}
\newcommand{\alzer}[1]{\mathsf{0}_{#1}}
\newcommand{\matR}[1]{\mathsf{R}_{#1}}

\newcommand{\sfc}{\mathsf{c}}

\newcommand{\dsN}{\mathds{N}}

\newcommand{\mN}{\mathbb{N}}
\newcommand{\mNp}{\mathbb{N}^{+}}

\newcommand{\mZ}{\mathbb{Z}}
\newcommand{\mK}{\mathcal{K}}
\newcommand{\mA}{\mathcal{A}}

\newcommand{\mR}{\mathbb{R}}
\newcommand{\mX}{\mathcal{X}}
\newcommand{\mS}{\mathcal{S}}
\newcommand{\mQ}{\mathbb{Q}}
\newcommand{\mU}{\mathcal{U}}

\newcommand{\mT}{\mathcal{T}}

\newcommand{\evec}{\mathbf{e}}
\newcommand{\wvec}{\mathbf{w}}
\newcommand{\xvec}{\mathbf{x}}
\newcommand{\Xvec}{\mathbf{X}}
\newcommand{\Yvec}{\mathbf{Y}}
\newcommand{\Wvec}{\mathbf{W}}

\newcommand{\tDelta}{\tilde{\Delta}}

\newcommand{\tX}{\tilde{X}}

\newcommand{\bXk}{\bar{X}^{(k)}}
\newcommand{\bXn}{\bar{X}_n}
\newcommand{\bXnk}{\bar{X}_n^{(k)}}

\newcommand{\tW}{\tilde{W}}
\newcommand{\tY}{\tilde{Y}}
\newcommand{\bY}{\bar{Y}}
\newcommand{\bYnk}{\bar{Y}_n^{(k)}}

\newcommand{\ttnepsk}{\dbtilde{k}_{\eps,k}}
\newcommand{\tkx}{\tilde{k}_{X,k}}
\newcommand{\talpha}{\tilde{\alpha}}

\newcommand{\liminfk}{\mathop{\lim\!\inf}\limits_{k\rightarrow\infty}}
\newcommand{\liminfn}{\mathop{\lim\!\inf}\limits_{n\rightarrow\infty}}

\newcommand{\dbtilde}[1]{\accentset{\approx}{#1}}
\renewcommand{\liminf}{\mathop{\lim\!\inf}}
\newcommand{\cov}{\mbox{cov}}
\newcommand{\logdet}{\mbox{logdet}}
\newcommand{\Dg}{\Delta_g}
\newcommand{\tauepsk}{\tau_{\eps,k}}
\newcommand{\taunk}{\tau_{n,k}}

\newcommand{\btauepsk}{\bar{\tau}_{\eps,k}}
\newcommand{\Det}{{ \rm Det}}
\newcommand{\maxEig}{{ \rm maxEig}}
\newcommand{\maxDiag}{{ \rm maxDiag}}
\newcommand{\argmax}{{ \rm argmax}}
\newcommand{\Cmatxk}{{\Cmat}_{X^{(k)}}}
\newcommand{\Cmatxnk}{{\Cmat}_{X_n^{(k)}}}
\newcommand{\Cmatxepsk}{{\Cmat}_{X_{\eps}^{(k)}}}
\newcommand{\mCxk}{\mathcal{C}_{X^{(k)}}}

\newcommand{\myEps}{\epsilon}
\newcommand{\eps}{\epsilon}

\newcommand{\Wc}{W_{\rm c}}
\newcommand{\Weps}{W_{\myEps}}

\newcommand{\var}{\mbox{var}}

\newcommand{\Wnk}{W_n^{(k)}}

\newcommand{\Wepsk}{W_{\epsilon}^{(k)}}
\newcommand{\Yeps}{Y_{\myEps}}

\newcommand{\Xscal}{X}

\newcommand{\Wn}{W_{n}}

\newcommand{\Yn}{Y_{n}}

\newcommand{\Yepsk}{Y_{\epsilon}^{(k)}}

\newcommand{\bYepsk}{\bar{Y}_{\epsilon}^{(k)}}
\newcommand{\Ynk}{Y_{n}^{(k)}}

\newcommand{\Xepsk}{X^{(k)}}
\newcommand{\xepsk}{x^{(k)}}
\newcommand{\Xnk}{X_{n}^{(k)}}

\newcommand{\Cwc}{c_{\Wc}}
\newcommand{\Cweps}{c_{\Weps}}

\newcommand{\Tc}{T_{\rm pw}}
\newcommand{\Tsamp}{T_{\rm s}}

\newcommand{\Td}{p}

\newcommand{\barx}{\bar{x}}

\newcommand{\Memc}{\lambda_m}
\newcommand{\Memd}{\tau_{\rm m}}

\newcommand{\cdf}[1]{F_{#1}}

\newcommand{\pdf}[1]{p_{#1}}

\newcommand{\TDelta}{{\bar{\Delta}}}

\newcommand{\TLambda}{\Lambda'}
\newcommand{\Capacity}{C}
\newcommand{\tk}{\tilde{k}}

\newcommand{\Tr}{\mathrm{Tr}}

\newcommand{\corrmatwe}{\Cmat_{\Wepsk}}
\newcommand{\corrmatye}{\Cmat_{\Yepsk}}
\newcommand{\corrmatxe}{\Cmat_{\Xepsk}}
\newcommand{\corrmatxeopt}{\Cmat_{\Xinkopt}}

\newcommand{\corrmatwn}{\Cmat_{\Wnk}}

\newcommand{\corrmatxn}{\Cmat_{\Xnk}}

\newcommand{\ttCepsk}{\tilde{\tilde{\Cmat}}_{\eps}^{(k)}}

\newcommand{\tvarepsk}{\tilde{\sigma}^2_{\eps,k}}
\newcommand{\dsE}{\mathds{E}}
\newcommand{\contcorr}{c_{W_{\sfc}}}
\newcommand{\disccorr}{c_{W_{\myEps}}}
\newcommand{\disccorrn}{c_{W_{n}}}
\newcommand{\DC}{t_{\rm dc}}
\newcommand{\Trise}{t_{\rm rf}}

\newcommand{\EqDist}{\mathop{=}\limits^{(dist.)}}
\newcommand\independent{\protect\mathpalette{\protect\independenT}{\perp}}
\def\independenT#1#2{\mathrel{\rlap{$#1#2$}\mkern2mu{#1#2}}}
\long\def\symbolfootnote[#1]#2{\begingroup\def\thefootnote{\fnsymbol{footnote}}\footnote[#1]{#2}\endgroup}

\acrodef{sdd}[SDD]{strictly diagonally dominant}
\acrodef{psd}[PSD]{power spectral density}
\acrodef{bps}[bps]{bits per second}
\acrodef{Mbps}[Mbps]{megabits per second}
\acrodef{cdf}[CDF]{cumulative distribution function}
\acrodef{dac}[DAC]{digital-to-analog converter}
\acrodef{adc}[ADC]{analog-to-digital converter}
\acrodef{pdf}[PDF]{probability density function}
\acrodef{rv}[RV]{random variable}
\acrodef{dt}[DT]{discrete-time} 
\acrodef{ct}[CT]{continuous-time} 
\acrodef{sss}[SSS]{strict-sense stationary}
\acrodef{sscs}[SSCS]{strict-sense cyclostationary}
\acrodef{wss}[WSS]{wide-sense stationary}
\acrodef{wscs}[WSCS]{wide-sense cyclostationary}
\acrodef{dc}[DC]{duty cycle}
\acrodef{wsacs}[WSACS]{wide-sense almost cyclostationary}
\acrodef{ofdm}[OFDM]{orthogonal frequency division multiplexing}
\acrodef{tdma}[TDMA]{time division multiple access}
\acrodef{cdma}[CDMA]{code division multiple access}
\acrodef{noma}[NOMA]{non-orthogonal multiple access}
\acrodef{asmcgn}[AS-MCGNC]{asynchronously-sampled memoryless cyclostationary Gaussian noise channel}
\acrodef{psdm}[P.S.D]{positive semidefinite}
\acrodef{cmt}[CMT]{continuous mapping theorem}
\acrodef{ecmt}[ECMT]{extended continuous mapping theorem}
\acrodef{cs}[CS]{cyclostationary}
\acrodef{dcd}[DCD]{decimated component decomposition}
\acrodef{iid}[i.i.d]{independent and identically distributed}
\acrodef{rdf}[RDF]{rate-distortion function}
\acrodef{mimo}[MIMO]{multiple input-multiple output}
\acrodef{siso}[SISO]{single-input single-output}
\acrodef{awgn}[AWGN]{additive white Gaussian noise}
\acrodef{csi}[CSI]{channel state information}
\acrodef{snr}[SNR]{signal to noise power ratio}
\acrodef{mac}[MAC]{multiple access channel}
\acrodef{bc}[BC]{broadcast channel}
\acrodef{su}[SU]{single-user}
\acrodef{cp}[CP]{cyclic prefix}
\acrodef{qam}[QAM]{quadrature amplitude modulation}
\acrodef{mu}[MU]{multi-user}
\acrodef{plc}[PLC]{power line communications}
\acrodef{jscc}[JSCC]{joint source-channel coding}
\acrodef{aacs}[ACS]{almost cyclostationary}
\acrodef{cf}[CF]{compress-and-forward}
\acrodef{psd}[PSD]{power spectral density}
\acrodef{cpsd}[CPSD]{cyclic power spectral density}
\acrodef{tpsd}[TPSD]{time-varying power spectral density}
\acrodef{ptp}[PtP]{point-to-point}
\acrodef{lti}[LTI]{linear, time-invariant}
\acrodef{soi}[SOI]{signal-of-interest}
\acrodef{caf}[CAF]{cyclic autocorrelation function}
\acrodef{osi}[OSI]{open systems interconnection}
\acrodef{lptv}[LPTV]{linear periodically time-varying}
\acrodef{dtgc}[DTGC]{discrete time Gaussian channel}
\acrodef{isi}[ISI]{inter-symbol interference}
\acrodef{ncgc}[NCGC]{N-circular Gaussian channel}
\acrodef{ofdma}[OFDMA]{orthorgonal frequency division multiple access}
\acrodef{dsl}[DSL]{digital subscriber line}
\acrodef{dms}[DMS]{discrete memoryless source}
\acrodef{dmc}[DMC]{discrete memoryless channel}
\acrodef{aep}[AEP]{asymptotic equipartition property}
\acrodef{caf}[CAF]{cyclic autocorrelation function}
\acrodef{ic}[IC]{interference cancellation}
\acrodef{sic}[SIC]{successive interference cancellation}
\acrodef{ia}[IA]{interference alignment}
\acrodef{marc}[MARC]{multiple-access relay channels}
\acrodef{mabrc}[MABRC]{multiple-access broadcast relay channels}
\acrodef{pam}[PAM]{pulse amplitude modulation}
\acrodef{rhs}[RHS]{right-hand side}
\acrodef{lhs}[LHS]{left-hand side}
\acrodef{mse}[MSE]{mean squared error}
\acrodef{cdf}[CDF]{cumulative distribution function}
\acrodef{pdf}[PDF]{probability density function}
\acrodef{rv}[RV]{random variable}
\acrodef{dt}[DT]{discrete-time} 
\acrodef{ct}[CT]{continuous-time} 
\acrodef{wscs}[WSCS]{wide-sense cyclostationary}
\acrodef{wsacs}[WSACS]{wide-sense {\em almost} cyclostationary}
\acrodef{ofdm}[OFDM]{orthogonal frequency division multiplexing}
\acrodef{tdma}[TDMA]{time division multiple access}
\acrodef{cdma}[CDMA]{code division multiple access}
\acrodef{noma}[NOMA]{non-orthogonal multiple access}
\acrodef{asmcgn}[AS-MCGNC]{asynchronously-sampled memoryless cyclostationary Gaussian noise channel}
\acrodef{cmt}[CMT]{continuous mapping theorem}
\acrodef{cir}[CIR]{channel impulse response}
\acrodef{lpf}[LPF]{lowpass filter}
\acrodef{acgn}[ACGN]{additive \ac{wscs} Gaussian noise}
\acrodef{tf}[TF]{transfer function}